\documentclass[preprint2]{aastex}

\usepackage{natbib}
\usepackage{graphicx} 
\usepackage{subfigure}

\bibpunct{(}{)}{;}{a}{}{,}

\shorttitle{\texttt{MASSCLEAN} - MASSive CLuster Evolution and
ANalysis Package} \shortauthors{B. Popescu & M.M. Hanson}

\begin{document}


\title{\texttt{MASSCLEAN} - MASSive CLuster Evolution and ANalysis Package - Description and Tests}

\normalsize

\author{Bogdan Popescu and M. M. Hanson}
\affil{University of Cincinnati, Physics Department, 400 G/P, PO Box 210011, Cincinnati, OH 45221-0011}
\email{popescb@email.uc.edu, margaret.hanson@uc.edu}

\begin{abstract}

We present \texttt{MASSCLEAN}, a new, sophisticated and robust stellar cluster image and photometry simulation package. This visualization tool is able to create color-magnitude diagrams and standard FITS images in any of the traditional optical and near-infrared bands based on cluster characteristics input by the user, including but not limited to distance, age, mass, radius and extinction.  At the limit of very distant, unresolved clusters, we have checked the integrated colors created in \texttt{MASSCLEAN} against those from other simple stellar population models with consistent results.  We have also tested models which provide a reasonable estimate of the field star contamination in images and color-magnitude diagrams.  We demonstrate the package by simulating images and color-magnitude diagrams of well known massive Milky Way clusters and compare their appearance to real data. Because the algorithm populates the cluster with a discrete number of tenable stars, it can be used as part of a Monte Carlo Method to derive the probabilistic range of characteristics (integrated colors, for example) consistent with a given cluster mass and age.  Our simulation package is available for download and will run on any standard desktop running UNIX/Linux. Full documentation on installation and its use is also available. Finally, a web-based version of \texttt{MASSCLEAN} which can be immediately used and is sufficiently adaptable for most applications is available through a web interface.

\vskip 0.5cm
\small
\noindent
Submitted to {\it Astronomical Journal}
\normalsize
\vskip 0.5cm

\end{abstract}


\section{Introduction}

Stellar clusters provide among the most critical observational benchmarks for testing the physics of stellar and galactic evolution and galactic structure.  Stellar clusters are often exploited by assuming their constituent stars all formed from the same interstellar material (similar initial chemistry) and at the same time (same age). \citeauthor*{tinsley1978} \citeyearpar{tinsley1978} pioneered the application of using stellar evolutionary codes to model the observed characteristics of unresolved stellar clusters and galaxies, referred to as evolutionary population synthesis (\citeauthor*{tinsley1976} \citeyear{tinsley1976}). The present day version of such methods is seen in the application of {\it simple stellar population} (hereafter SSP) models to unresolved systems.

Owing to their great utility, a vast array of SSP models have been developed by researchers in the past two decades: \citeauthor*{leitherer1999} \citeyear{leitherer1999}; \citeauthor*{hurley2000} \citeyear{hurley2000}; \citeauthor*{schulz2002} \citeyear{schulz2002}; \citeauthor*{bruzual2003} \citeyear{bruzual2003}; \citeauthor*{vazquez2005} \citeyear{vazquez2005}; \citeauthor*{maraston2005} \citeyear{maraston2005} to name just a few of the most widely used.  As these models are applied to a multitude of galactic and extragalactic applications, they each have slightly different techniques in how they include the input physics (beyond differences in which stellar evolutionary codes they choose).  For example, the most common method derives the emergent properties by integrating along a single stellar isochrone - {\it isochrone synthesis} - (\citeauthor*{charlot1991} \citeyear{charlot1991}), while another technique follows the cluster evolution based on {\it fuel consumption} (\citeauthor*{maraston2005} \citeyear{maraston2005}).  This can lead to perceivable differences in predicted outputs for the same input cluster characteristic (age, metallicity, stellar mass function) from one SSP code to the next, even when the same stellar evolutionary codes are used (see the tests presented in \citeauthor*{beasley2002} \citeyear{beasley2002}, \citeauthor*{pessev2008} \citeyear{pessev2008}).  

Among the variety of SSP codes already available, we are unaware of any that take advantage of the situation when individual stars are fully or at least partially resolved. For example, the GALEV evolutionary synthesis models of Fritze and collaborators (\citeauthor*{schulz2002} \citeyear{schulz2002}; \citeauthor*{galev} \citeyear{galev}; \citeauthor*{galev2009} \citeyear{galev2009}) can produce very useful synthetic CMDs for discrete stellar clusters (see Fig. A3 in \citeauthor*{galev2009} \citeyear{galev2009}) though they have not gone so far as to attempt to make simulated images of stellar clusters. The need for a stellar cluster image simulation, based on the tenets of the traditional SSP code, drove us to develop a new analysis tool. Our motivation to develop an image simulation code was to apply it to the search and analysis of deeply embedded massive open clusters lying in the inner Milky Way. However, the code can be applied to Local Group galaxies where massive clusters can be partially resolved with some telescopes.

Our visualization and analysis tool, called \texttt{MASSCLEAN}, provides image simulations and thus can be used to answer entirely different, important questions in stellar astronomy and galactic structure than previous SSP codes, both within the Galaxy and in nearby external galaxies. In this paper we present the details and current coding of the first release of our new, SSP-like imaging and photometric  simulation code, \texttt{MASSCLEAN}. In \S \ref{generaldescription}, we provide a quantitative description of the computational algorithms used in the routine. In \S \ref{sampleruns} we provide test runs of the simulations comparing the integrated colors and magnitudes against those of currently used SSP codes.  We also provide example images with an eye towards accurately modeling well known massive young clusters. In \S \ref{discussion}, we discuss both the promise and limitations of \texttt{MASSCLEAN} for a variety of astrophysical applications.

\section{General description of \texttt{MASSCLEAN}} \label{generaldescription}

Our simulation package, \texttt{MASSCLEAN}\footnote{ \texttt{MASSCLEAN} (\textbf{MASS}ive \textbf{CL}uster \textbf{E}volution and \textbf{AN}alysis) package}, uses a nominal number of input parameters: mass, initial mass function (hereafter, IMF), metallicity, extinction, distance, spatial distribution parameters and stellar field density.  The predicted characteristics are computed for a range of cluster ages from $10^6$ to a few $10^9$ years.  The user can also choose to include a parameter that allows for mass segregation as the cluster ages. Many other features are described in the next sections. 

The simulation code is built using numerous well established theoretical and empirical models for stars and stellar clusters, beginning with the Kroupa-Salpeter IMF for stellar mass distribution \citep{Kroupa2002,Salpeter1955}.  The user has the option to choose between two stellar evolution models: the Geneva Models (\citeauthor*{geneva1} \citeyear{geneva1}) or the Padova Models (\citeauthor*{padova2008}\citeyear{padova2008}).  The extinction model based on \citeauthor*{ccm89} \citeyear{ccm89}, the King Model for spatial distribution \citep{King1962} and the SKY Model for the stellar field (\citeauthor*{Wainscoat1992} \citeyear{Wainscoat1992}; \citeauthor*{cohen1994} \citeyear{cohen1994}; \citeauthor*{cohen1995} \citeyear{cohen1995}; \citeauthor*{bahcall1984} \citeyear{bahcall1984}) are further used. This is reviewed in more detail below. 

It was our intent to make \texttt{MASSCLEAN} user friendly and versatile. The package is designed to allow the user significant latitude and flexibility in how they will apply it.  \texttt{MASSCLEAN} is written as a series of independently run sub-routines, performing individual calculations at various stages.  This allows users to substitute their own inputs at any stage or to skip or perform their own calculations in lieu of those provided within the \texttt{MASSCLEAN} package. Information is passed from one routine to the next through read/write of ASCII files.  While these files become quite large (gigabyte is not uncommon for the entire output of hundred of files), this allows users of the \texttt{MASSCLEAN} routines to easily check, edit and or substitute the output at any stage in the calculations. Although it can take some time to generate images, the code for the photometric simulation is very fast. The package is freely available under GNU General Public License at \url{http://www.physics.uc.edu/\textasciitilde bogdan/}. Downloadable documentation on installing and running the code is also available. Finally, \texttt{MASSCLEAN}{\bf .web}, a web-based interface, is immediately available\footnote{\url{http://www.physics.uc.edu/\textasciitilde bogdan/massclean/}} and can be used for many basic applications.

\subsection{The Mass Distribution} \label{massdist}

   The number of stars formed in the $(M,M+\mathrm{d}M)$ range is:

   \begin{equation} \label{eq:1}
    \mathrm{d}N=\xi(M) \mathrm{d}M
   \end{equation}

    The multi-part power law $\xi(M)$ derived from Kroupa-Salpeter IMF \citep{Kroupa2002,Salpeter1955} is :
\begin{equation}
    \xi (M) = k \left \{ 
    \begin{array}{lcc} 
    \left (  \frac{M}{m_{1}} \right )^{-\alpha_{1}} &,& m_{0}<M \leq m_{1} \\
    \left (  \frac{M}{m_{1}} \right )^{-\alpha_{2}} &,& m_{1}<M \leq m_{2} \\
    \left (  \frac{m_{2}}{m_{1}} \right )^{-\alpha_{2}} \left (  \frac{M}{m_{2}} \right )^{-\alpha_{3}} &,& m_{2}<M \leq m_{3} 
    \end{array}
    \right.  
   \end{equation}
   with mass expressed in $M_{\sun}$ units and :
\begin{equation}
   \begin{array}{lcc} 
   \alpha_{1}=+0.3 \pm 0.7 &,& 0.01 \leq M/M_{\sun}<0.08 \\
   \alpha_{2}=+1.3 \pm 0.5 &,& 0.08 \leq M/M_{\sun}<0.50 \\
   \alpha_{3}=^{+2.3 \pm 0.3} _{+2.7 \pm 0.3}  &,& 0.50 \leq M/M_{\sun}<m_{3} 
   \end{array}
   \end{equation}
and $m_{3}=\infty$ or $m_{3}=m_{up}$ (for an IMF with upper mass cutoff \citep{Oey2005}).
Using $\xi (M)/k = \xi_{i} (M)$ (with $i=1,2,3$ respectively), the total mass of the cluster can be written :

\begin{equation}
    M_{total}= \int_{0}^{N_{total}} M(N) \mathrm{d}N 
   \end{equation}
   
\begin{equation}
    M_{total}=  \int_{m_{0}}^{m_{3}} M \frac{\mathrm{d}N }{\mathrm{d}M } \mathrm{d}M = \int_{m_{0}}^{m_{3}} \xi (M) M \mathrm{d}M 
   \end{equation}   
   
\begin{equation}
    M_{total}=\sum_{i=1}^{3} \left( k \int_{m_{i-1}}^{m_{i}} \xi_{i} (M) M \mathrm{d}M  \right)
\end{equation}   
   
The normalization constant :   
\begin{equation} \label{eq:2}
   k =\frac{ M_{total}} {\sum_{i=1}^{3} \left(    \int_{m_{i-1}}^{m_{i}} \xi_{i} (M) M \mathrm{d}M  \right)}
\end{equation}   

From the equations (\ref{eq:1}) and (\ref{eq:2}) : 
\begin{equation} \label{eq:3}
   N_{i}(M,M+\mathrm{d} M) =\frac{ M_{total} \int_{M}^{M+\mathrm{d} M} \xi_{i}(M) \mathrm{d}M} {\sum_{i=1}^{3} \left( \int_{m_{i-1}}^{m_{i}} \xi_{i} (M) M \mathrm{d}M  \right)}
\end{equation}

Our package uses equation (\ref{eq:3}) to compute the mass distribution of stars based on the total mass of the cluster $M_{total}$ and the IMF, described by the three $\xi_{i}(M)$. The mass range to be included in the distribution is chosen by the user. The values $\alpha_{i}$ and $m_{i}$ are also input parameters. In this way, a Kroupa IMF can become a Salpeter IMF when $\alpha_{1}=\alpha_{2}=\alpha_{3}=2.35$. The versatility of the package allows us to use a first-order, second-order or third-order power law model (Kroupa-Salpeter type) for the IMF. The use of a truncated IMF is optional. All the mass bins $(M,M+\mathrm{d} M)$ are computed such that the value of $N_{i}(M,M+\mathrm{d} M)$ is an integer (up to some tolerance, also chosen by the user).  The program allows fluctuations in the computed mass distribution, such that the distribution is different for every run. The user can turn this feature on or off.

\begin{figure}[ht] 
\begin{center}
\subfigure[$\alpha_{1}=0.3$, $\alpha_{2}=1.3$, $\alpha_{3}=2.3$]{\includegraphics[angle=270,width=6cm]{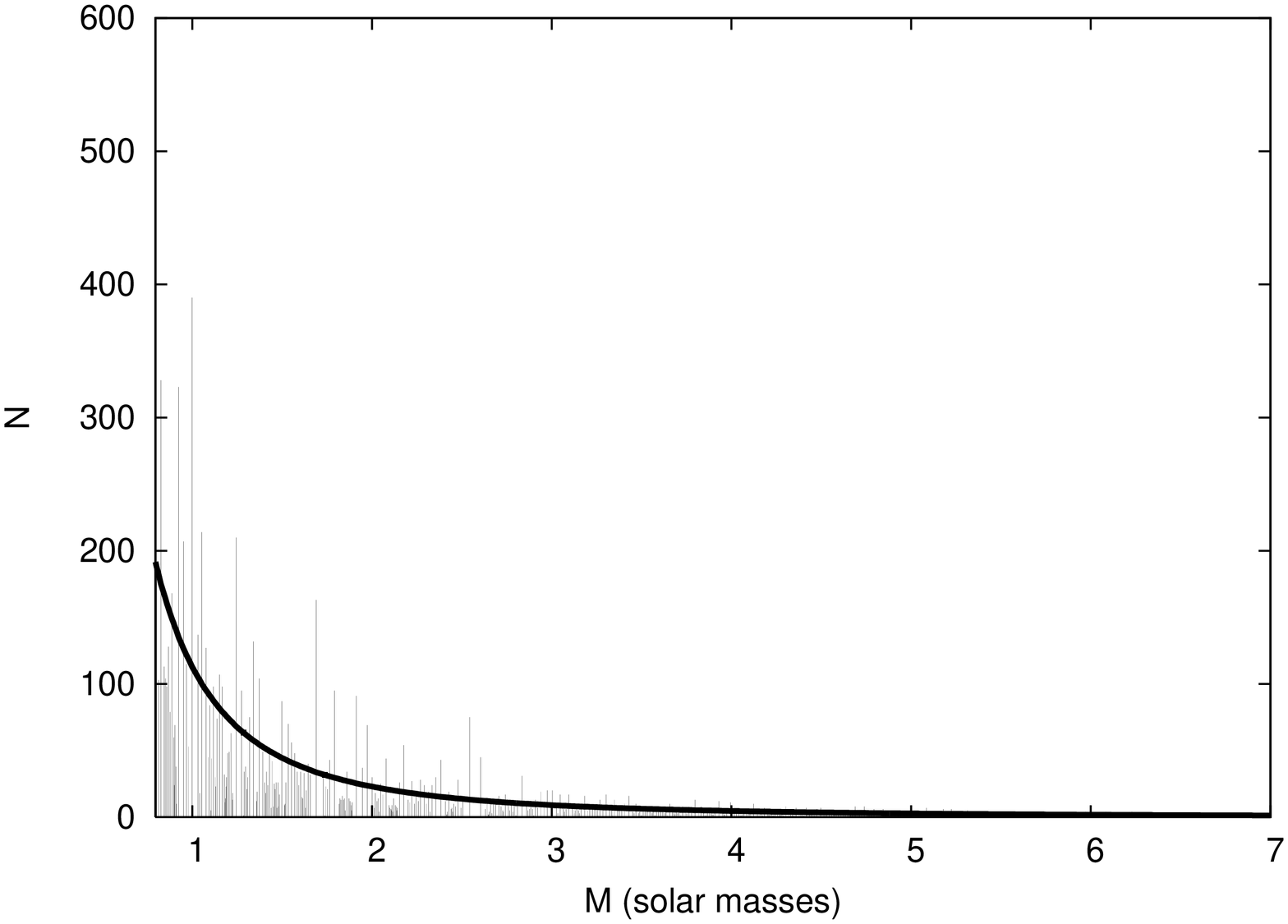}} 
\subfigure[$\alpha_{1}=0.3$, $\alpha_{2}=1.3$, $\alpha_{3}=2.6$]{\includegraphics[angle=270,width=6cm]{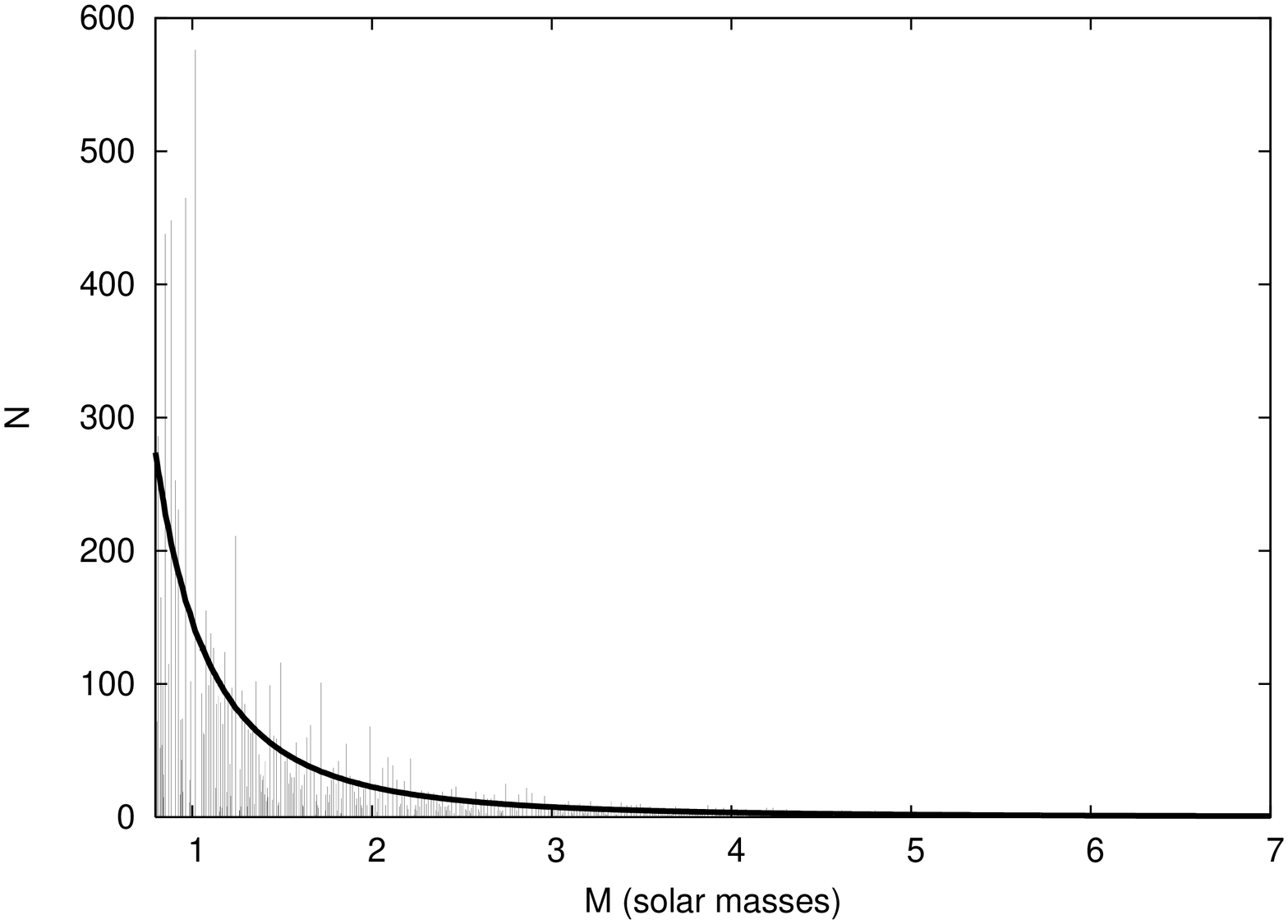}}
\caption{\small Mass distribution for $M_{total}=5\times 10^{4} M_{\sun}$ using Kroupa IMF. Because all the bins above $7 M_{\sun}$ contain only one star, only the $[0.8,7.0]$ interval is displayed. The solid line shows the mathematical form of the distribution for constant width bins.\normalsize}\label{fig:nm}
\end{center}
\end{figure}

The characteristics of the stellar mass distribution, whether truncated or tied to the cluster mass, etc., are an important component of a stellar cluster for a variety of investigations. This is why the \texttt{MASSCLEAN} package offers a wide array of options to build the stellar mass distribution. Sample computed mass distributions are presented in Figure \ref{fig:nm}.

\subsection{Evolutionary Models}

A variety of evolutionary tracks have been used in simulation codes to define the relationship between age and metallicity and integrated broad-band colors or spectral features (\citeauthor*{charlot1996} \citeyear{charlot1996}, \citeauthor*{brocato2000} \citeyear{brocato2000}, \citeauthor*{bruzual2003} \citeyear{bruzual2003}, \citeauthor*{leitherer1999} \citeyear{leitherer1999}). Selecting which models to use is influenced by the goals of that simulation. Presently, the \texttt{MASSCLEAN} package supports two different evolutionary models and the user can add their own.  In this first version of the \texttt{MASSCLEAN} package,  we have chosen to include the Geneva Models (\citeauthor*{geneva1} \citeyear{geneva1}) and also the recently released Padova Models (\citeauthor*{padova2008} \citeyear{padova2008}) The Geneva models provide excellent treatment of the evolutionary properties of high mass stars and ample time sampling to closely follow the evolution of a young cluster (age $\ge 10^7$ years).  This is important in the study of young clusters where evolution occurs quickly and the light from a few massive stars can dominate the light of the cluster.  The Padova models are considered superior in their treatment of older stars, as it more carefully considers the evolution of AGB populations and several peculiarities present at the onset of the thermally-pulsating asymptotic giant branch phase (TP-AGB).  The package can easily be switched to use a different set of isochrones and tracks as the user sees fit, such as evolutionary models taylored to old stars, or low metallicity, etc. One can also expand the current set of photometric bands to include the ultraviolet and mid-infrared. 

 \subsection{CCM Extinction Model} \label{ccmmethod}

\begin{figure}[ht] 
\begin{center}
\includegraphics[angle=270,width=7.0cm]{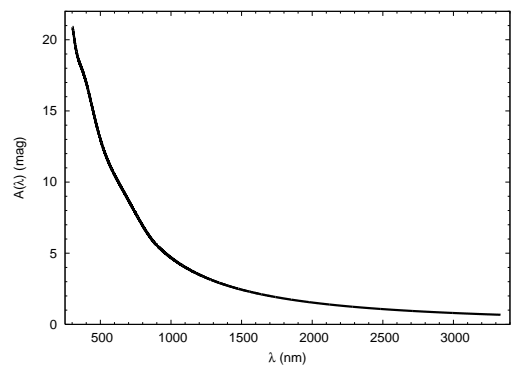} 
\caption{\small CCM Extinction Model, $A_{V}=11.6$, $R_{V}=3.1$ \normalsize}\label{fig:ccm}
\end{center}
\end{figure}

We wish to use our simulations in the study of Milky Way clusters as well as more distant extragalactic clusters. In both applications, we must pay close attention to extinction effects. For the extinction, the user can enter the value in two ways: the exact extinction value can be entered manually (in the configuration file) for every band or the extinction in each band can be computed using an inputted $R_{V}$ and $A_{V}$ and applying the CCM Extinction Model (\citeauthor*{ccm89}, \citeyear{ccm89}). A sample computed extinction curve is presented in Figure \ref{fig:ccm}.

  \subsection{Spatial Distribution} \label{king}

The \texttt{MASSCLEAN} package is unique to previous SSP codes in that it produces simulated images of the stellar clusters. Thus, the code is concerned with selecting appropriate spatial parameters defined by the angular size (linear scale and distance to the cluster) and for the first time, intrinsic stellar density. To accomplish this we have introduced the King Model Distribution \citep{King1962}, given by :

\begin{equation}
f(r)=k \left[ \left({\sqrt{1+\left(\frac{r}{r_{c}}\right)^{2}}}\right)^{-1}  - \left({\sqrt{1+\left(\frac{r_{t}}{r_{c}}\right)^{2}}}\right)^{-1} \right]^{2}    
   \end{equation}

An anisotropic spatial distribution can even be generated such that it still obeys the radial King profile. An ellipsoid in a prolate or oblate projection can be generated according to the parameters entered by the user in the configuration file. The rotation angle is also selected by the user.

A simple linear mass segregation feature allows more massive stars to fall toward the cluster's center as the cluster ages (this feature can be turned on or off).  

The file containing the spatial distribution can be replaced by another file provided by the user, for example, based on coordinates from a real image.

\begin{figure}[htb] 
\begin{center}
\subfigure[Isotropic Distribution]{\includegraphics[angle=270,width=3.75cm]{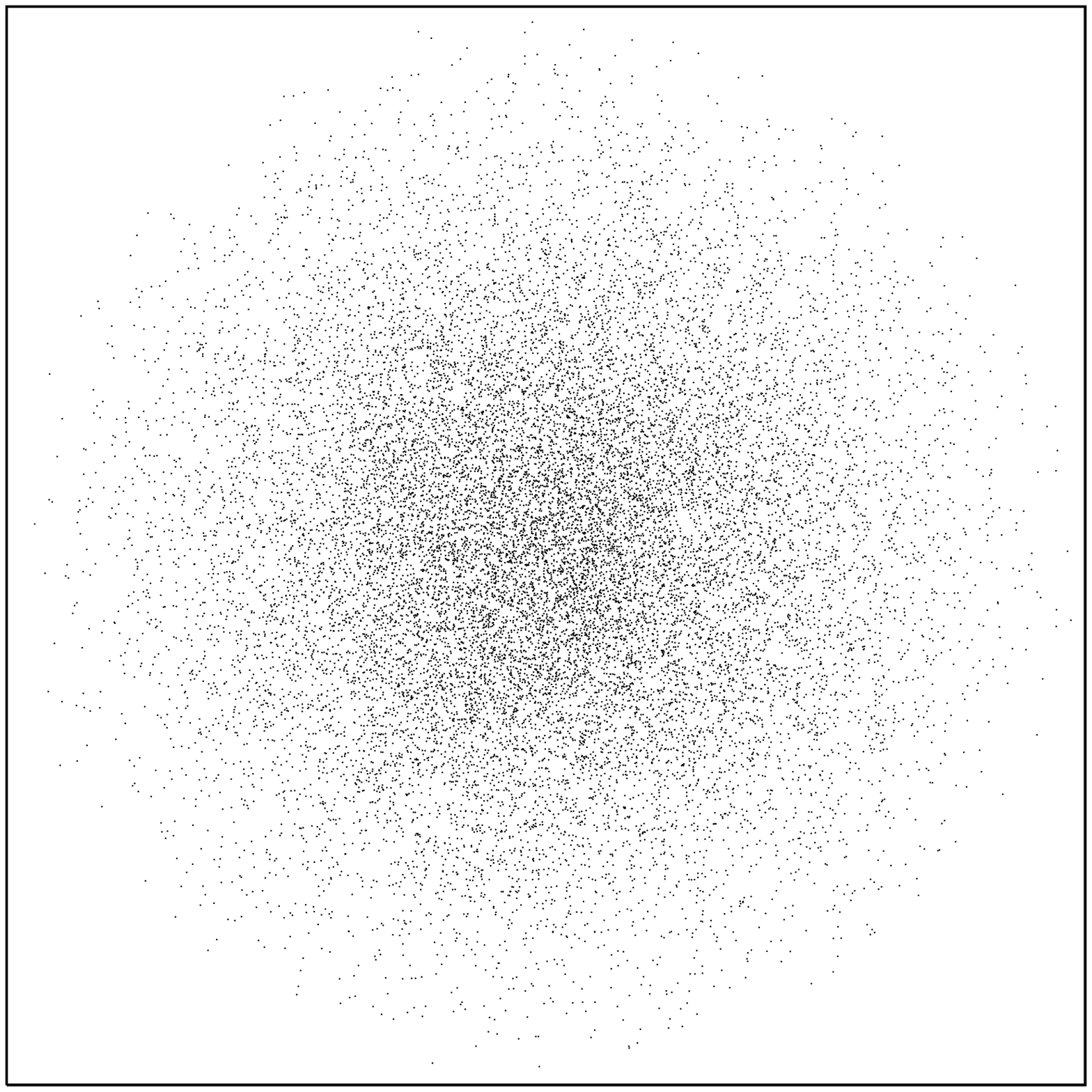}} 
\subfigure[Anisotropic Distribution]{\includegraphics[angle=270,width=3.75cm]{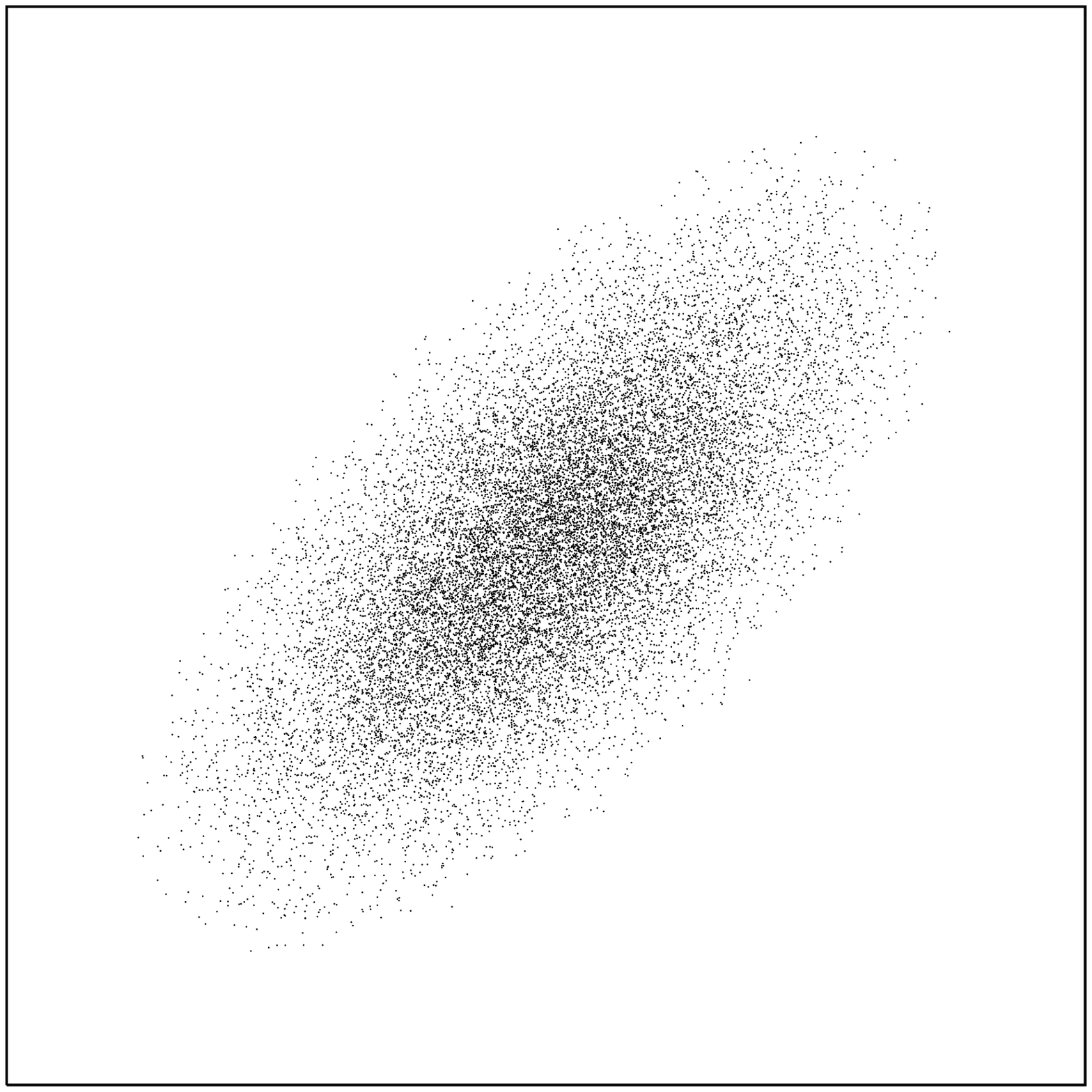}}
\caption{\small King Distribution, $r_{c}=0.50$ pc, $r_{t}=0.72$ pc, $M_{total}=10^{5} M_{\sun}$ \normalsize}\label{fig:king1}
\end{center}
\end{figure}

Sample computed spatial distributions are presented in Figure \ref{fig:king1}.

\subsection{Description of the Stellar Field} \label{stellarfield}

The \texttt{MASSCLEAN} package has the option to include a field star population. This may be a simulated one, using the SKY Model for the stellar field (\citeauthor*{Wainscoat1992} \citeyear{Wainscoat1992}; \citeauthor*{cohen1994} \citeyear{cohen1994}; \citeauthor*{cohen1995} \citeyear{cohen1995}; \citeauthor*{bahcall1984} \citeyear{bahcall1984}) or a real one. 

\begin{figure}[ht] 
\begin{center}
\includegraphics[angle=270,width=7.0cm]{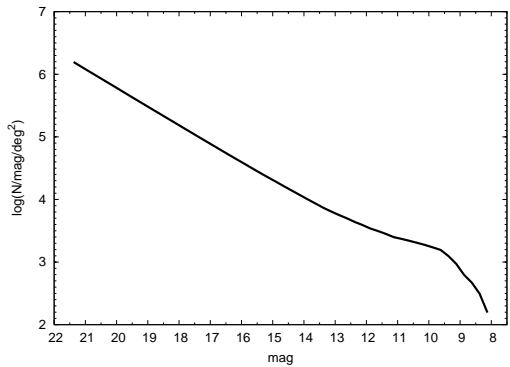} 
\caption{\small Cumulative number of stars vs. magnitude \normalsize}\label{fig:field}
\end{center}
\end{figure}

In the first case, starting from the total number of stars brighter than the selected magnitude limit (which is an input parameter), the distribution is generated using the slope of the cumulative numbers of stars (which is also an input parameter), such as shown in Figure \ref{fig:field}.  
 Colors are computed based on the Geneva Models (\citeauthor*{geneva1} \citeyear{geneva1}) and the {\it BaSeL}$-2.2$ grid (\citeauthor*{basel22} \citeyear{basel22}), and Padova Models (\citeauthor*{padova2008} \citeyear{padova2008}) and the ATLAS9-ODFNEW grid (\citeauthor*{castelli2003} \citeyear{castelli2003}), respectivelly. Extinction is also included and the user has the option to provide the values for every band. The program used to generate Figure \ref{fig:field} can compute the necessary parameters for a simulated stellar field based on the properties derived from a real one. 

The second option is to use a real stellar field. Since the file format is the same as the one used by {\sc SkyMaker} (\citeauthor*{ber01} \citeyear{ber01}; \citeauthor*{ber07} \citeyear{ber07}) and {\sc SExtractor} (\citeauthor*{sextractor} \citeyear{sextractor}), from the computational point of view there is no difference between a real and a simulated stellar field. \texttt{MASSCLEAN} can use a real stellar field provided by {\sc SExtractor}. The user can also choose not to include any stellar field at all.

\subsection{Creating the FITS Images and HR Diagrams}

Using all of the above models, \texttt{MASSCLEAN} computes actual mass, absolute and apparent magnitude (UBVRIJHK), color indexes, temperature, luminosity and position for all the stars and all the ages included in the evolutionary database (which ever isochrones one employs). The default age range is $log(age/yr)=6.0-9.5$, but the option to run it only for a few selected values is also available. These outputs are directly used to generate the color-magnitude diagrams and images. There are available relations of transformation from UBVRIJHK to different bandpasses for the Geneva and Padova models.  It is expected that the bandpasses for HST, 2MASS and Spitzer will be directly available soon (and will be included in a future upgrade). 

The FITS images in each of the broad bands are generated using {\sc SkyMaker} (\citeauthor*{ber01} \citeyear{ber01}; \citeauthor*{ber07} \citeyear{ber07}). \texttt{MASSCLEAN} writes all the necessary scripts to run {\sc SkyMaker} that generates the images. The configuration files for galactic and extragalactic clusters are also provided, so no knowledge about {\sc SkyMaker} is required. The description of the point spread function (PSF) is available in these configuration files and can be changed by the user. In addition to that, {\sc SkyMaker} can work with a different PSF provided by the user in a separate file.
The users can make any plot from the package's output using the plotting program of their choice, but \texttt{MASSCLEAN} also writes a script which generates HR and color-magnitude diagrams.

\section{Tests of the \texttt{MASSCLEAN} Package} \label{sampleruns}

Before demonstrating the unique utility of \texttt{MASSCLEAN} and directly comparing its output to real clusters, we will first provide a few logical tests.  Our first test will be to derive from \texttt{MASSCLEAN}, the same kind of values which come from other stellar population simulation programs which are widely used in the field to ensure that \texttt{MASSCLEAN} gives consistent results.  \texttt{MASSCLEAN} does not generate spectral features, such as provided by {\it Starburst99} (\citeauthor*{leitherer1999} \citeyear{leitherer1999}). However, it can be made to generate integrated magnitudes and colors.  This is achieved by simply summing up the flux over all stars in the final simulated cluster.

\subsection{Integrated colors as a function of age and metallicity}

The most commonly-used discriminator in the study of extragalactic super cluster studies is integrated colors (\citeauthor*{holtzman1992} \citeyear{holtzman1992}, \citeauthor*{fusibecci} \citeyear{fusibecci}). This is because it is typically the only information obtainable for very distant, unresolved star clusters (\citeauthor*{whitmore1995} \citeyear{whitmore1995}).  Among the most widely-used SSP models for interpreting integrated colors are those given by \citeauthor*{bruzual2003} \citeyearpar{bruzual2003} (GALAXEV), \citeauthor*{maraston2005} \citeyear{maraston2005}, GALEV 2003 (\citeauthor*{schulz2002} \citeyear{schulz2002} and \citeauthor*{galev} \citeyear{galev}), GALEV 2009 (\citeauthor*{galev2009} \citeyear{galev2009}), and Padova 2008 (\citeauthor*{padova2008} \citeyear{padova2008}). We shall provide a side by side comparison of \texttt{MASSCLEAN} against these three modern SSP codes.

\begin{figure*}[ht] 
\begin{center}
\subfigure[]{\includegraphics[angle=0,width=8.0cm]{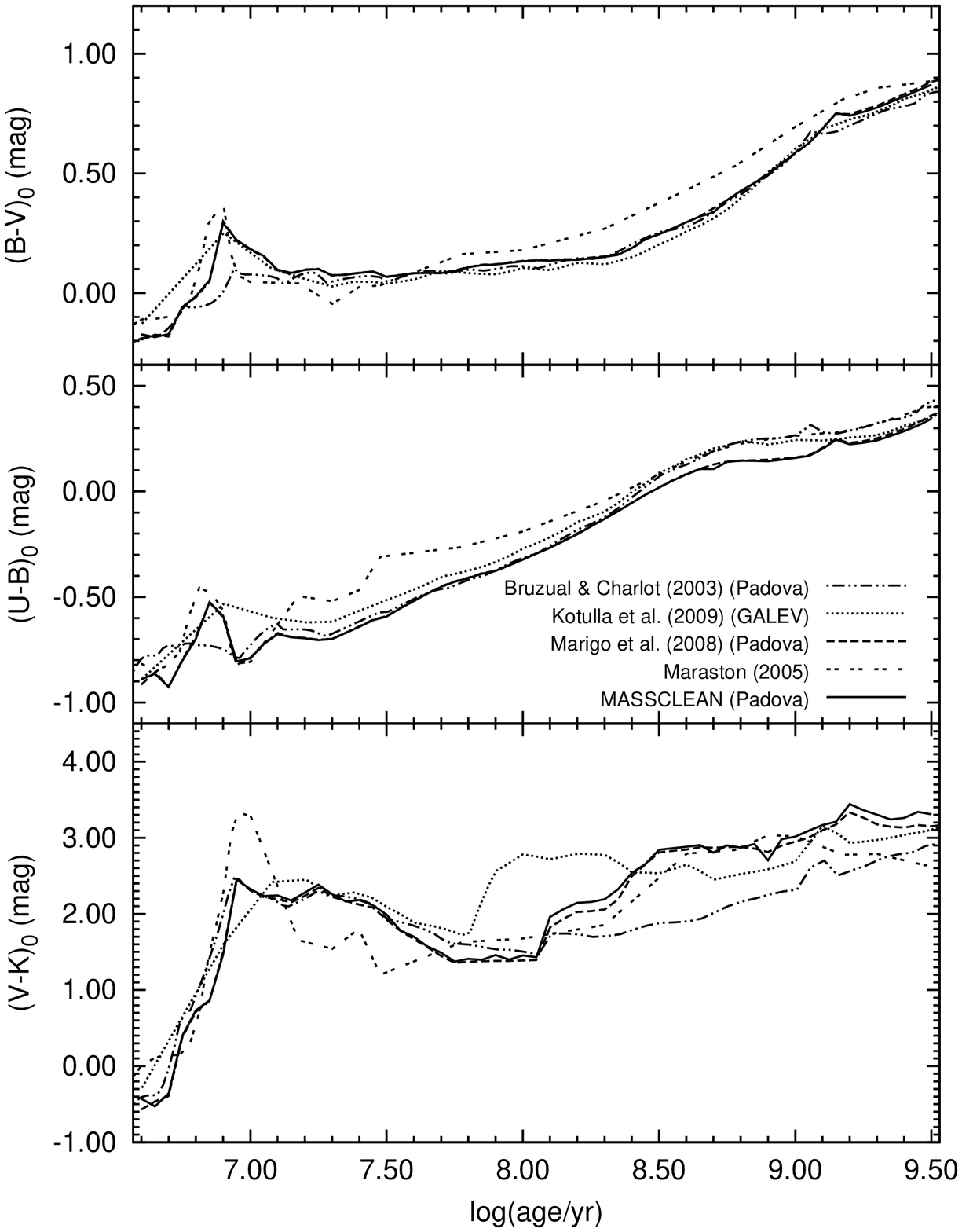}} 
\subfigure[]{\includegraphics[angle=0,width=8.0cm]{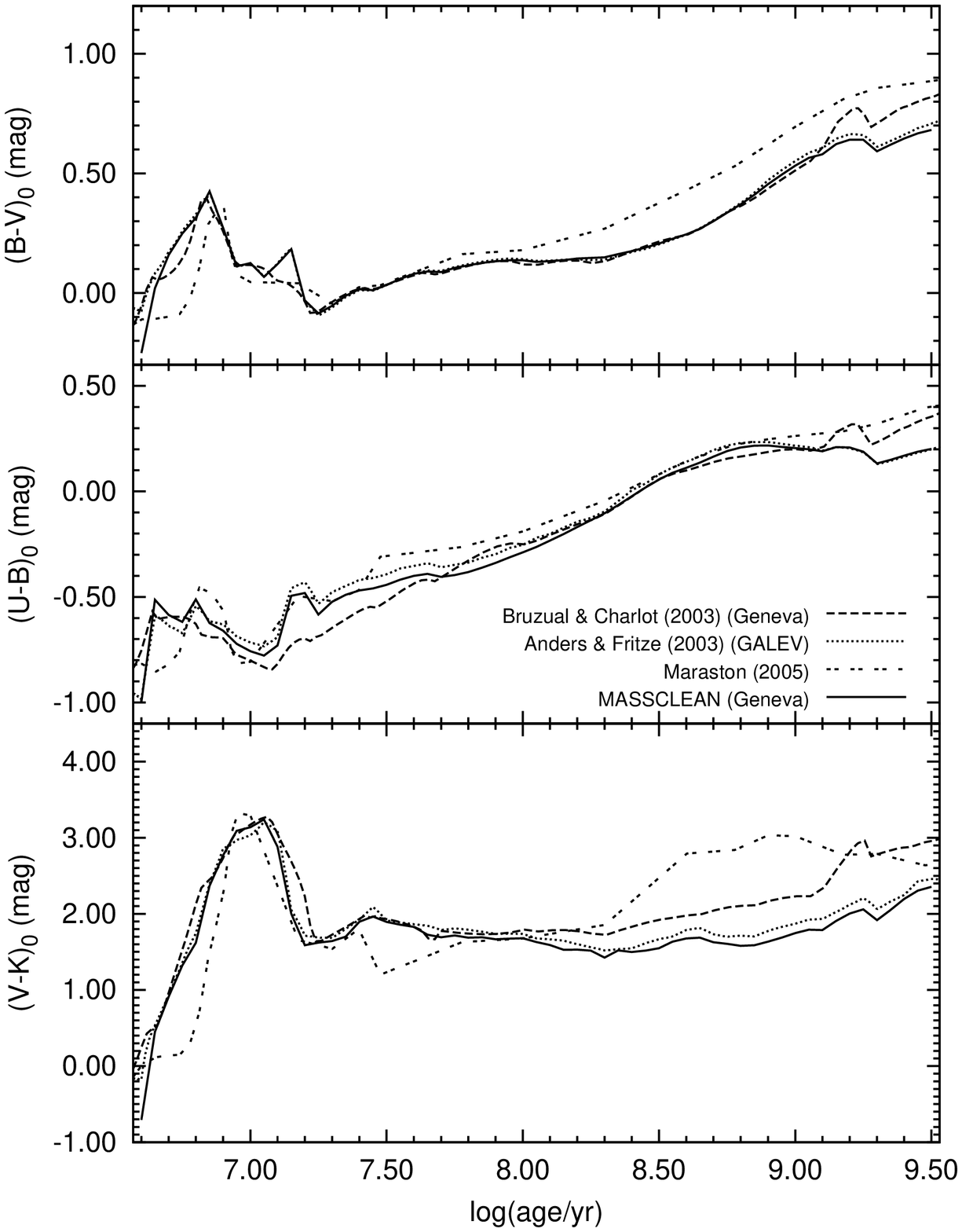}}
\caption{\small Integrated colors for different SSP Models as a function of age for solar metallicity: (a) Padova 2008 (\texttt{MASSCLEAN}) - solid line; GALEV (Kotulla et al. 2009) - dotted line; GALAXEV (Bruzual \& Charlot 2003) with Padova 1994 - dashed-double dotted line; Maraston (2005) - double dotted line; and Padova 2008, Mariago et al.\ 2008 is the dashed line. (b) Geneva (\texttt{MASSCLEAN}) - solid line; GALAXEV (Bruzual \& Charlot 2003) with Geneva - dashed line; GALEV (Anders \& Fritze 2003) - dotted line; Maraston (2005) - double dotted line. \normalsize}\label{fig:multiplot1}
\end{center}
\end{figure*}

\begin{figure*}[ht] 
\begin{center}
\subfigure[]{\includegraphics[angle=0,width=8.0cm]{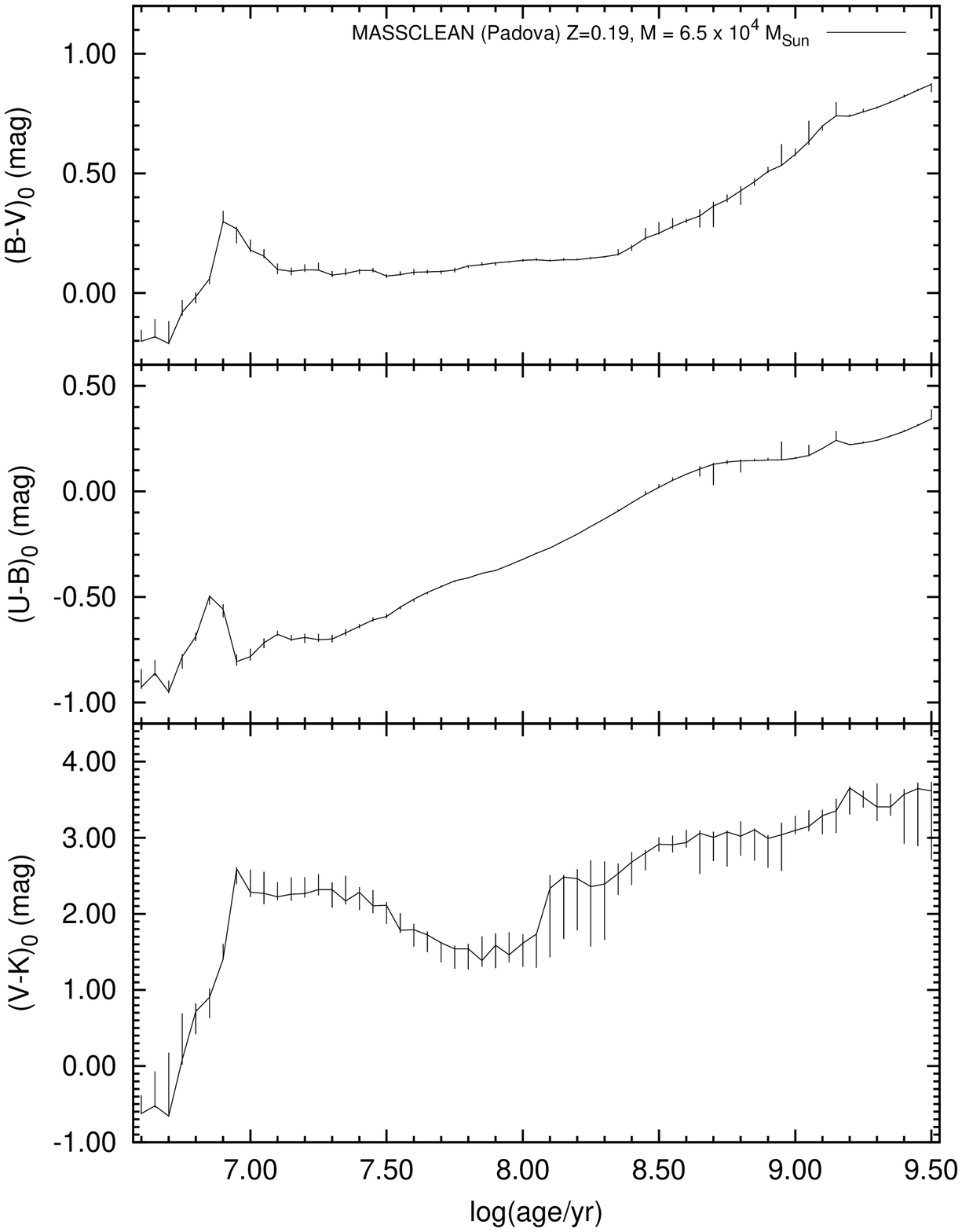}} 
\subfigure[]{\includegraphics[angle=0,width=8.0cm]{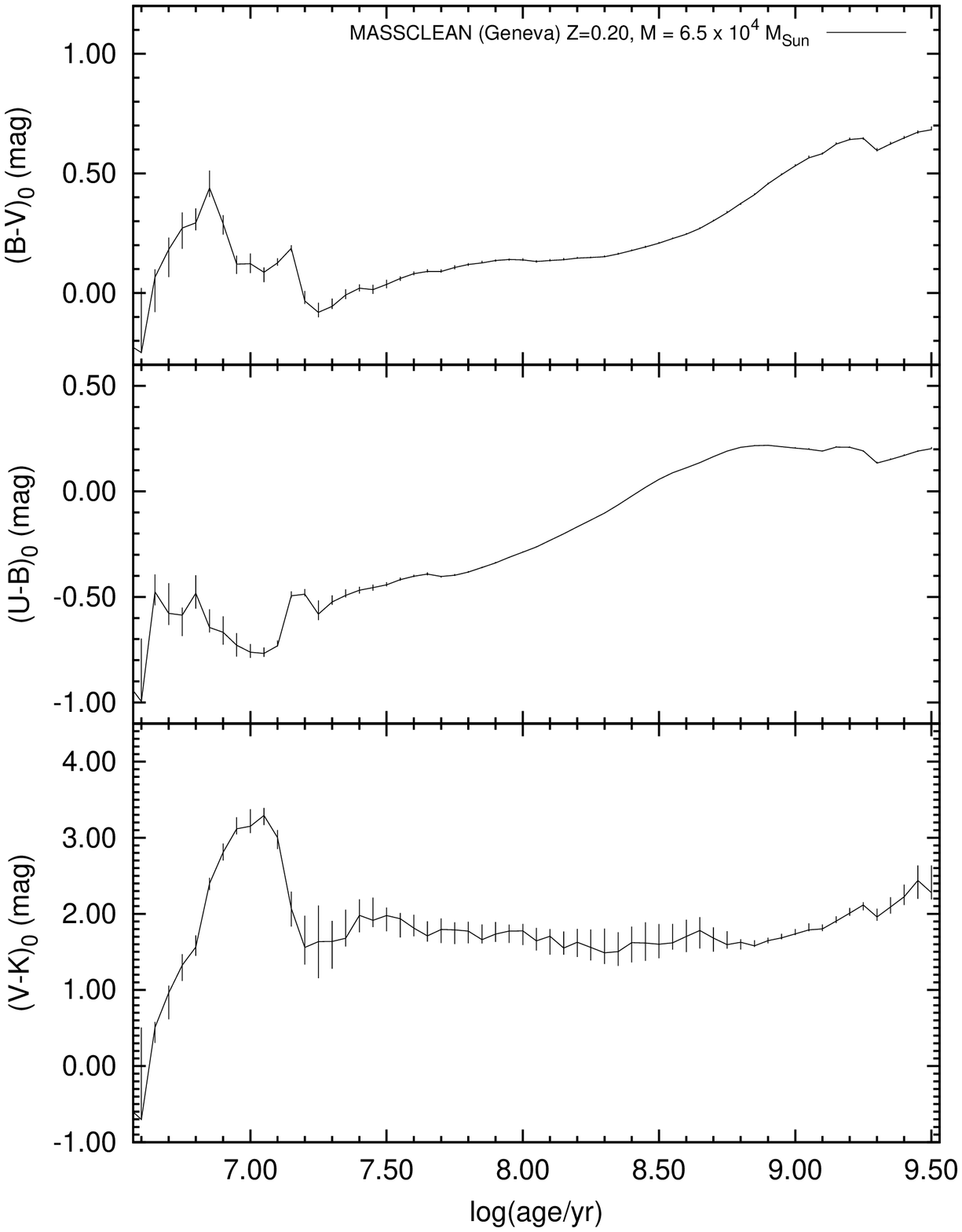}}
\caption{\small Sample stochastic fluctuations as a function of colors,  (a) Padova 2008 (\texttt{MASSCLEAN})  (b) Geneva (\texttt{MASSCLEAN}).   The bar heights represent the range of observed values seen with a Monte Carlo run of 1000 clusters.  \normalsize}\label{fig:multiplot004}
\end{center}
\end{figure*}

The default age range for \texttt{MASSCLEAN} is $[6.0,9.5]$ in the logarithmic scale. In Figures \ref{fig:multiplot1}  though \ref{fig:multiplot005} we use a smaller age range in order to accomodate the data from all the available models. In Figure \ref{fig:multiplot1} we plot the variation of integrated colors with age for different SSP models.   Models using solar metallicity are presented here. In Figure \ref{fig:multiplot1} (a) the results from \texttt{MASSCLEAN} using Padova 2008 is shown as the solid line and compared with GALEV 2009 Models (Padova) in the dotted line, GALAXEV models (Padova 1994) are shown in the dashed-double dotted line, Maraston (2005) is shown in the double dotted line and Padova 2008 (Marigo et al. 2008) is shown in a dashed line. Specifically, the \texttt{MASSCLEAN} colors follow quite closely to the colors given by GALAXEV (Padova) and are essentially identical to Padova (2008), as it should be. In Figure \ref{fig:multiplot1} (b) the results from \texttt{MASSCLEAN} using the Geneva evolutionary models is shown as the solid line and compared with GALEV 2003 Models (Geneva) in dotted line, GALAXEV (Geneva) are shown in the dashed line, and Maraston (2005) is again shown in the double dotted line.  Figure \ref{fig:multiplot1} (a) and (b) demonstrates that our code, based on a finite stellar generation algorithm, gives the same integrated color results of other SSP codes that use the same evolutionary models but using a statistically weighted mass distribution. The \texttt{MASSCLEAN} integrated colors were computed in the high mass limit ($\sim 10^{6} M_{\Sun}$), which allows for isochrones to be fully populated for all masses.



\begin{figure*}[ht] 
\begin{center}
\subfigure[]{\includegraphics[angle=0,width=8.0cm]{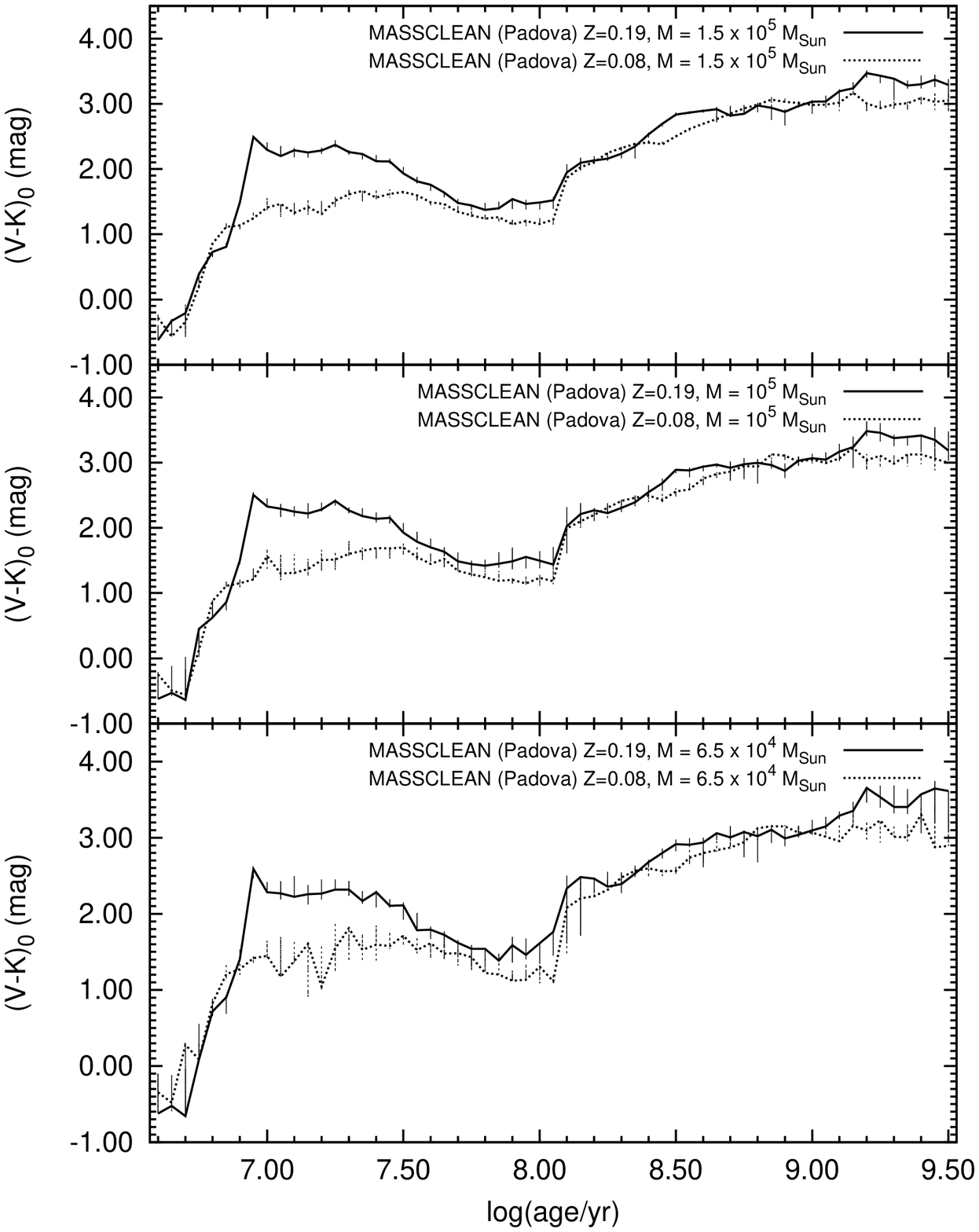}} 
\subfigure[]{\includegraphics[angle=0,width=8.0cm]{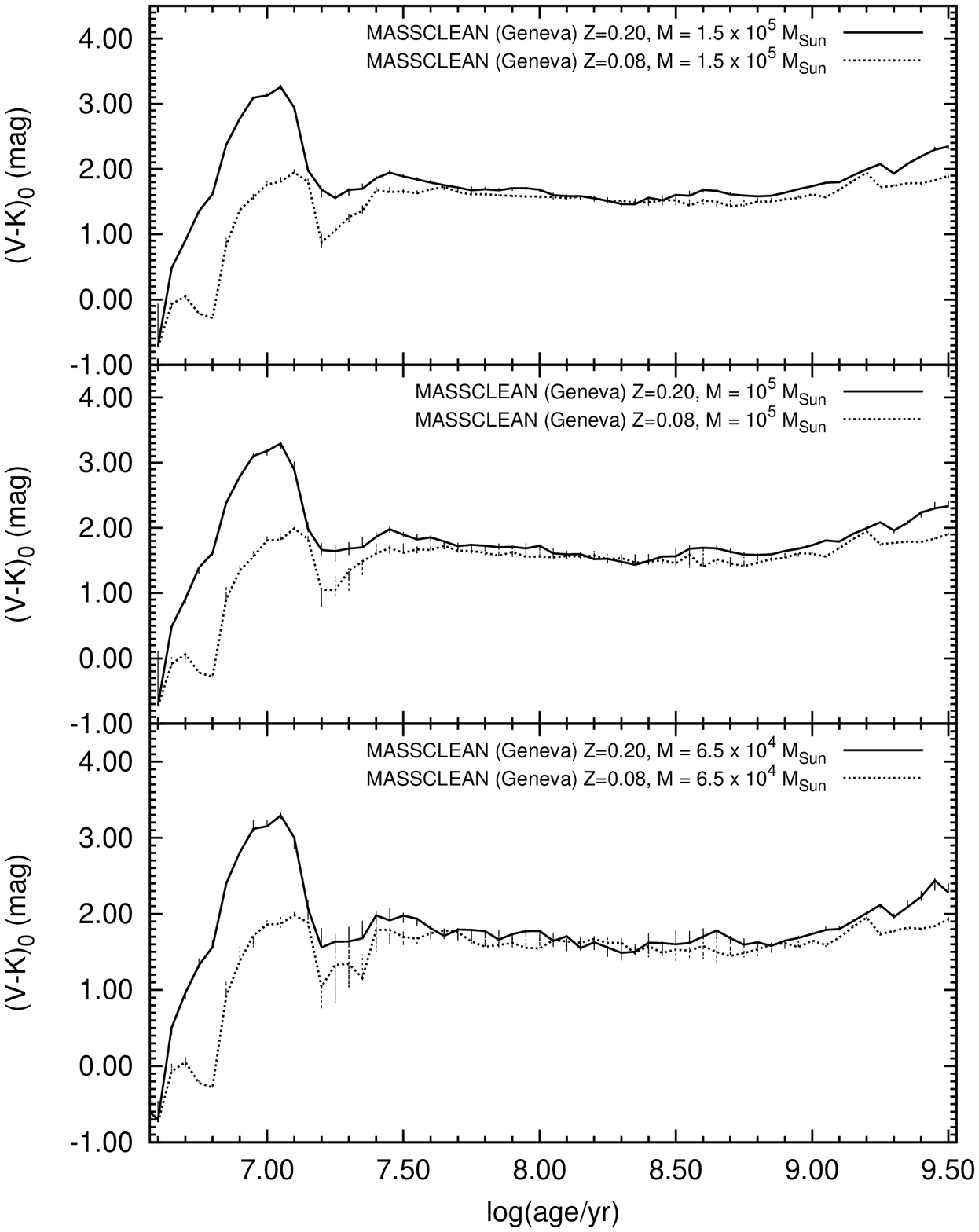}}
\caption{\small The range of variation in V-K for three mass intervals: $50-75K$ ({\it bottom panel}), $75-125K$ ({\it middle panel}), $125-175K$ ({\it upper panel}). (a) Padova 2008 (\texttt{MASSCLEAN})  (b) Geneva (\texttt{MASSCLEAN}).   The height of the bar shows the observed range of output colors from 1000 Monte Carlo runs where input mass was allowed to vary over the range listed. \normalsize}\label{fig:multiplot003}
\end{center}
\end{figure*}

In Figure \ref{fig:multiplot004} is presented the variation of integrated colors with age computed by \texttt{MASSCLEAN} for a moderate cluster mass ($6.5 \times 10^{4}M_{\Sun}$). Figure \ref{fig:multiplot004} (a) is using Padova 2008, and (b) is using the Geneva models.  This is a demonstration the other SSP codes are not able to make since total mass is not available as an input.  The bars represent the range observed in the output integrated colors from 1000 randomly generated cluster simulations (Monte Carlo runs of the stellar mass function).   What is shown here is an increase in observed deviation in color with longer wavelength colors.  This is further explored in Figure \ref{fig:multiplot003}.  Here, we've concentrated on the mass effect seen for the V-K colors only, as this band is where the observed color range is known to be greatest.  These data in this figure were constructed by again creating 1000 randomly generated clusters, now over a variety of masses.  The figure provides the results from three general mass ranges.  As cluster mass decreases, the observed range in integrated colors increases.  

\begin{figure}[htbp] 
\begin{center}
\includegraphics[angle=0,width=8.0cm]{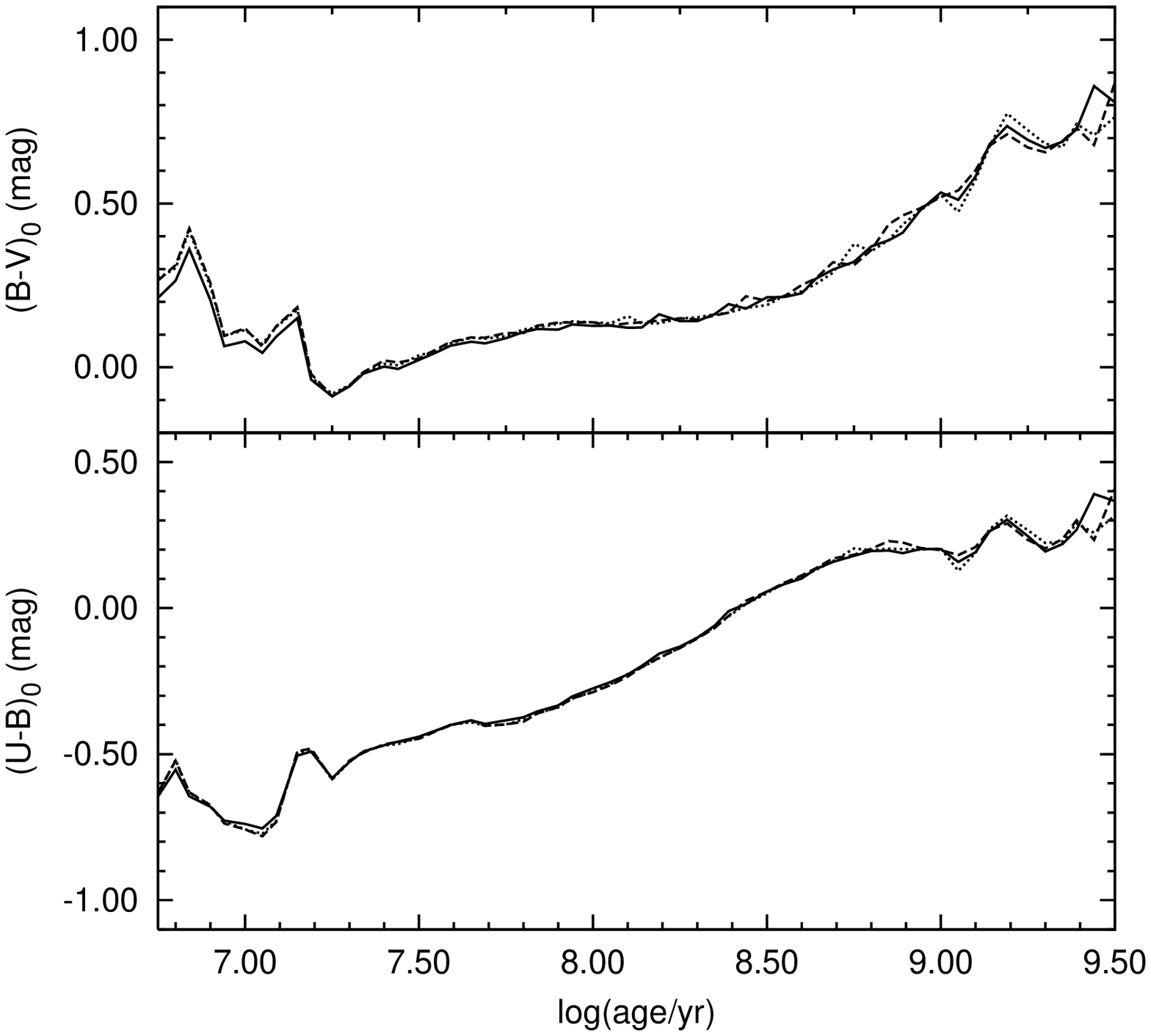} 
\caption{\small Integrated colors for different IMFs as a function of age computed by \texttt{MASSCLEAN} (Geneva): Kroupa IMF $\alpha_{3}=2.6$ - solid line; Kroupa IMF $\alpha_{3}=2.3$ - dashed line; Salpeter IMF - dotted line \normalsize}\label{fig:multiplot3}
\end{center}
\end{figure}

\begin{figure*}[hpt] 
\begin{center}
\includegraphics[angle=0,width=15cm]{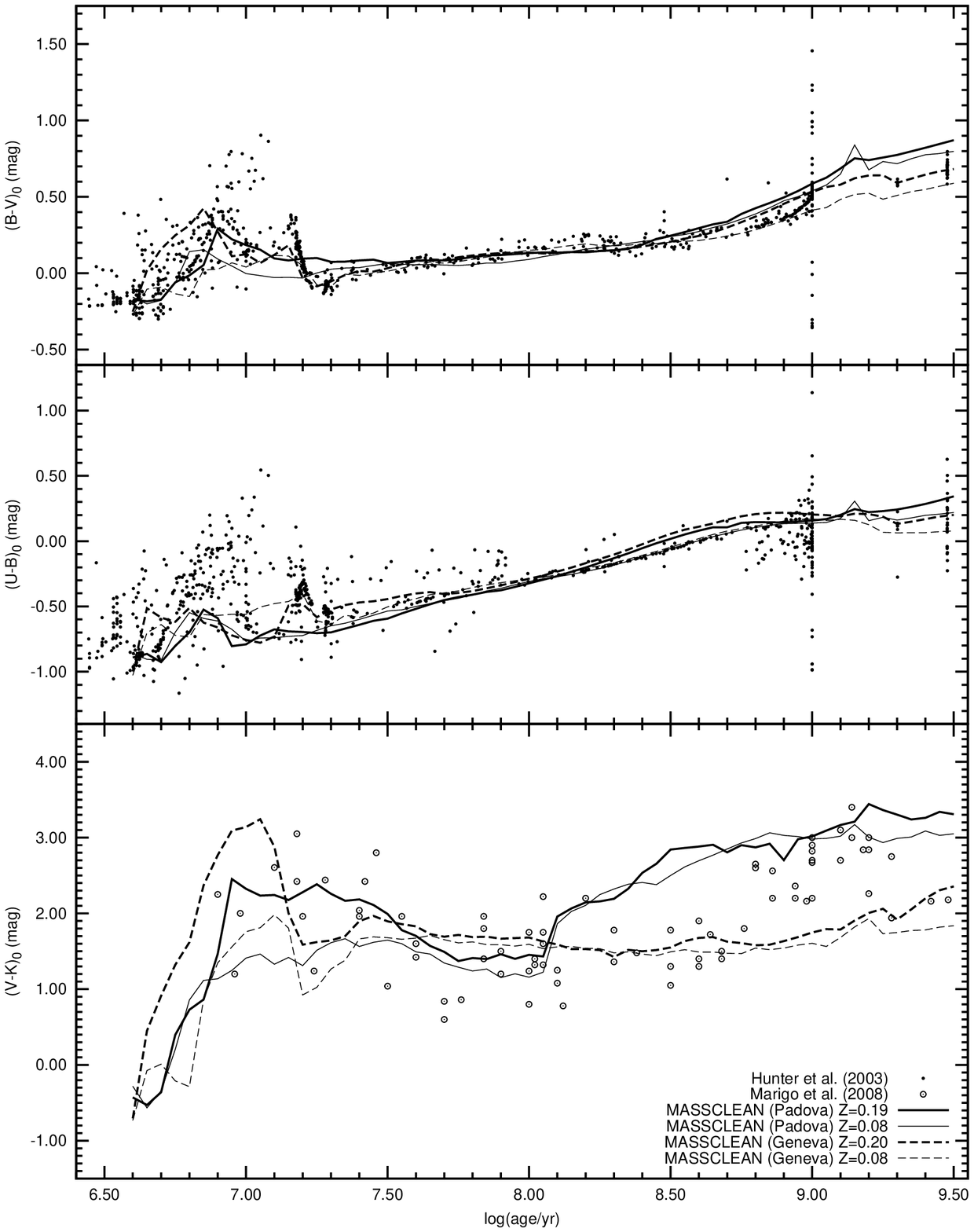} 
\caption{\small \texttt{MASSCLEAN} integrated colors for differing evolutionary models and metallicities displayed with observed photometric data taken  in the Magellanic Clouds from \citeauthor*{hunter2003} \citeyearpar{hunter2003}, the solid dots, and \citeauthor*{padova2008} \citeyearpar{padova2008}, the solar symbols.  \normalsize}\label{fig:multiplot005}
\end{center}
\end{figure*}

Finally, the effect of changing the metallicity in the stellar evolution models (in (a) for Padova, in (b) for Geneva) is also displayed in Figure \ref{fig:multiplot003}.   The variation of integrated colors for different IMFs as computed by \texttt{MASSCLEAN} are presented in Figure \ref{fig:multiplot3}.  Previous studies have noted the very weak sensitivity of the IMF on the integrated colors of stellar clusters over this age range (see, for example, Figure 4 of \citeauthor*{bruzual2003} \citeyear{bruzual2003}).

In Figure \ref{fig:multiplot005} we show a full scale diagram of the integrated broadband colors, U-B, B-V, and V-K derived with \texttt{MASSCLEAN} models using Padova and Geneva, and for two metallicities.  Here we also show observed data, taken from \citeauthor*{hunter2003} \citeyearpar{hunter2003}, for stellar clusters in the Magellanic Clouds in the upper two panels.  In the lower panel, we have borrowed the observed photometry given in \citeauthor*{padova2008} \citeyearpar{padova2008}, Fig. 9, to show the observed variation in V-K color for LMC clusters.  The most obvious deviations are seen with the prediction of fairly red clusters in the age range from log(age) = 8 to 8.8 by the Padova evolutionary models (this was first noted by \citeauthor*{padova2008} \citeyear{padova2008}) and the prediction of fairly blue clusters with log(age) $>$ 9.0 by the Geneva models.

The color dispersion presented in Figures \ref{fig:multiplot004} and \ref{fig:multiplot003}, along with the photometric data presented in Figure \ref{fig:multiplot005}, show the advantage of the \texttt{MASSCLEAN} approach with respect to real clusters. As described by \citeauthor*{padova2008} \citeyearpar{padova2008} (\S 5.4 and Figure 9), the continuous and smooth SSP lines are not able to describe accurately the observational data. A more accurate description of the color dispersion as a function of the mass of the cluster will be presented in a subsequent paper (Popescu \& Hanson, in prep.).

\begin{figure*}[htbp] 
\begin{center}
\subfigure[]{\includegraphics[angle=0,width=7.0cm]{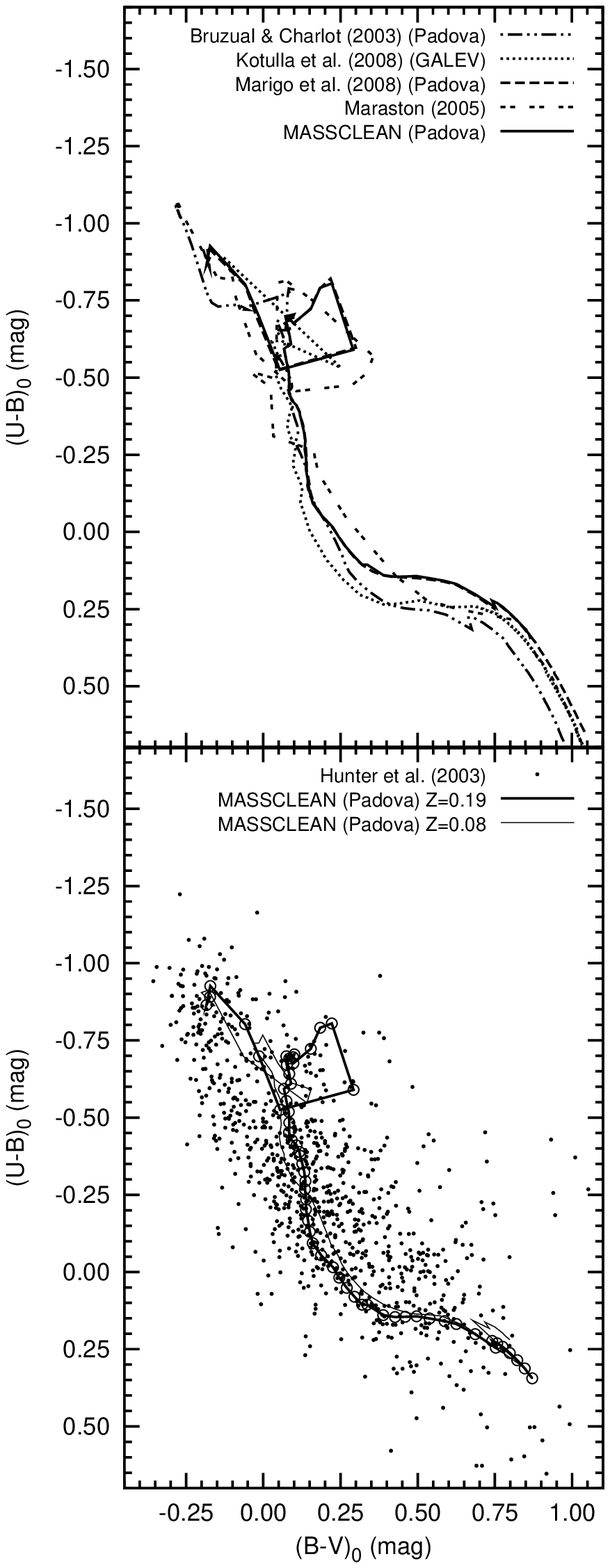}} 
\subfigure[]{\includegraphics[angle=0,width=7.0cm]{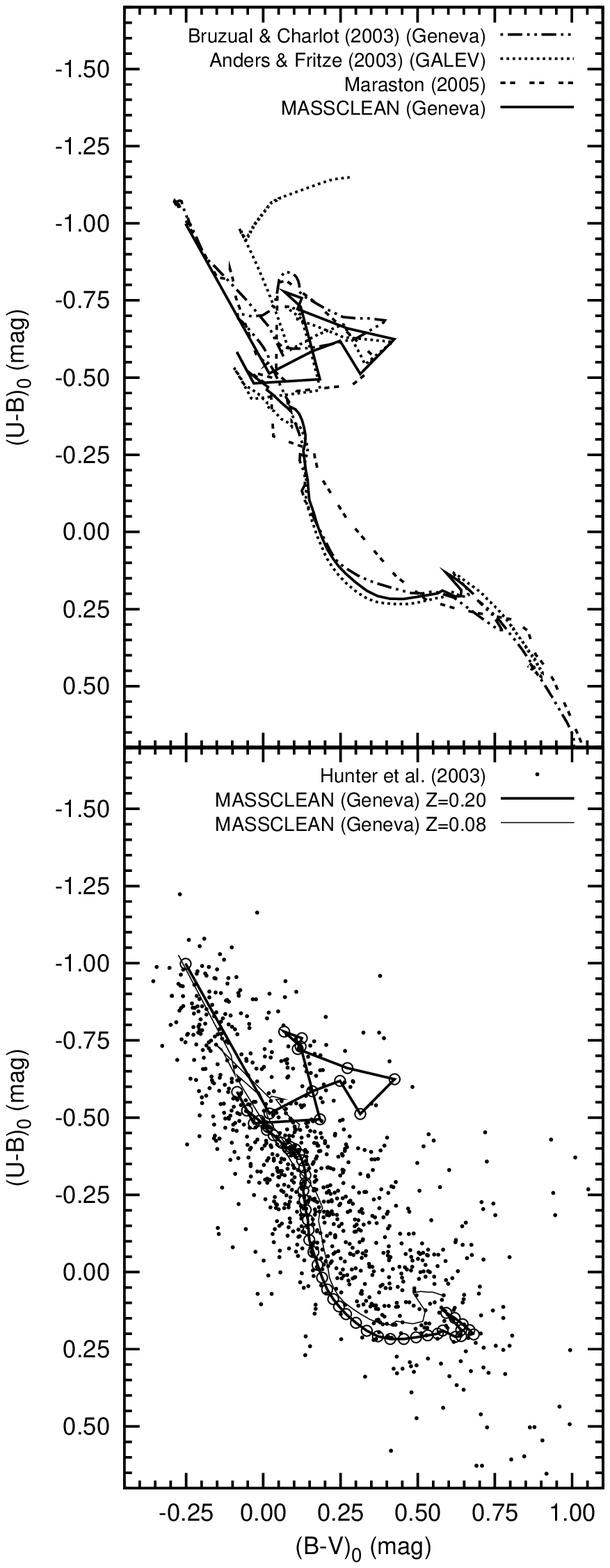}}
\caption{\small Integrated colors $(U-B)_{0}$ vs $(B-V)_{0}$. The circles along the solar metallicity path in the lower panels correspond to the $\Delta$$log (age/yr) = 0.05$.\normalsize}\label{fig:multiplot4}
\end{center}
\end{figure*}

Finally, in the top panel of Figure \ref{fig:multiplot4} we present the evolution of integrated colors $(U-B)_{0}$ vs $(B-V)_{0}$ for several different SSP models along side predictions coming from \texttt{MASSCLEAN} (solid line).  All simulations were made using solar metallicity. The \texttt{MASSCLEAN} colors compare well to real data as shown already with previous work (\citeauthor*{girardi1995} \citeyear{girardi1995}). The lower panel shows predicted color evolution using \texttt{MASSCLEAN}, but showing a variety of metallicities.   As in Figure \ref{fig:multiplot005}, the solid dots in the lower two panels represent observed photometric data for stellar clusters in the Magellanic Clouds from \citeauthor*{hunter2003} \citeyearpar{hunter2003}.

\subsection{Color-magnitude diagrams} \label{cmdetal}

\begin{figure}[htbp] 
\begin{center}
\subfigure[]{\includegraphics[angle=0,width=3.75cm]{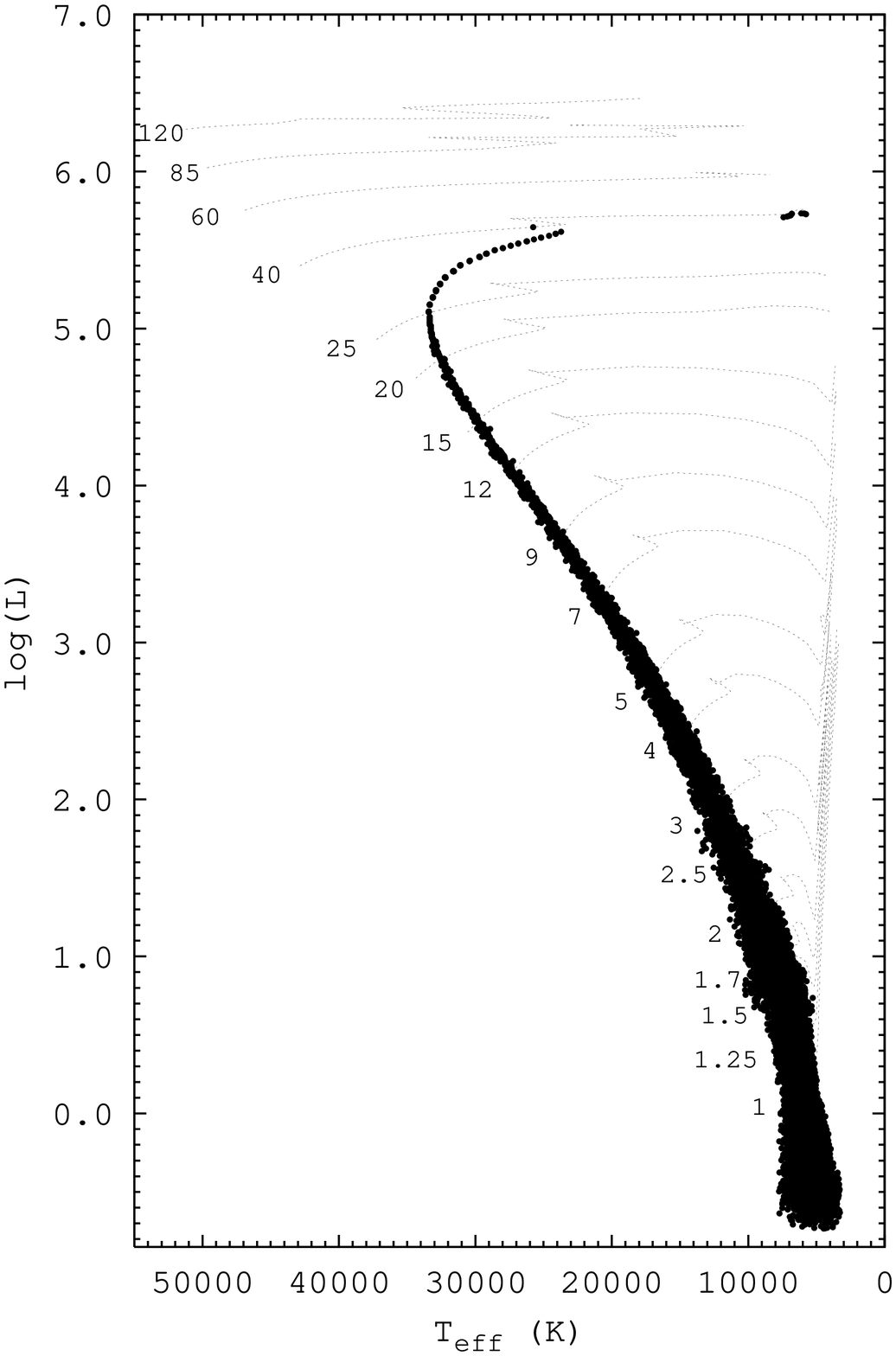}}
\subfigure[]{\includegraphics[angle=0,width=3.75cm]{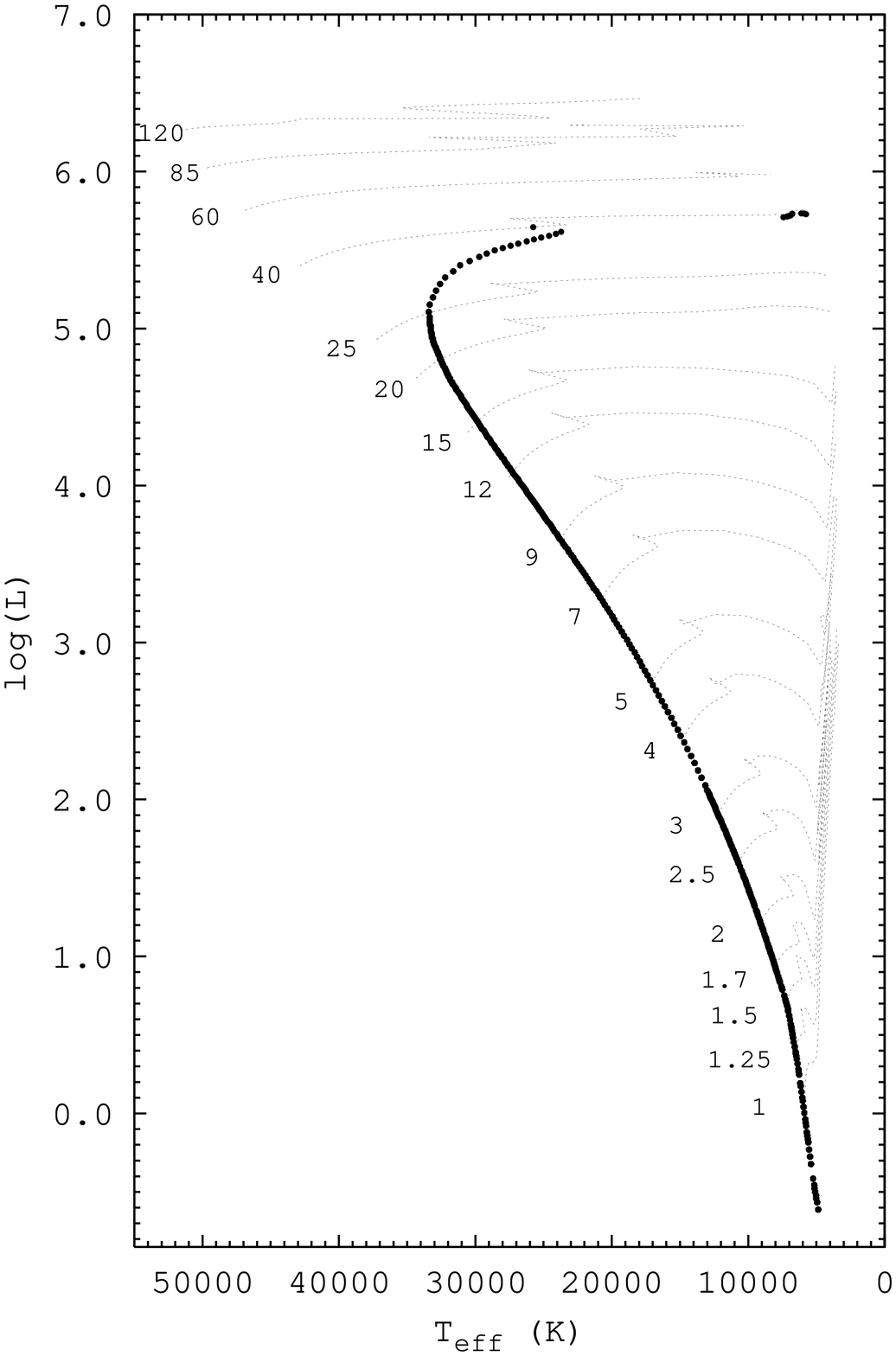}}
\caption{\small MASSCLEAN simulated HR Diagram, for cluster properties $\log (age) = 6.65$, $M_{total}=10^{5}M_{\sun}$. In (a), a rebinning has been done to give the stars a more realistic range of values.  In (b), one sees the stars following a theoretical track from the Geneva Database.  Both figures cover the stellar masses from $120 M_{\sun}$ to $1 M_{\sun}$. \normalsize}\label{fig:hr}
\end{center}
\end{figure}

\begin{figure}[htbp] 
\begin{center}
\subfigure[]{\includegraphics[angle=0,width=3.75cm]{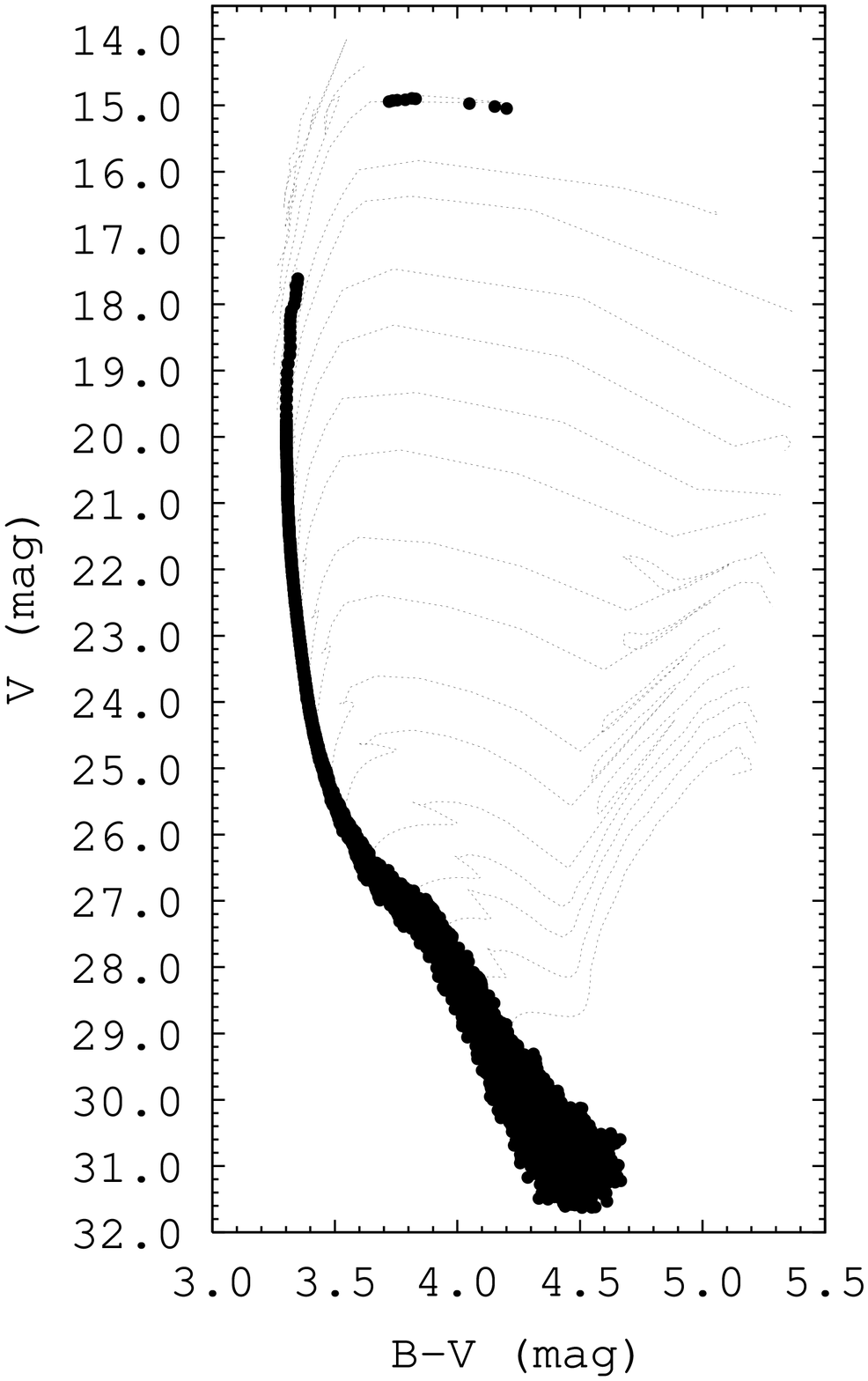}}
\subfigure[]{\includegraphics[angle=0,width=3.75cm]{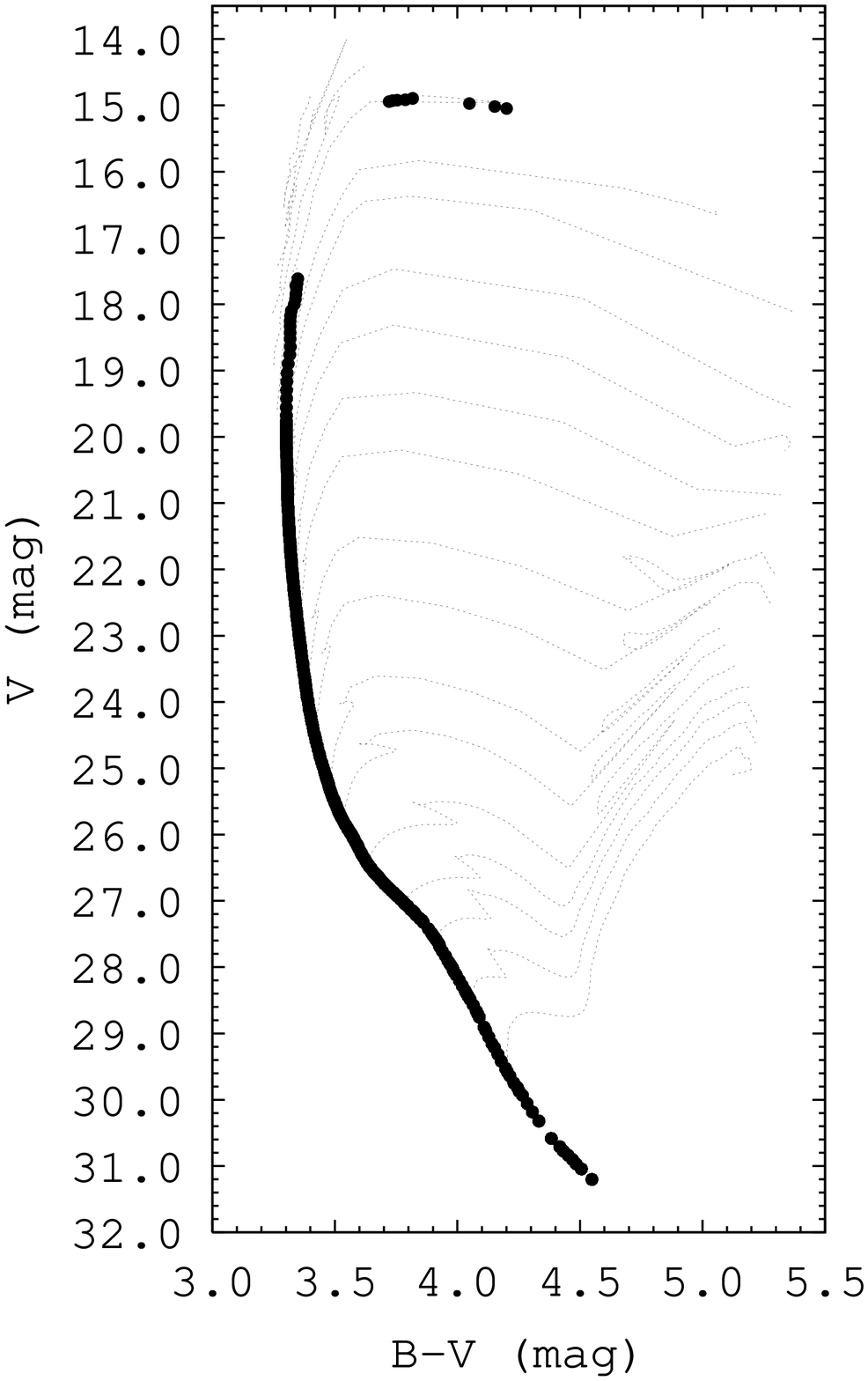}}
\caption{\small The $V$ vs $(B-V)$ color-magnitude diagram, $A_{V}=11.6$, $\log (age) = 6.65$, $M_{total}=10^{5}M_{\sun}$. The tracks correspond to the same values of mass as in Figure \ref{fig:hr}.\label{fig:cmd1} \normalsize}
\end{center}
\end{figure}

Our next test will be to simulate HR and color-magnitude diagrams of clusters using \texttt{MASSCLEAN}. As a first demonstration, we provide in Figure \ref{fig:hr}, an HR Diagram for a stellar cluster with a total mass of $10^{5}$ $M_{\odot}$ and a $log(age/yr) = 6.65$. 
While the individual stars in the cluster are clearly seen at the high mass end, crowding prevents one from seeing anything but a broad blur of stars below about $40$ $M_{\odot}$.  Naturally, this figure looks no different from the Geneva isochrones used for that same age.  In Figure \ref{fig:hr}(a) the stars have been rebinned to take on a more natural range of values; in Figure \ref{fig:hr}(b) the stars follow a pure single isochrone.  We also provide an example of a color-magnitude diagram, V versus B-V for this same simulated cluster, in Figure \ref{fig:cmd1}. At the low mass end, again a small variation has been introduced to give the main sequence some width in Figure \ref{fig:cmd1}(a). The widening in Figure \ref{fig:hr}(a) and Figure \ref{fig:cmd1}(a) is a user's choice, the maximum width is an input parameter (this option can also be turned off) and could be set up to correspond to the photometric error.

The cluster mass and age shown in these two figures were selected rather explicitly to represent a known cluster.  It matches the estimated mass and age of the Milky Way cluster, \textit{Westerlund 1}. The parameters used for the CMD simulation are based on \citeauthor*{Westerlund1961} \citeyearpar{Westerlund1961}, \citeauthor*{Clark2002} \citeyearpar{Clark2002}, \citeauthor*{Clark2005} \citeyearpar{Clark2005} and \citeauthor*{Figer2006} \citeyearpar{Figer2006}. We adopted the following values:  $M_{total}=10^{5} M_{\sun}$, solar metallicity, $A_{V}=11.6$ mag, $R_{V}=3.1$, $r_{t} = 0.72$ pc and $r_{c} = 0.50$ pc. 

There is some uncertainty in the distance to \textit{Westerlund 1}. We adopted a value of $d = 4$ kpc (distance modulus $13.01$ mag). A Kroupa IMF has been used, with $\alpha_{1} = 0.3$, $\alpha_{2} = 1.3$ and $\alpha_{3} = 2.3$. Our simulation shows that the best agreement with the actual data corresponds to the isochrone $\log (age) = 6.65$. This value also agrees with the recent results of  \citeauthor*{bradner} \citeyearpar{bradner}.

\begin{figure}[htbp] 
\begin{center}
\includegraphics[angle=0,width=7.0cm]{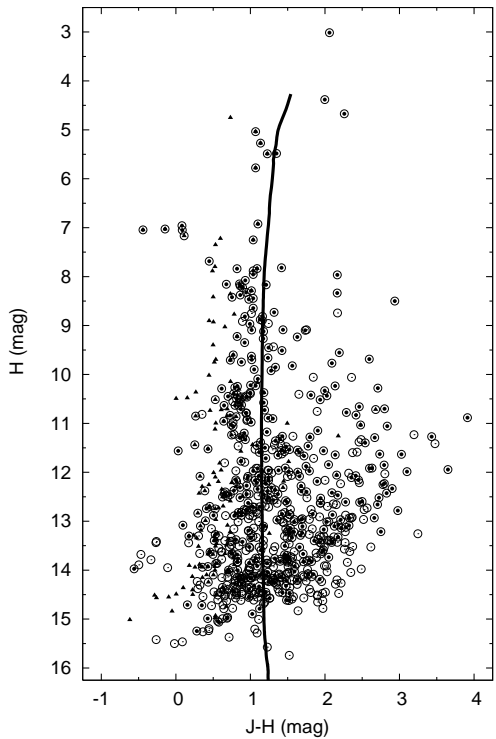} 
\caption{\small $H$ vs. $J-H$ color-magnitude diagram for {\it Westerlund 1}. Circles - 2MASS source catalog $4.8'$ radius; dots - 2MASS source catalog $2.0'$ radius; triangles - NOMAD catalog $4.8'$ radius. The solid lines correspond to the Geneva isochrones ($log (age/yr) = 6.65, 6.75$), and the dotted lines  correspond to the Padova isochrones ($log (age/yr) = 6.70, 6.75$). \normalsize}\label{fig:wd1obs1}
\end{center}
\end{figure}

\begin{figure}[htbp] 
\begin{center}
\includegraphics[angle=0,width=7.0cm]{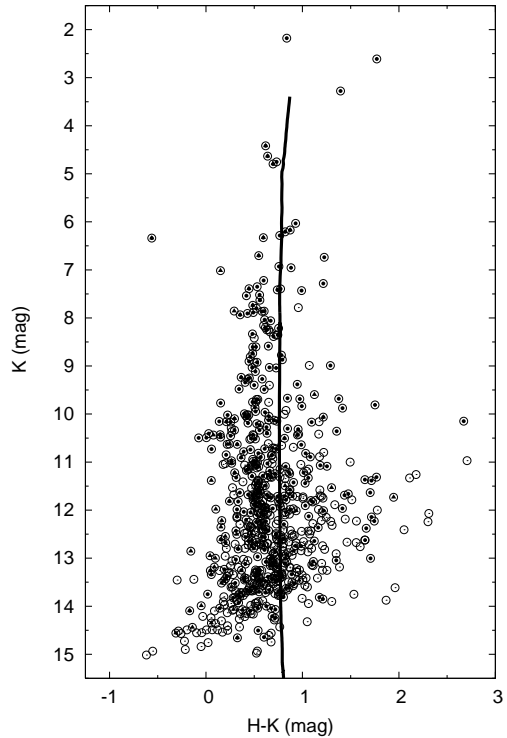} 
\caption{\small $K$ vs. $H-K$ color-magnitude diagram for {\it Westerlund 1}. Circles - 2MASS source catalog $4.8'$ radius; dots - 2MASS source catalog $2.0'$ radius; triangles - NOMAD catalog $4.8'$ radius. The solid lines correspond to the Geneva isochrones ($log (age/yr) = 6.65, 6.75$), and the dotted lines  correspond to the Padova isochrones ($log (age/yr) = 6.70, 6.75$). \normalsize}\label{fig:wd1obs2}
\end{center}
\end{figure}

In Figure \ref{fig:wd1obs1} and Figure \ref{fig:wd1obs2}, we present near-infrared color-magnitude diagrams for $Westerlund$ $1$.   \texttt{MASSCLEAN} isochrones are overlayed with real photometry for the cluster.  We cannot present optical color-magnitude diagrams for this cluster because of the enormous incompleteness due to high extinction ($A_V = 11$).  The photometric data in  Figure \ref{fig:wd1obs1} and Figure \ref{fig:wd1obs2} was taken from the 2MASS and NOMAD (\citeauthor{nomad} \citeyear{nomad}) catalogs.  

If \texttt{MASSCLEAN} is to be used to constrain the properties of observed clusters, a method for determining the best simulation inputs must still be developed. There is unlikely to be a straightforward way to determine the best fitting simulated image. Rather, we expect comparisons between simulated and real data will need to be done within the CMDs, comparing photometry. This is presently being worked on, but we expect to base such a goodness of fit on observed versus simulated cumulative distribution functions in various photometric bands and using a Kolmogorov-Smirnov test to select the closest fitting simulation (Popescu \& Hanson, in prep.).  

\subsection{Image simulations of Galactic clusters}

Our most visual demonstration will be the image simulations provided by \texttt{MASSCLEAN}.  The \texttt{MASSCLEAN} package has been used to simulate several well known young Milky Way clusters: $NGC$ $3603$, $h$ and $\chi$ \textit{Persei} (\textit{NGC 869}, \textit{NGC 884}), as well as \textit{Westerlund 1}.  This corresponds to a mass range from $4.3 \times 10^{3} M_{\sun}$ to $10^{5} M_{\sun}$.

\begin{figure}[htbp] 
\begin{center}
\subfigure[Simulated Image]{\includegraphics[angle=0,width=6.5cm]{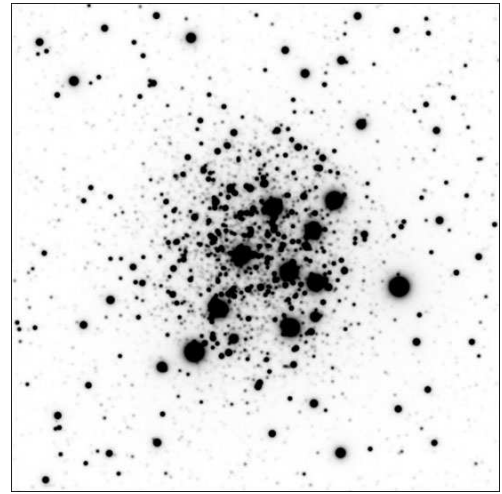}} 
\subfigure[2MASS Image]{\includegraphics[angle=0,width=6.5cm]{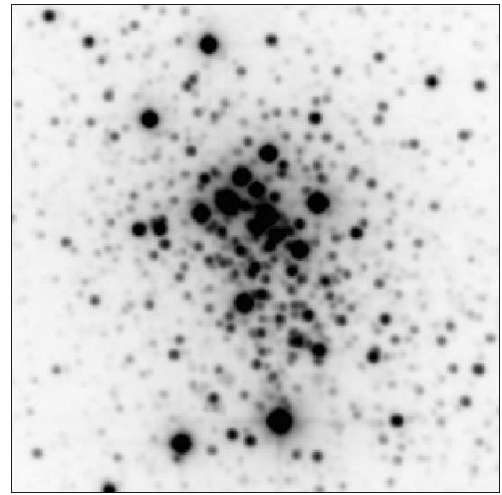}}
\end{center}
\caption{\small \texttt{MASSCLEAN} simulated images of \textit{Westerlund 1}. In the upper panel (a) we provide the simulated image in the $J$ Band.  A simulated field star population has been included.  The lower panel comes from 2MASS and is also in the $J$ Band.  Both fields shown are for a $4.8' \times 4.8'$ image. \normalsize} \label{fig:wd1}
\end{figure}

Using the input characteristics given in \S \ref{cmdetal}, we have created a simulation of \textit{Westerlund 1} in the J-band and provided in Figure \ref{fig:wd1} (a).  This can be compared directly with the 2MASS\footnote{ This publication makes use of data products from the Two Micron All Sky Survey, which is a joint project of the University of Massachusetts and the Infrared Processing and Analysis Center/California Institute of Technology, funded by the National Aeronautics and Space Administration and the National Science Foundation.}(\citeauthor{2MASS} \citeyear{2MASS}) J-band image of \textit{Westerlund 1} shown as Figure \ref{fig:wd1} (b). In looking at this comparison, one might also consider Figs. \ref{fig:hr}(a) and \ref{fig:cmd1}(a).  Presently, the simulated image has a non-realistic 'flatness' in the stellar brightnesses due to several RSGs exhibiting near identical magnitude and color.  Some amount of rebinning to randomize slightly the properties of the RSGs (something the cluster will have done more natually with the non-zero cluster age distribution) may be desirable to include in future updates and releases of the code.

\begin{figure}[htbp] 
\begin{center}
\subfigure[Simulated Image]{\includegraphics[angle=0,width=6.5cm]{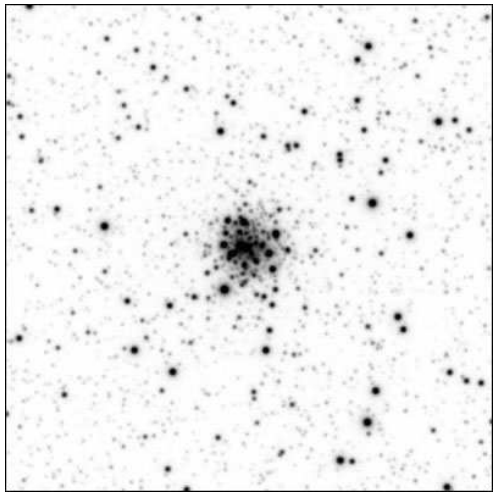}} 
\subfigure[2MASS Image]{\includegraphics[angle=0,width=6.5cm]{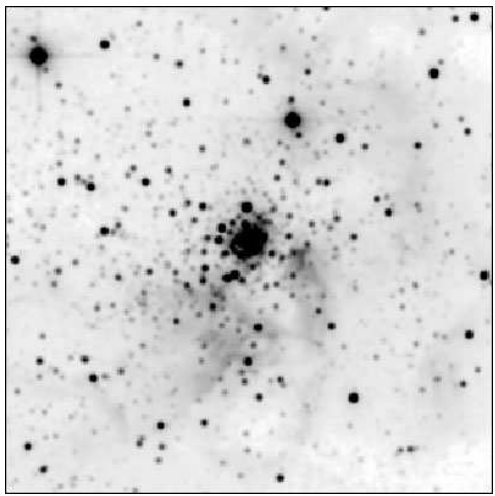}}
\end{center}
\caption{\small \texttt{MASSCLEAN} simulated image of the young Milky Way cluster, \textit{NGC 3603} is given in (a).  The real image from 2MASS is given in (b).  Both images represent a $4.4' \times 4.4'$ $J$ Band Image \normalsize} \label{fig:ngc3603}
\end{figure}

Using the data from \citeauthor*{stolte1} (\citeyear{stolte1}, \citeyear{stolte2}) we simulated \textit{NGC 3603} using: $\log (age) = 6.00$, $M_{total}=10^{4} M_{\sun}$ and solar metallicity, $A_{V}=4.5$ mag, $R_{V}=3.1$, distance $d = 6$ kpc (distance modulus $13.9$ mag). For the spatial distribution we used $r_{t} = 4.4'$ and $r_{c} = 0.4'$. 
We note that a single-power Salpeter IMF could not lead to an agreement with the actual photometric data. The best fit corresponds to Kroupa IMF with $\alpha_{1} = 0.3$, $\alpha_{2} = 1.3$ and $\alpha_{3} = 2.4$. The J-band image is shown in Figure \ref{fig:ngc3603}(a), and the 2MASS (\citeauthor{2MASS} \citeyear{2MASS}) image is shown in Figure \ref{fig:ngc3603}(b).  It should be clear to the reader we do not include nebulosity in the image simulations. 

We have built a suite of different cluster models by varying all of the input variables within the range of known measurements for these two clusters.  These resulting simulations were tested against the available catalogs of images and photometric data for the two clusters as a means of constraining the cluster properties.  The phase space of properties is far too large to enable \texttt{MASSCLEAN} to serve as an efficient means of modeling clusters.  However, if a reasonable guess can be made as to many of the cluster properties (age, distance, extinction) \texttt{MASSCLEAN} can work extremely effectively in a limited region of characteristics to determine the best model match to the observed data.  A complete analysis for \textit{NGC 3603} and \textit{Westerlund 1} will be presented in a subsequent paper, along with the development of a proper goodness of fit (Popescu \& Hanson, in prep.).

\subsection{Simulations of multiple starburst events or subclustering}

\begin{figure}[htbp]
\begin{center}
\includegraphics[angle=0,width=7.0cm]{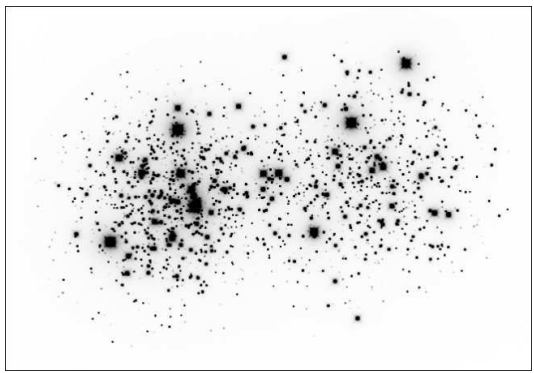} 
\end{center}
\caption{\small \texttt{MASSCLEAN} simulated image of the double Milky Way cluster, $h$ and $\chi$ \textit{Persei} (\textit{NGC 869}, \textit{NGC 884}). The image represents a $30' \times 30'$ field in the $V$ band.  No field star simulation is included to allow the separate cluster radii to be better viewed. \normalsize}\label{fig:persei}
\end{figure}

\texttt{MASSCLEAN} can be used to simulate clusters with multiple starbursting events.  As a demonstration of this, we present the twin cluster $h$ and $\chi$ \textit{Persei} (\textit{NGC 869}, \textit{NGC 884}), simulated based on measurements made by \citeauthor*{bragg} \citeyearpar{bragg}.  Our V-band simulation is shown in Figure \ref{fig:persei}.  Here, we applied the following inputs to our simulation: $\log (age_{h}) = 7.05$, $\log (age_{\chi}) = 7.09$, $M_{h}=5.5 \times 10^{3} M_{\sun}$, $M_{\chi}=4.3 \times 10^{3} M_{\sun}$.  For both clusters we assumed solar metallicity, $A_{V}=1.6$ mag, $R_{V}=3.1$, and a distance of $2$ kpc (distance modulus $11.5$ mag). The spatial distribution of the twin cluster is described by $r_{t} = 9.6'$ and $r_{c} = 7.01'$ for $h$ \textit{Persei}, and $r_{t} = 9.6'$ and $r_{c} = 8.86'$ for $\chi$ \textit{Persei}. 
A Kroupa IMF has been used for both clusters, with $\alpha_{1} = 0.3$, $\alpha_{2} = 1.3$ and $\alpha_{3} = 2.3$.  A field star population has not been applied in this image, making it easier to see the extent of the two clusters in the simulation.  

Because we use a King model for the spatial extent, it would not be possible to use our code to model complex OB associations.  However, it might be possible to contruct a reasonable OB association using multiple bursts with differing radii and age such as demonstrated for $h$ and $\chi$ \textit{Persei}.

\subsection{Image simulations of extragalactic clusters} \label{else}

Because the linear scale of the cluster (in parsecs) and distance to the cluster is determined by the user, one can use \texttt{MASSCLEAN} to simulate a cluster at any distance, even extragalactic stellar clusters.  We have created a series of simulations made using the same initial inputs as in Figure \ref{fig:wd1}(a) and Figure \ref{fig:ngc3603}(a) of \textit{Westerlund 1} and \textit{NGC 3603}, respectively.  Only now we have simulated these clusters over a range of ages $\log (age/years) = [6,8]$ and placed the clusters at the distance of M31.  The simulation is designed to represent the depth and resolution (0.14") expected from the {\it Hubble Space Telescope} using the {\it Advanced Camera for Surveys} instrument.  Both V-band and I-band simulated images are presented.  A model for the field stars has also been applied. We generated field stars close to the ones observed in real cluster images in M31 and to keep the cluster distinguishable from the stellar field for at least 10 million years. Beside each image we provide the current view of the CMD for that cluster. All CMDs are scaled to the same magnitude limits. 

\begin{figure*}[htbp] 
\begin{center}
\subfigure[log(age/yr)=6.00]{\includegraphics[angle=0,width=0.24\textwidth]{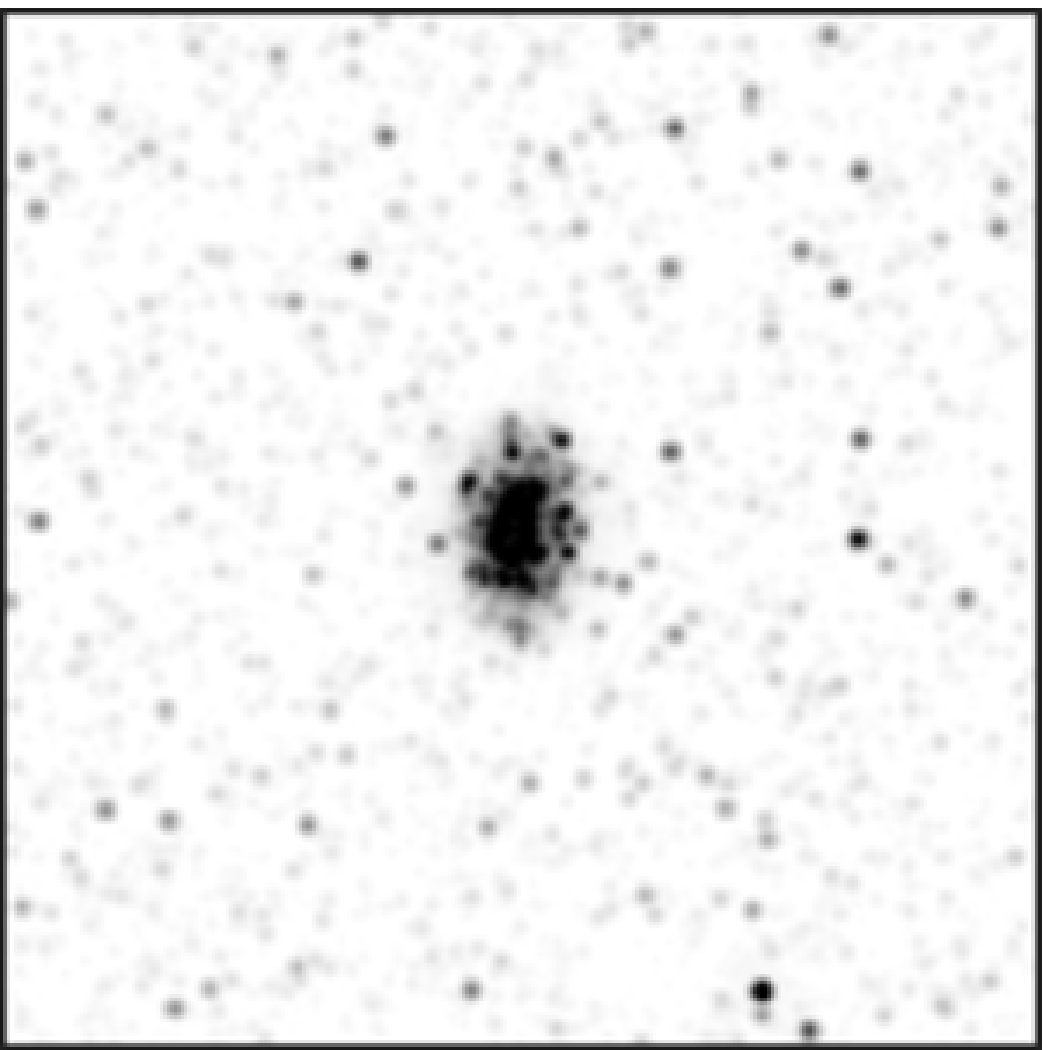}} 
\subfigure[log(age/yr)=6.00]{\includegraphics[angle=0,width=0.24\textwidth]{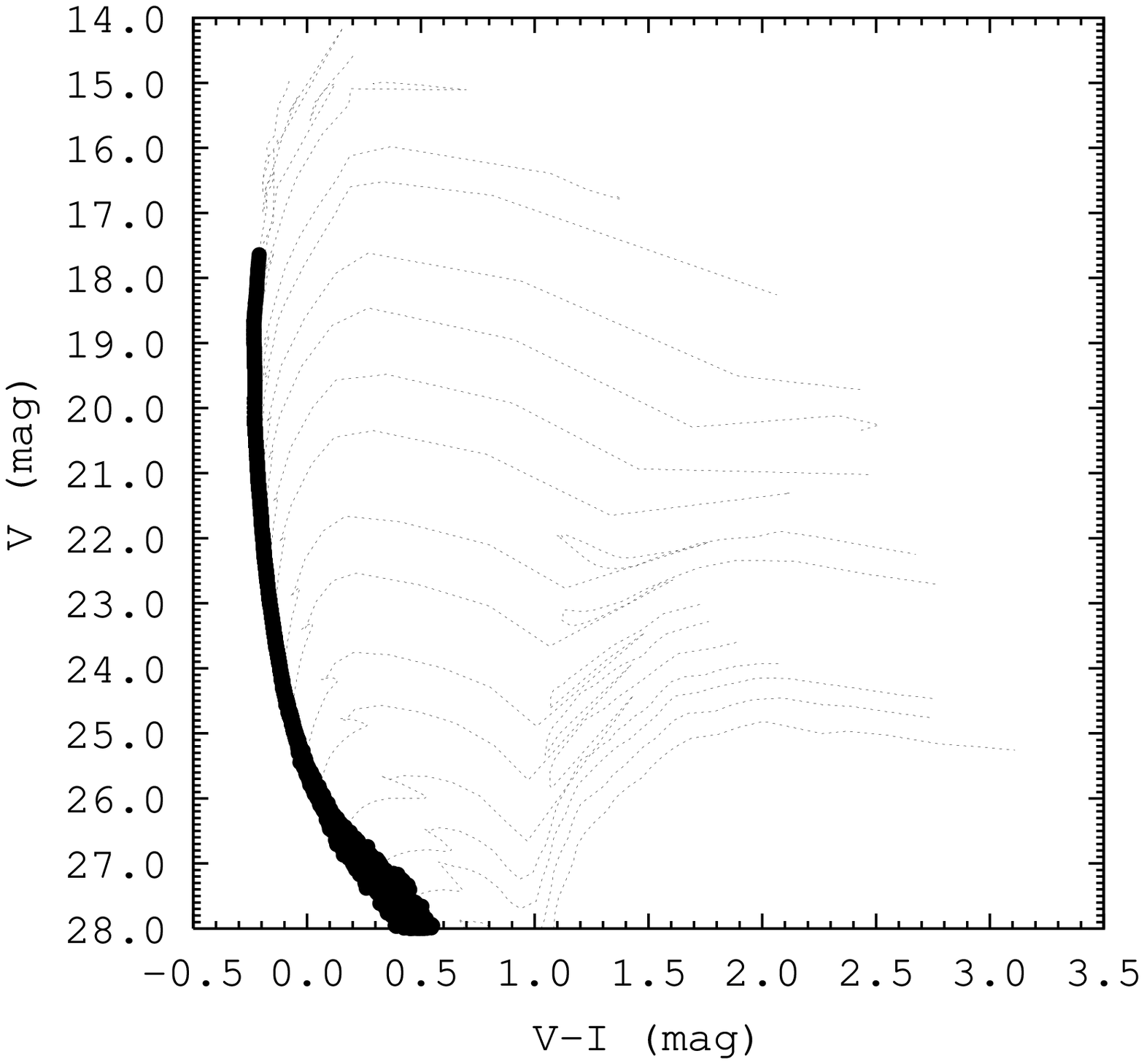}}
\subfigure[log(age/yr)=6.50]{\includegraphics[angle=0,width=0.24\textwidth]{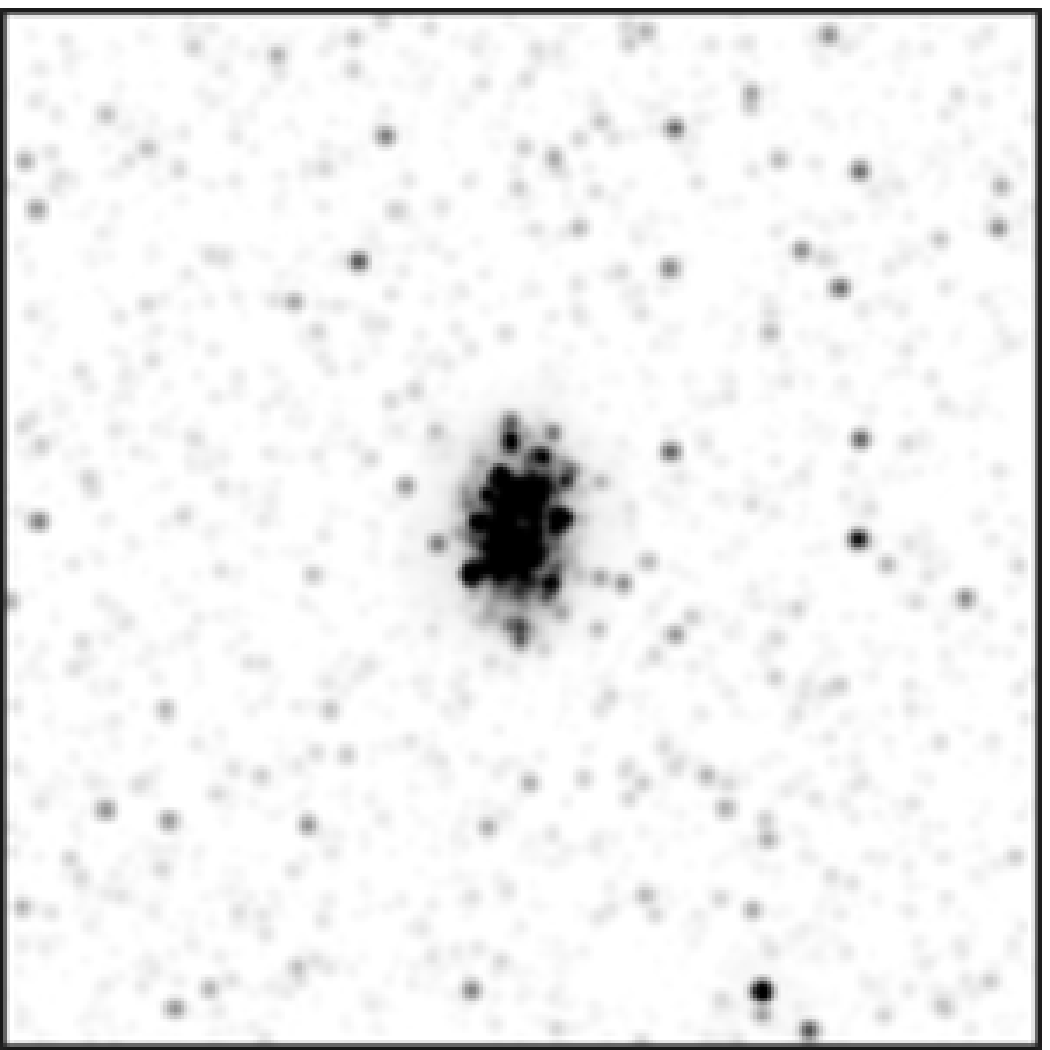}}
\subfigure[log(age/yr)=6.50]{\includegraphics[angle=0,width=0.24\textwidth]{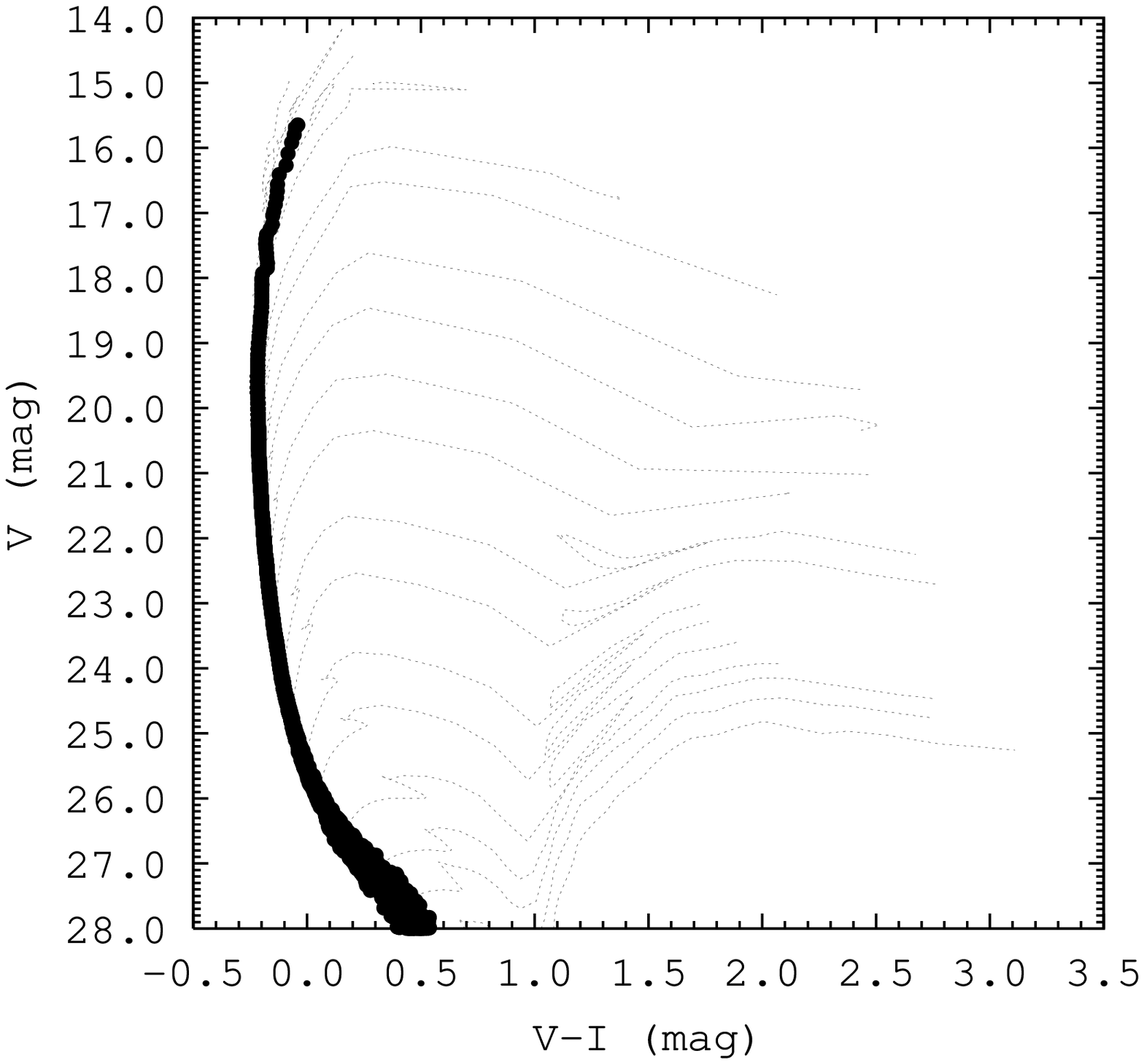}}
\subfigure[log(age/yr)=6.65]{\includegraphics[angle=0,width=0.24\textwidth]{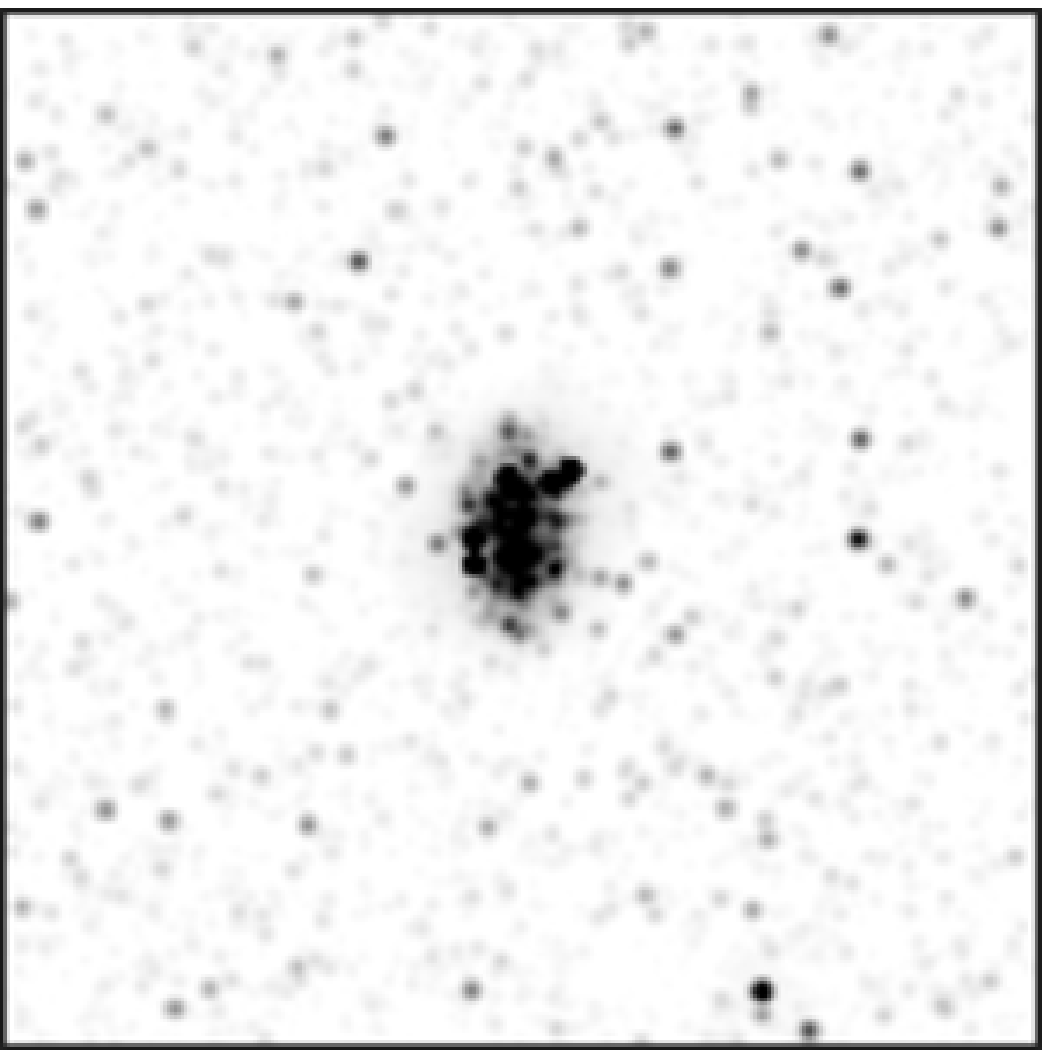}}
\subfigure[log(age/yr)=6.65]{\includegraphics[angle=0,width=0.24\textwidth]{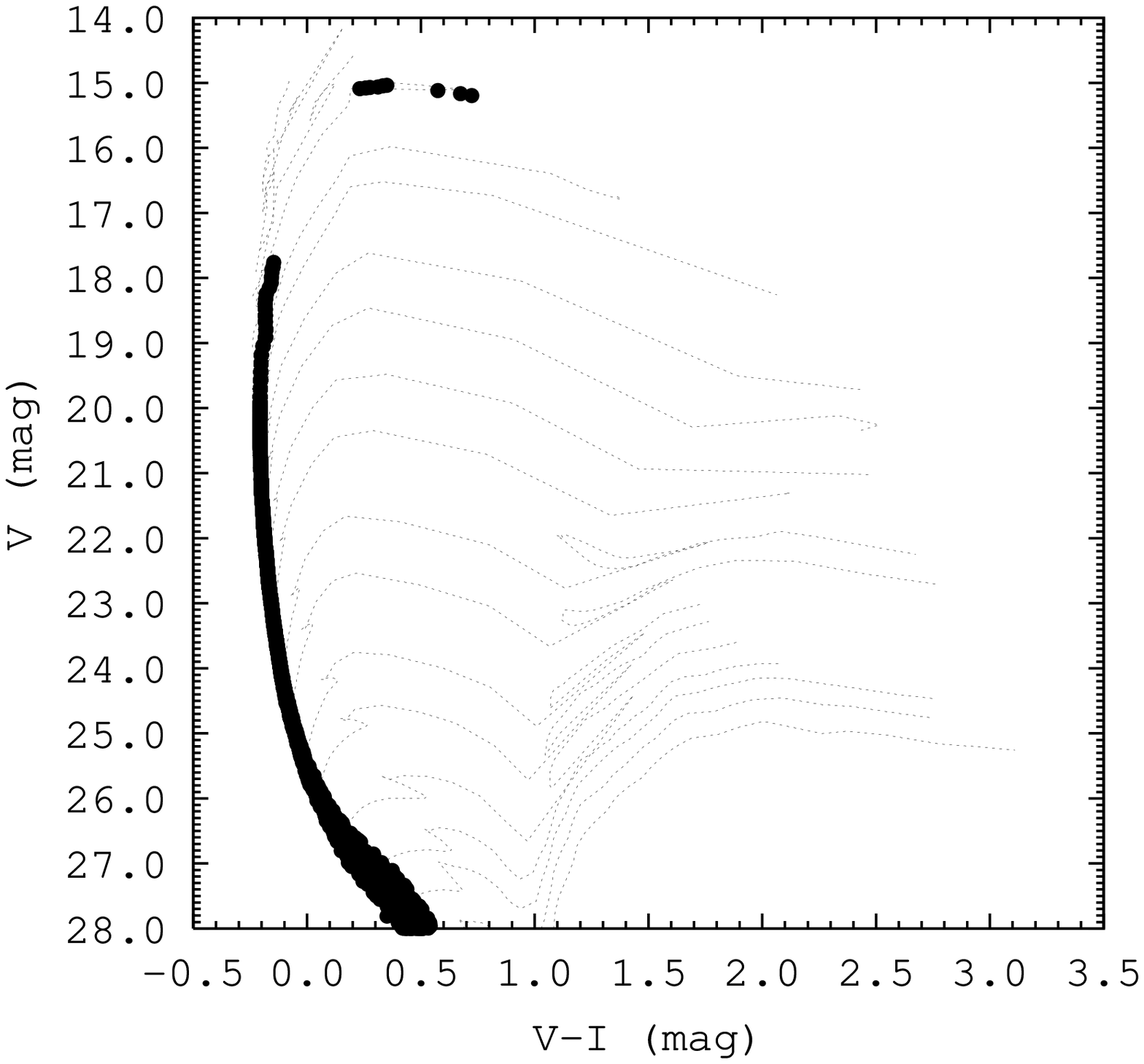}}
\subfigure[log(age/yr)=6.75]{\includegraphics[angle=0,width=0.24\textwidth]{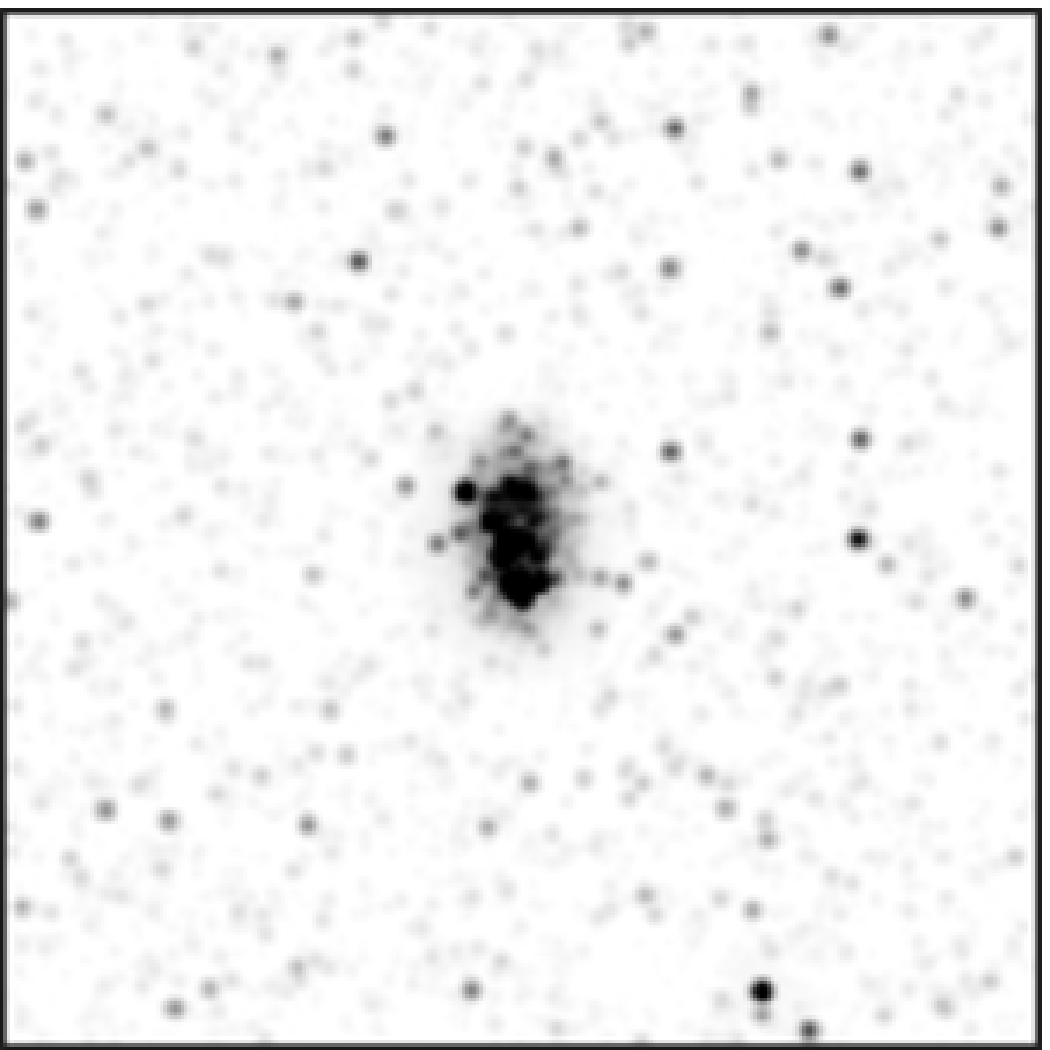}} 
\subfigure[log(age/yr)=6.75]{\includegraphics[angle=0,width=0.24\textwidth]{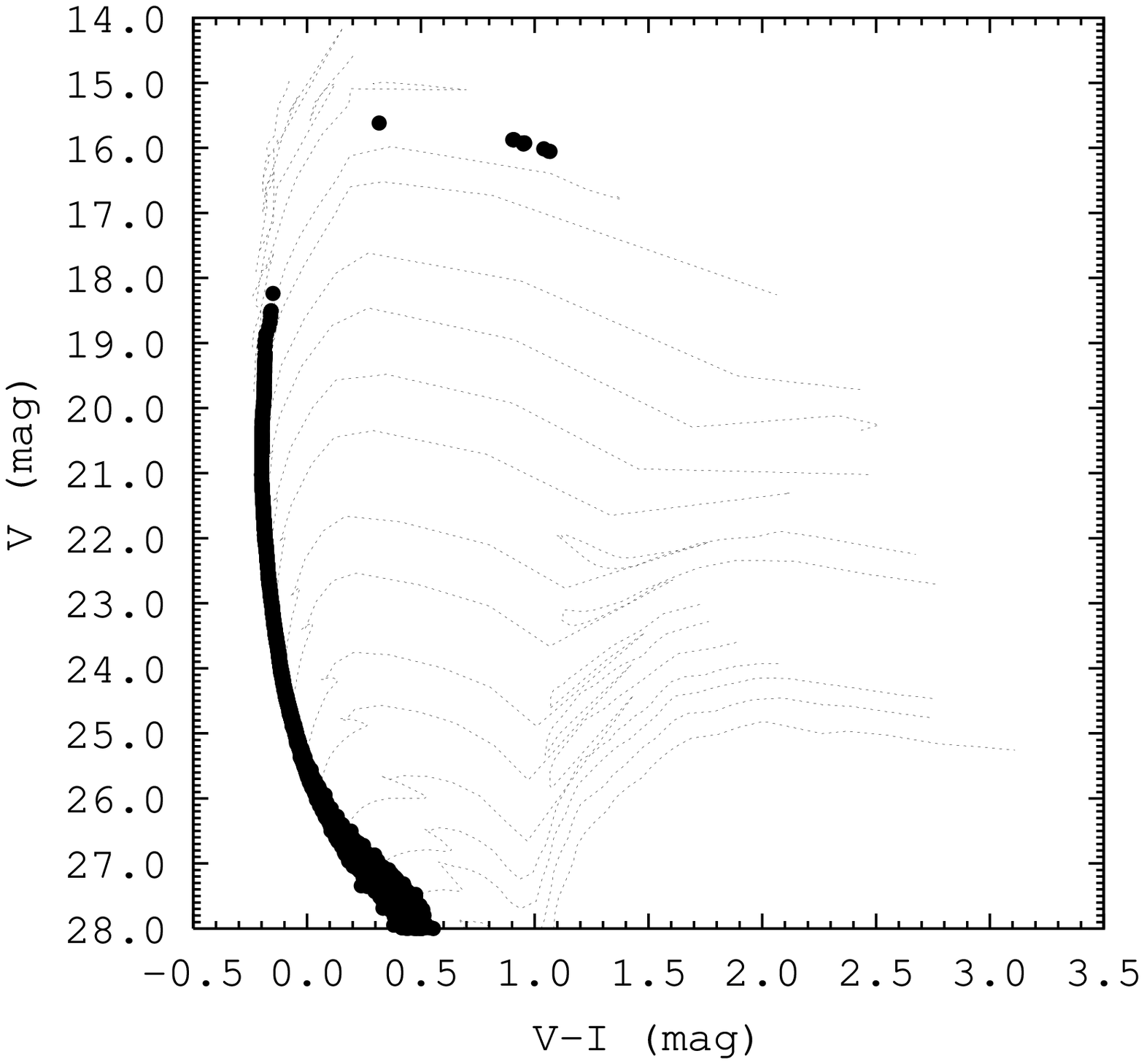}} 
\subfigure[log(age/yr)=7.00]{\includegraphics[angle=0,width=0.24\textwidth]{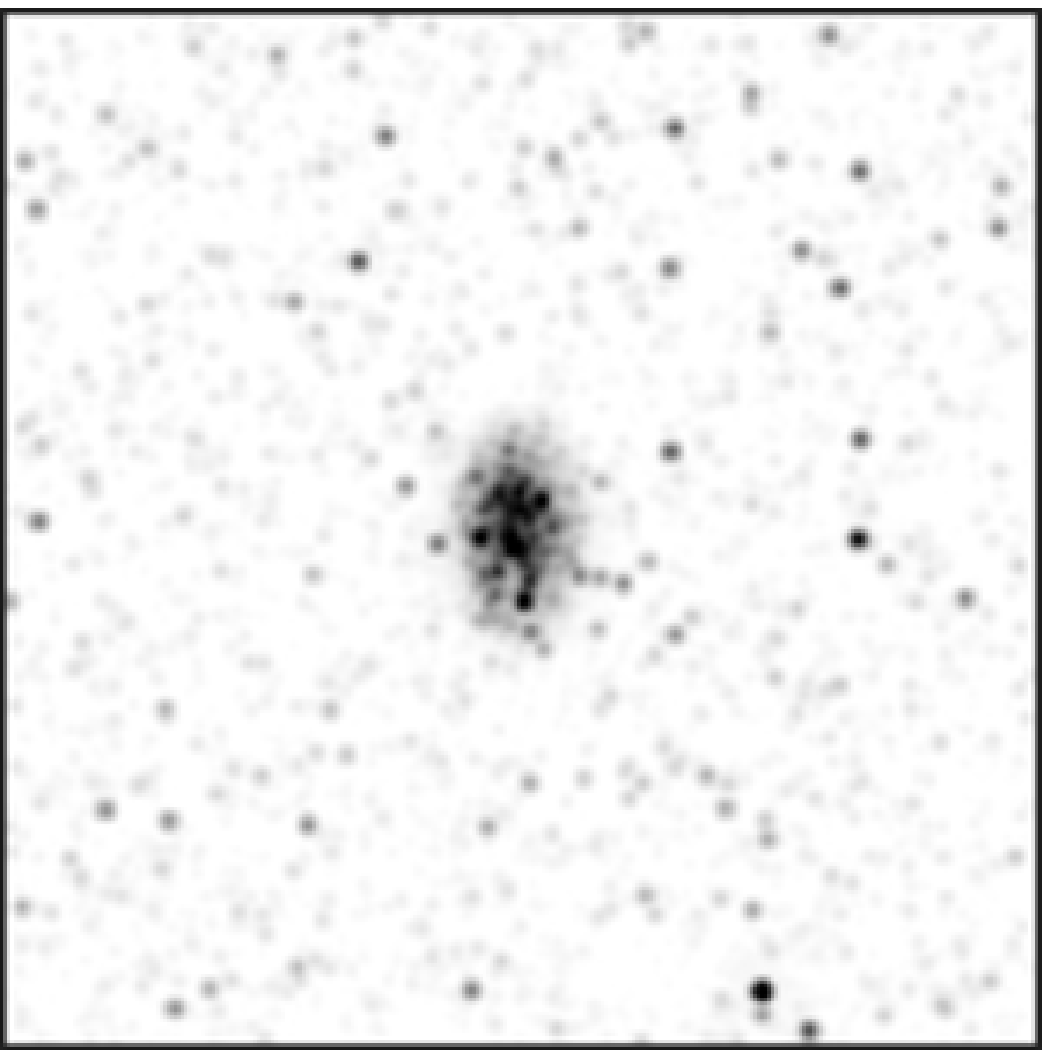}}
\subfigure[log(age/yr)=7.00]{\includegraphics[angle=0,width=0.24\textwidth]{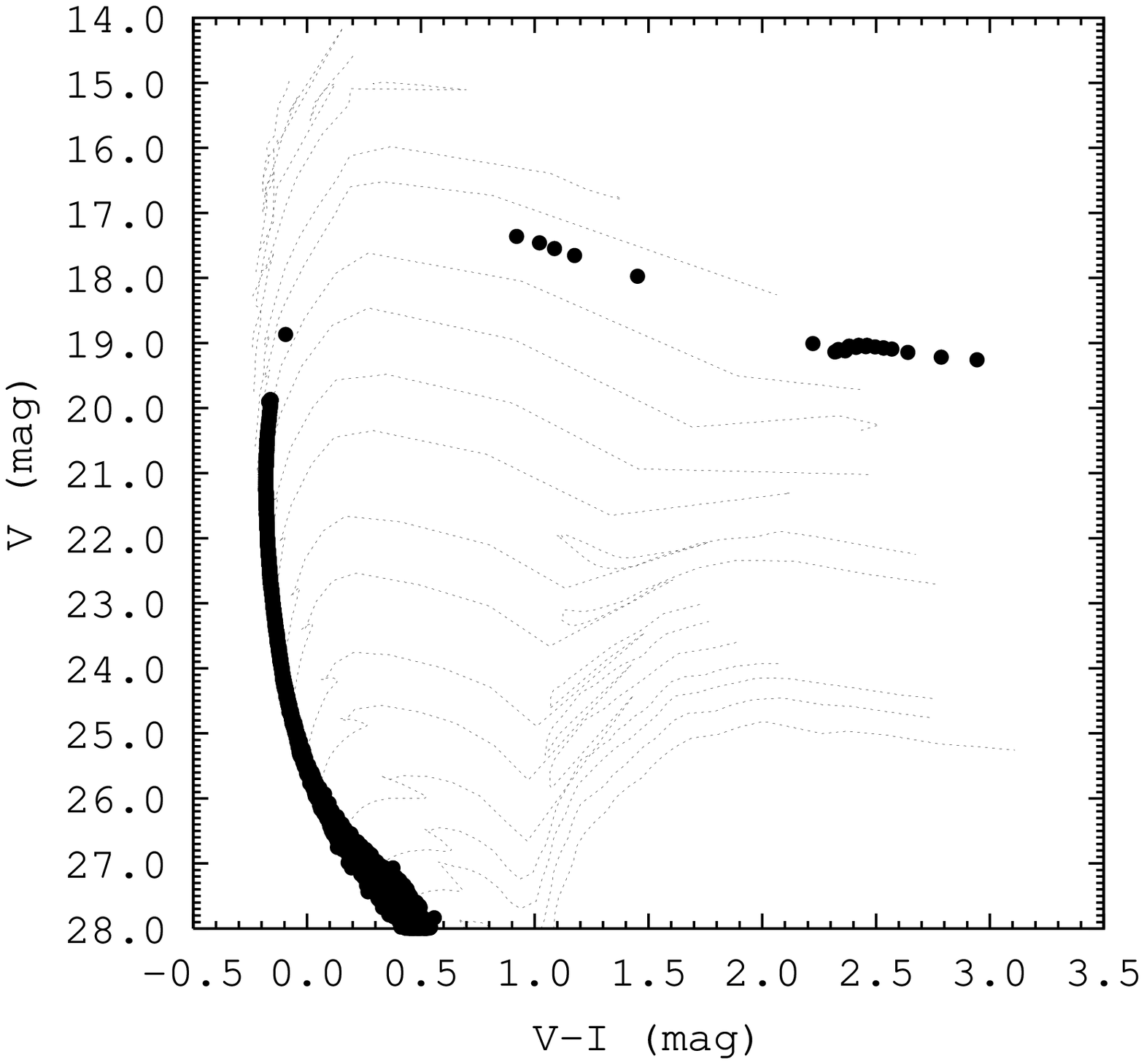}}
\subfigure[log(age/yr)=7.20]{\includegraphics[angle=0,width=0.24\textwidth]{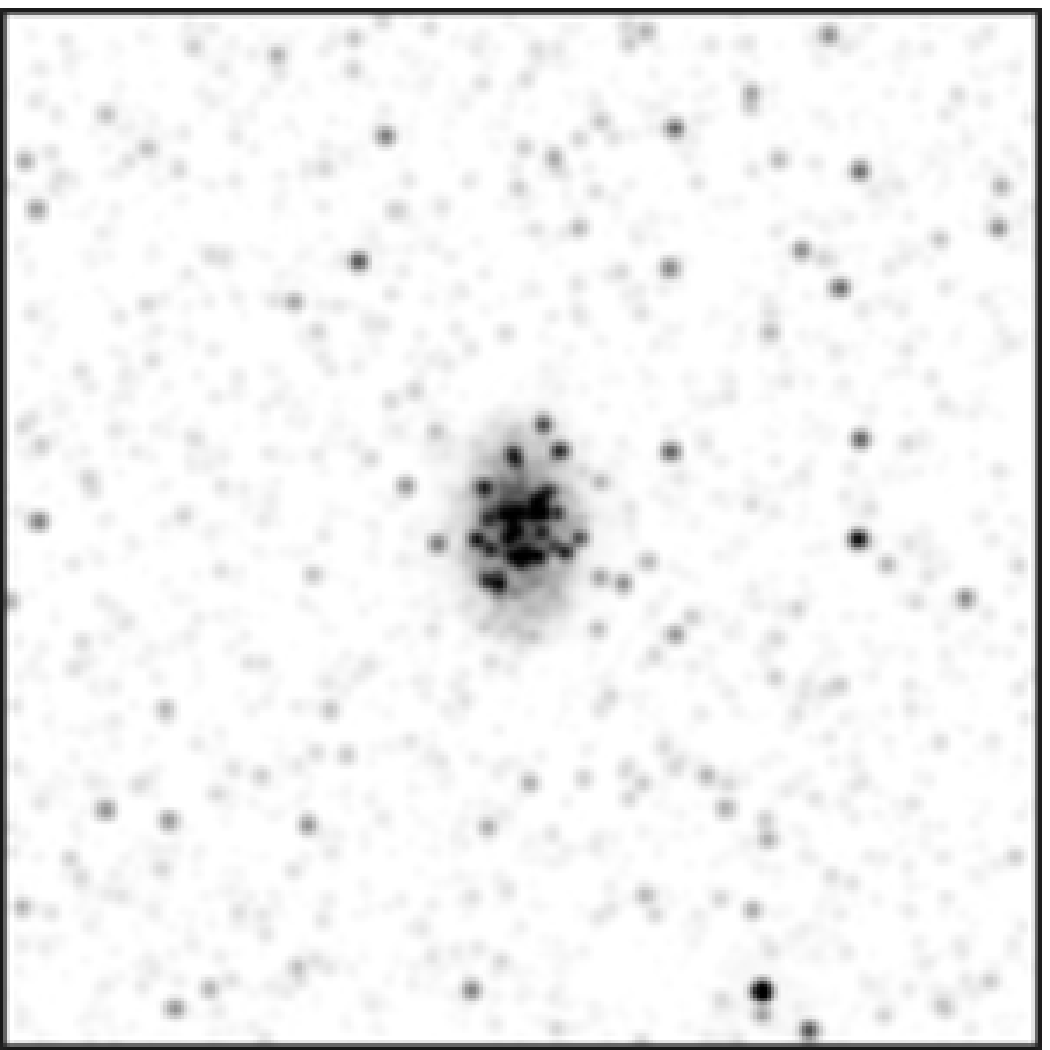}}
\subfigure[log(age/yr)=7.20]{\includegraphics[angle=0,width=0.24\textwidth]{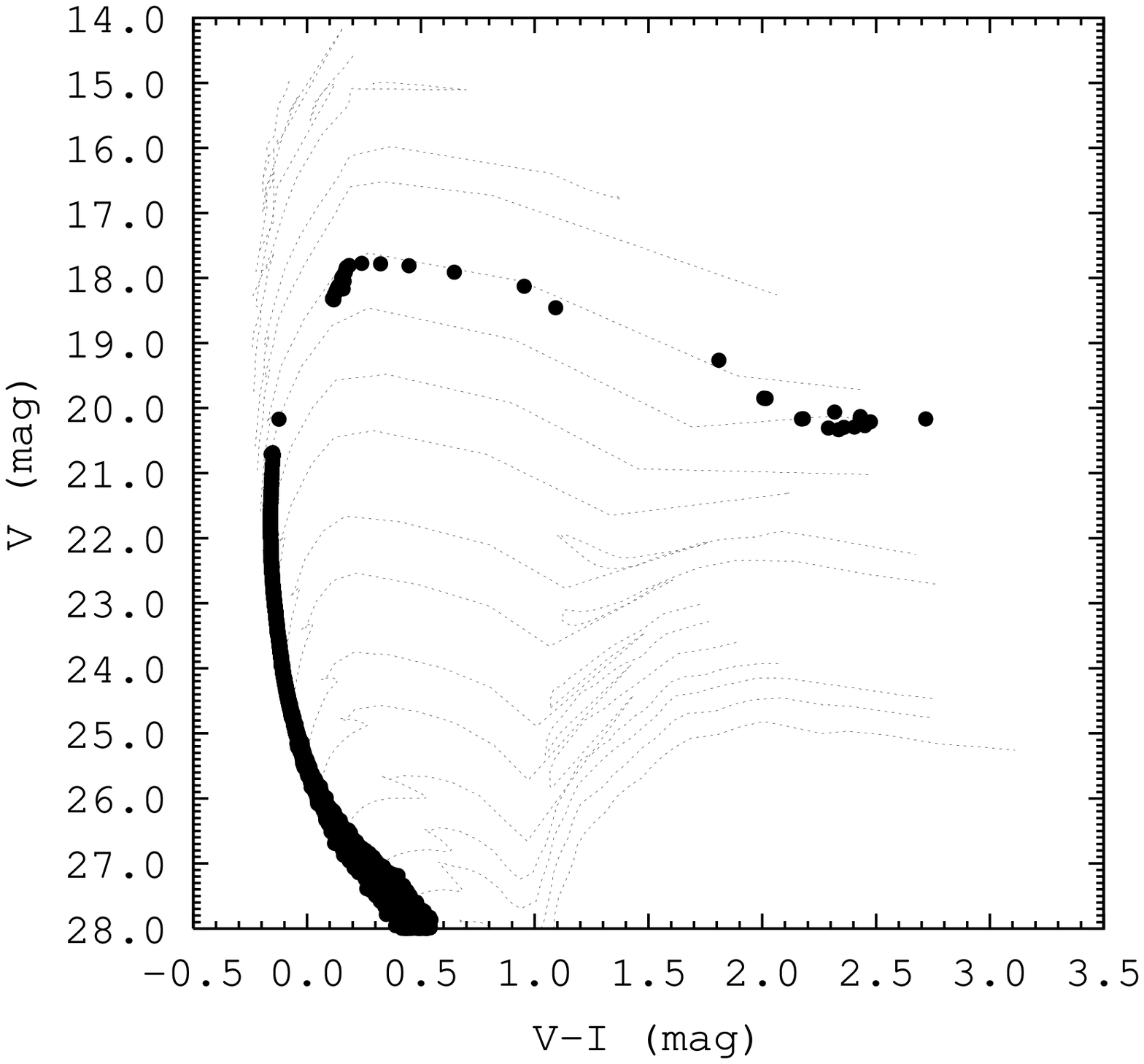}}
\subfigure[log(age/yr)=7.50]{\includegraphics[angle=0,width=0.24\textwidth]{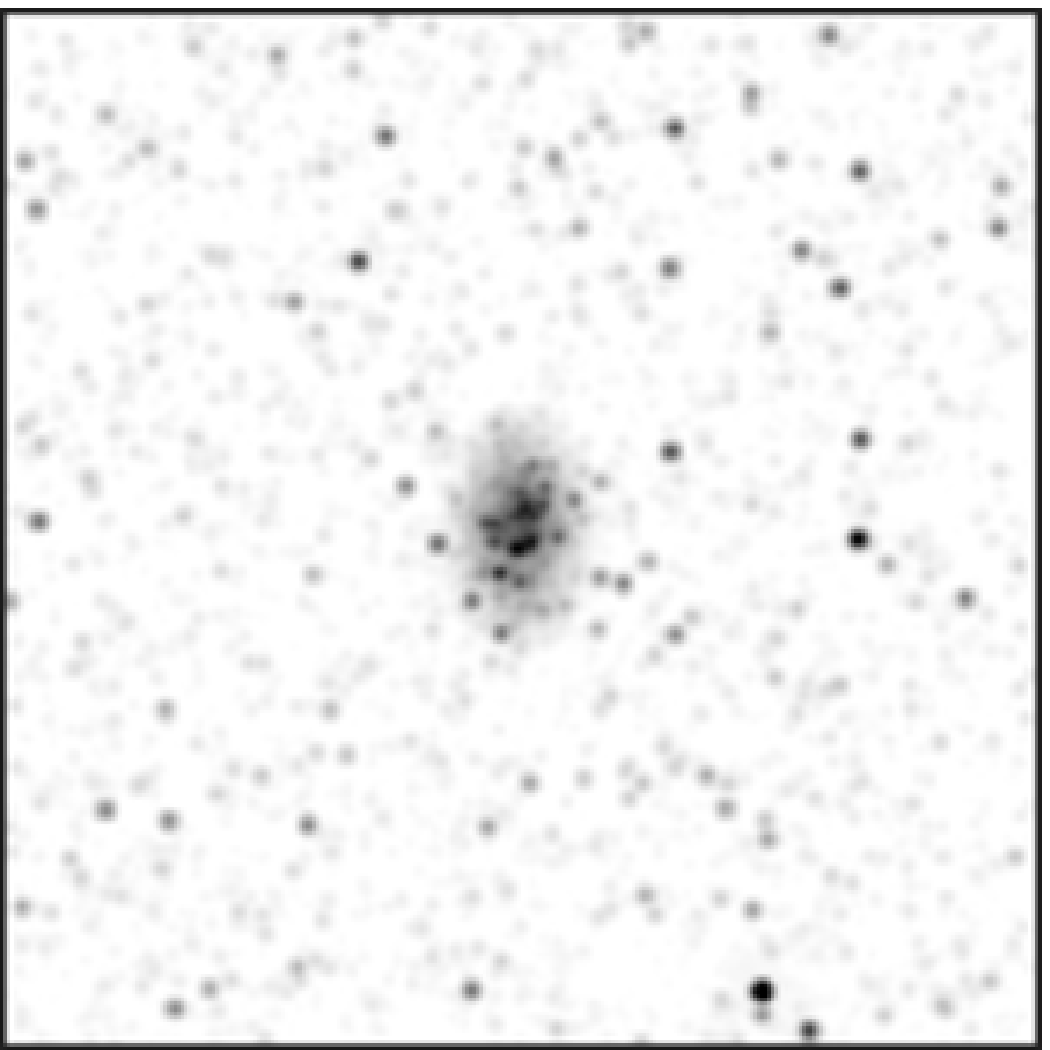}}
\subfigure[log(age/yr)=7.50]{\includegraphics[angle=0,width=0.24\textwidth]{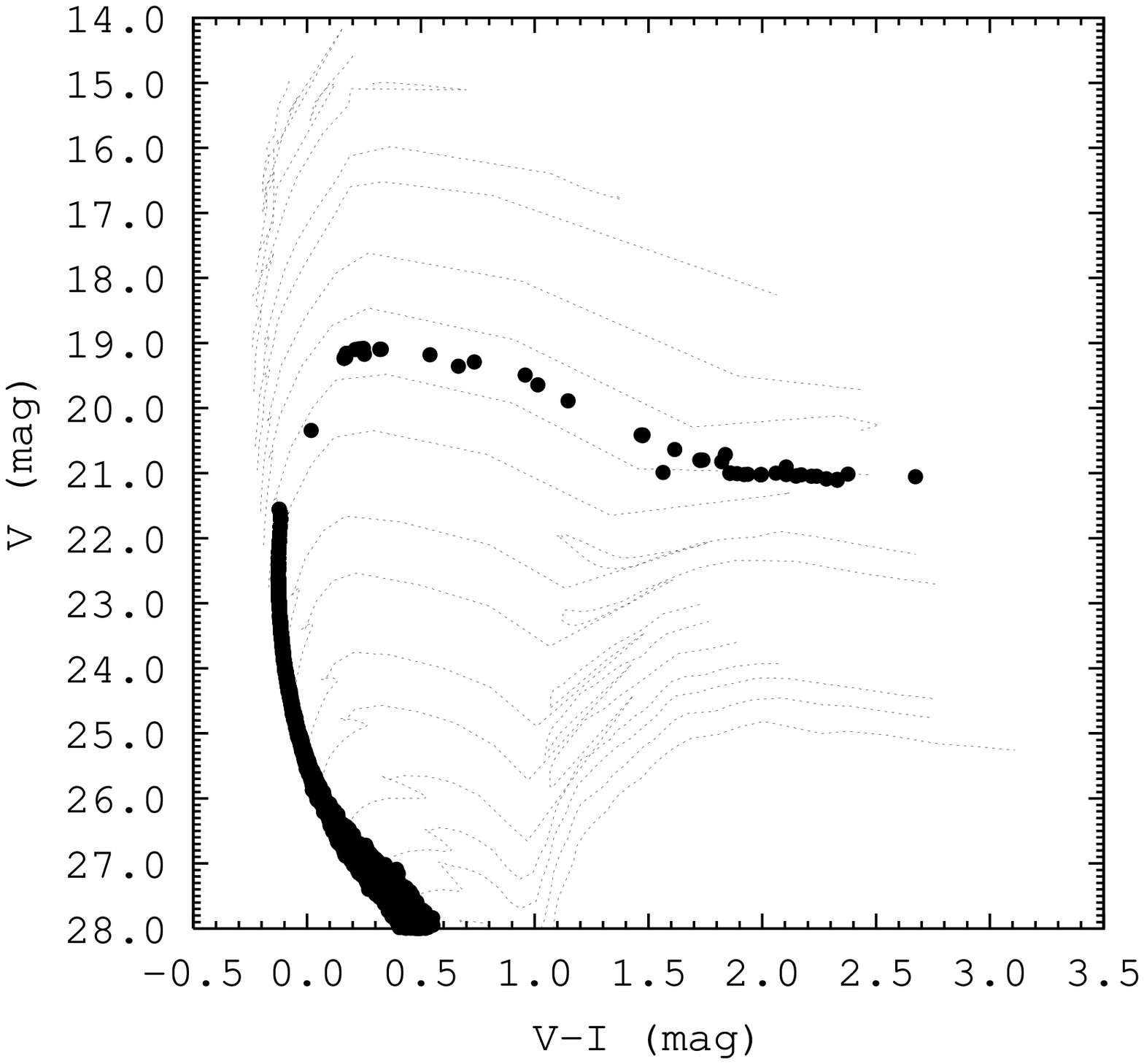}}
\subfigure[log(age/yr)=8.00]{\includegraphics[angle=0,width=0.24\textwidth]{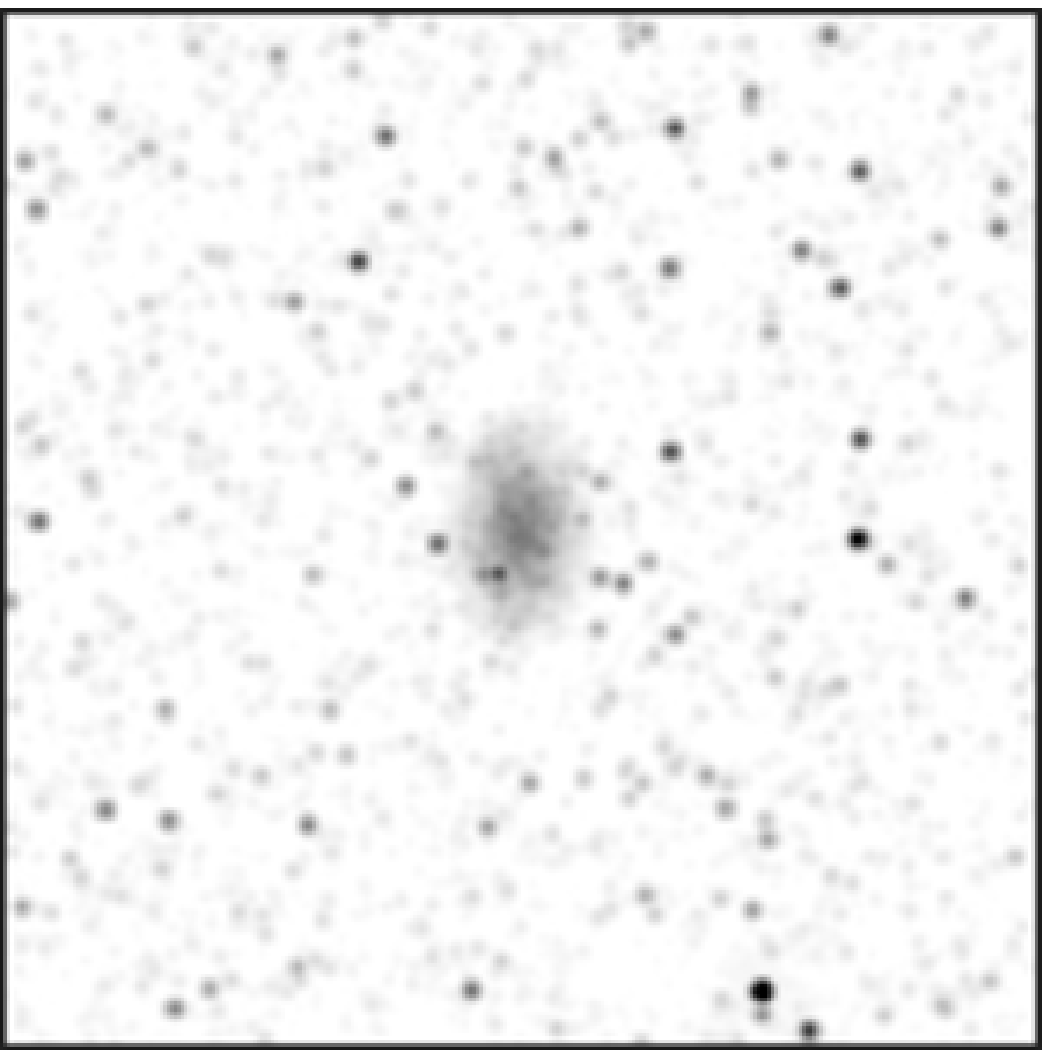}}
\subfigure[log(age/yr)=8.00]{\includegraphics[angle=0,width=0.24\textwidth]{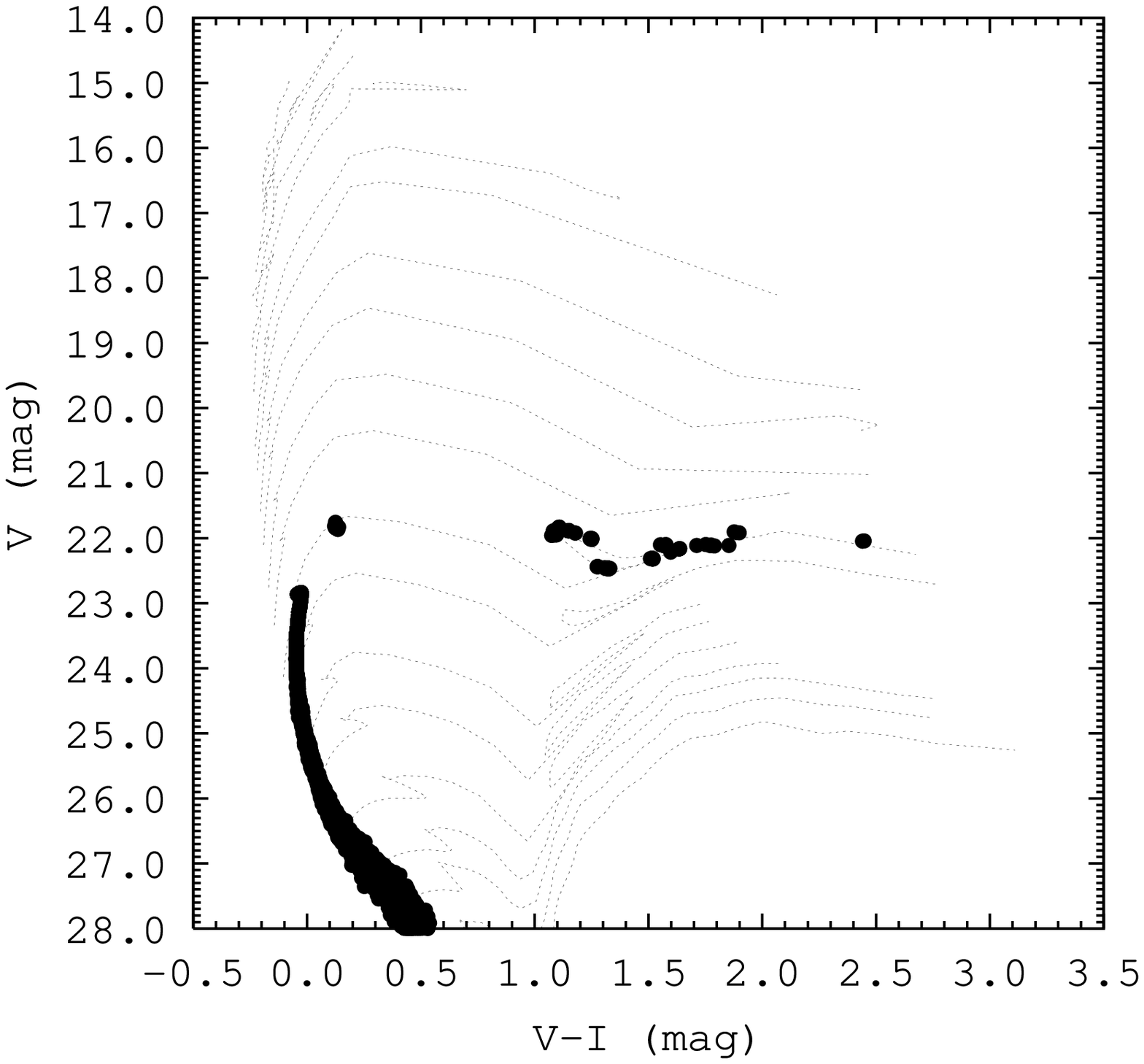}}
\caption{\small \textit{Westerlund 1} in M31 (V Band), simulated images and color-magnitude diagrams.\normalsize}\label{fig:tbd01}
\end{center}
\end{figure*}

\begin{figure*}[htbp] 
\begin{center}
\subfigure[log(age/yr)=6.00]{\includegraphics[angle=0,width=0.24\textwidth]{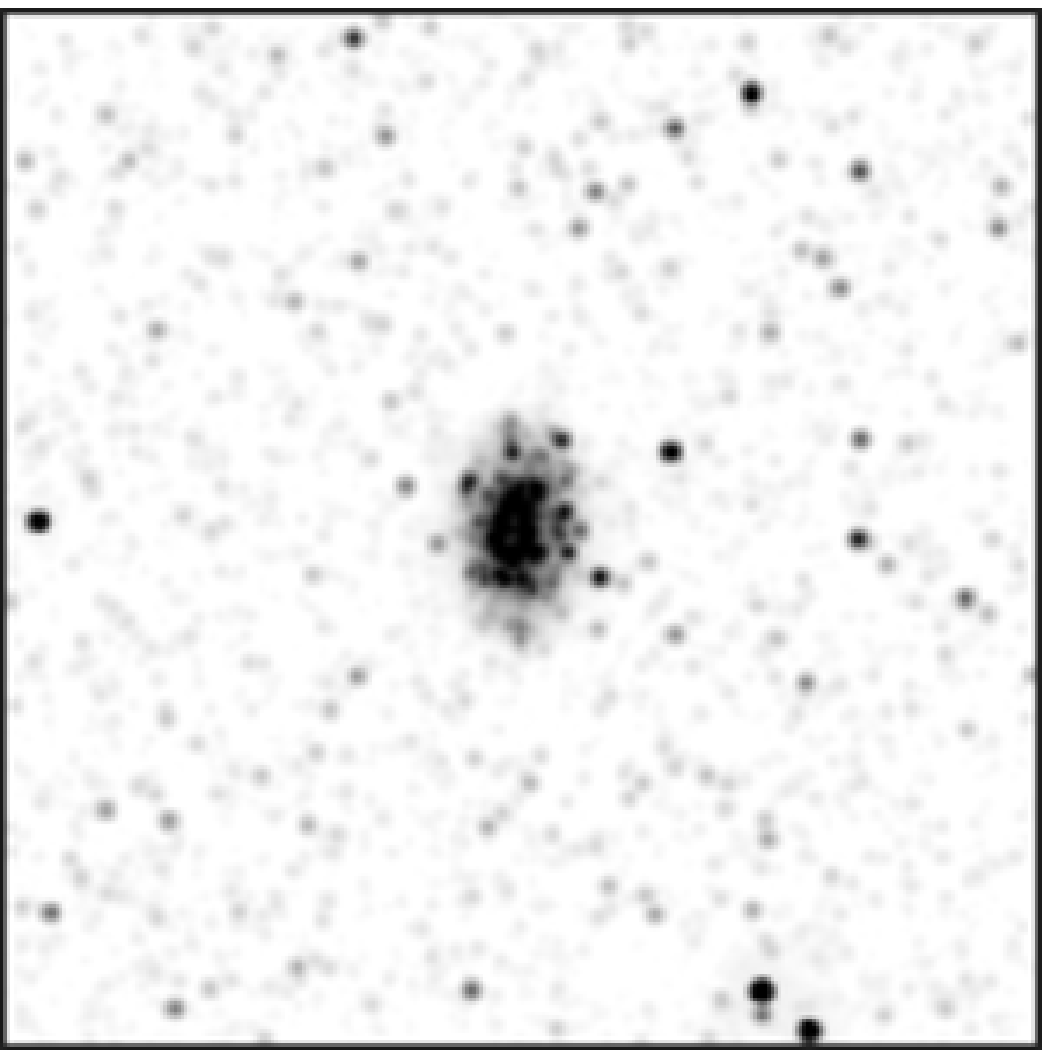}}
\subfigure[log(age/yr)=6.00]{\includegraphics[angle=0,width=0.24\textwidth]{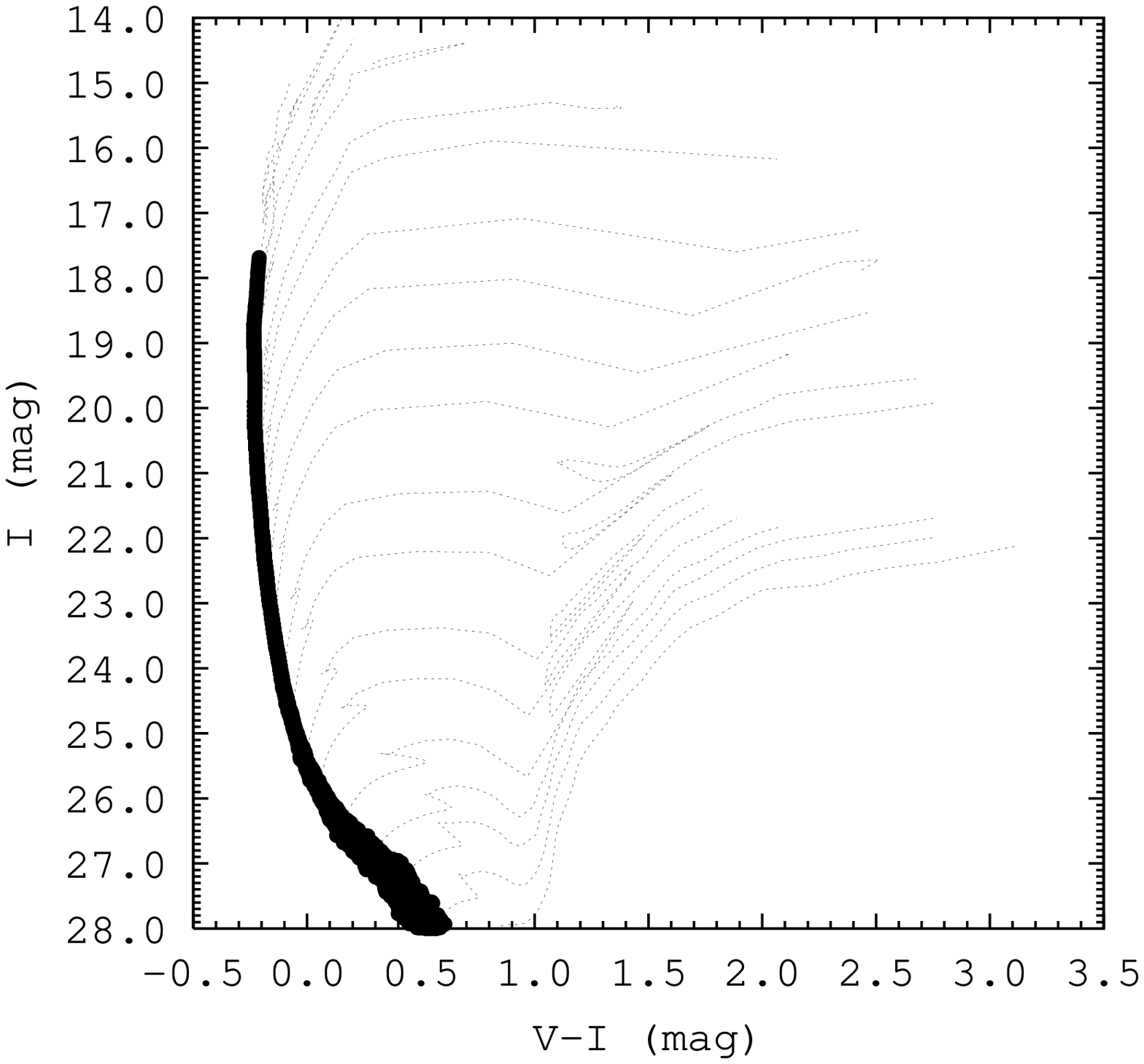}}  
\subfigure[log(age/yr)=6.50]{\includegraphics[angle=0,width=0.24\textwidth]{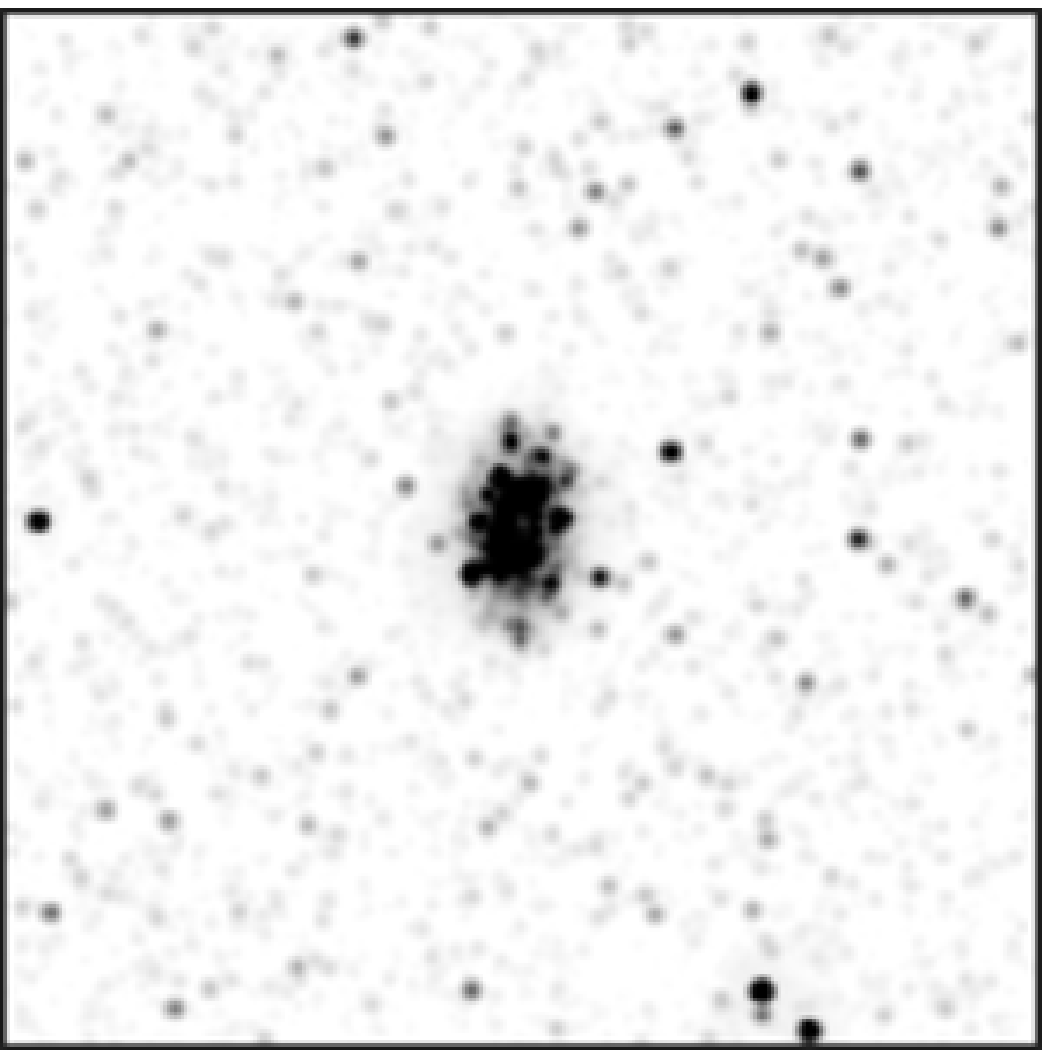}}
\subfigure[log(age/yr)=6.50]{\includegraphics[angle=0,width=0.24\textwidth]{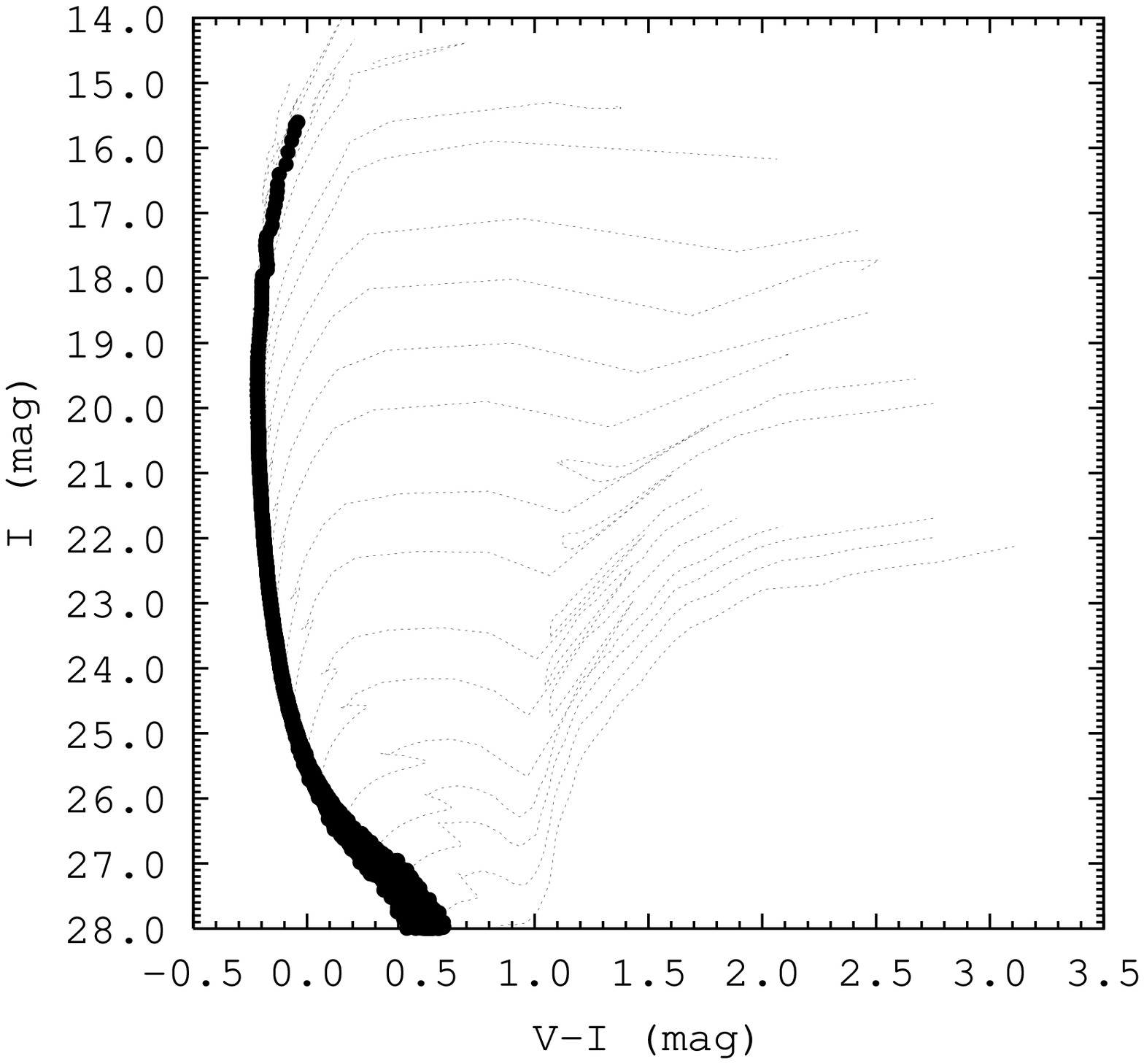}}
\subfigure[log(age/yr)=6.65]{\includegraphics[angle=0,width=0.24\textwidth]{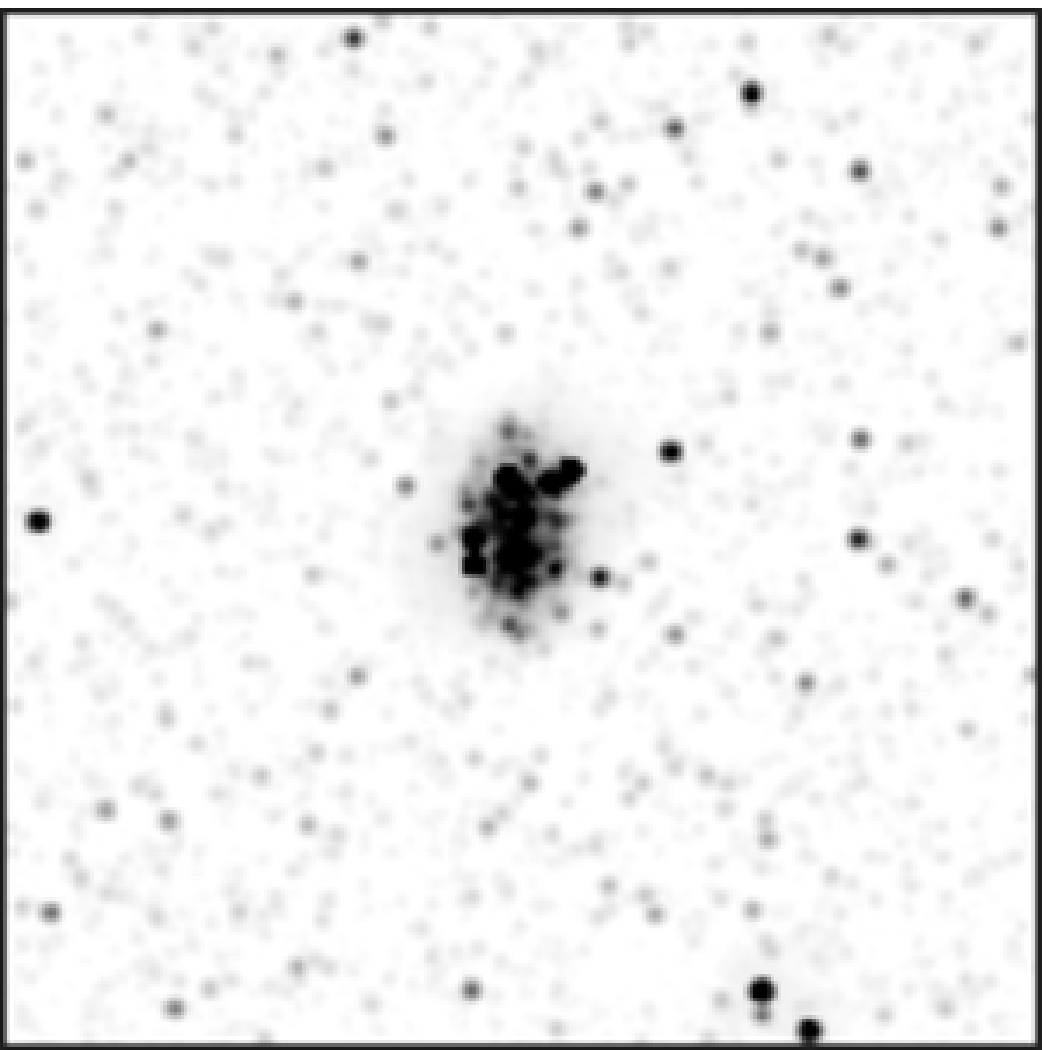}}
\subfigure[log(age/yr)=6.65]{\includegraphics[angle=0,width=0.24\textwidth]{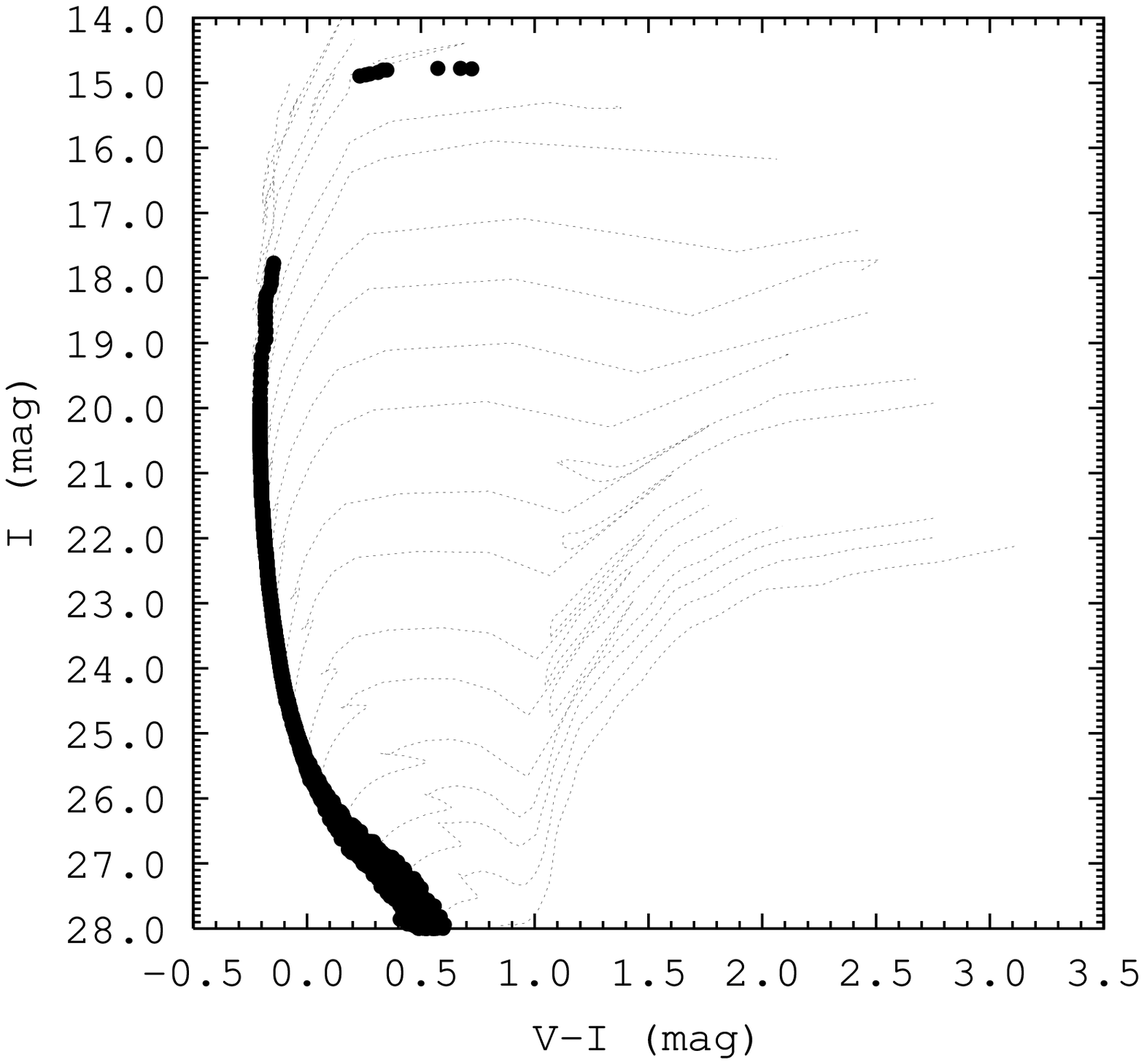}}
\subfigure[log(age/yr)=6.75]{\includegraphics[angle=0,width=0.24\textwidth]{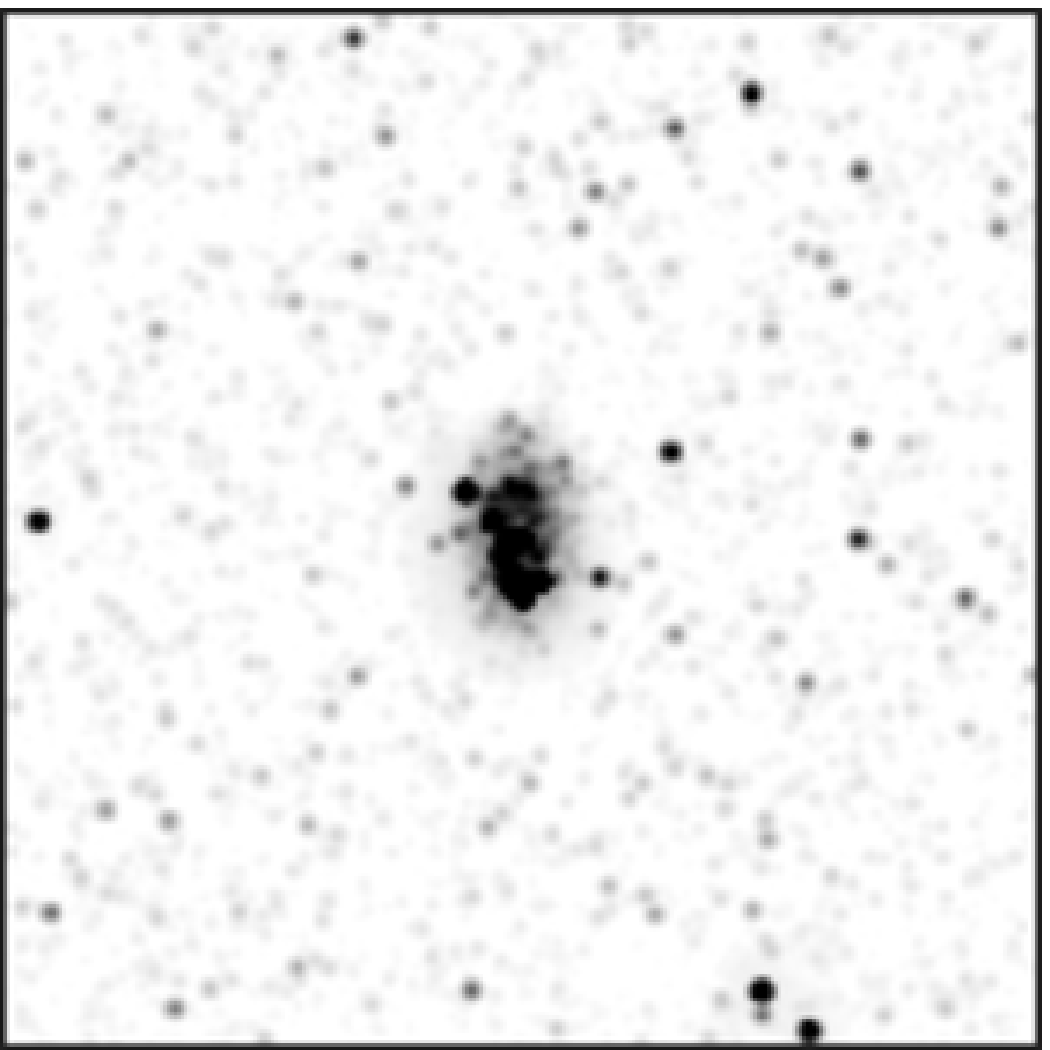}}
\subfigure[log(age/yr)=6.75]{\includegraphics[angle=0,width=0.24\textwidth]{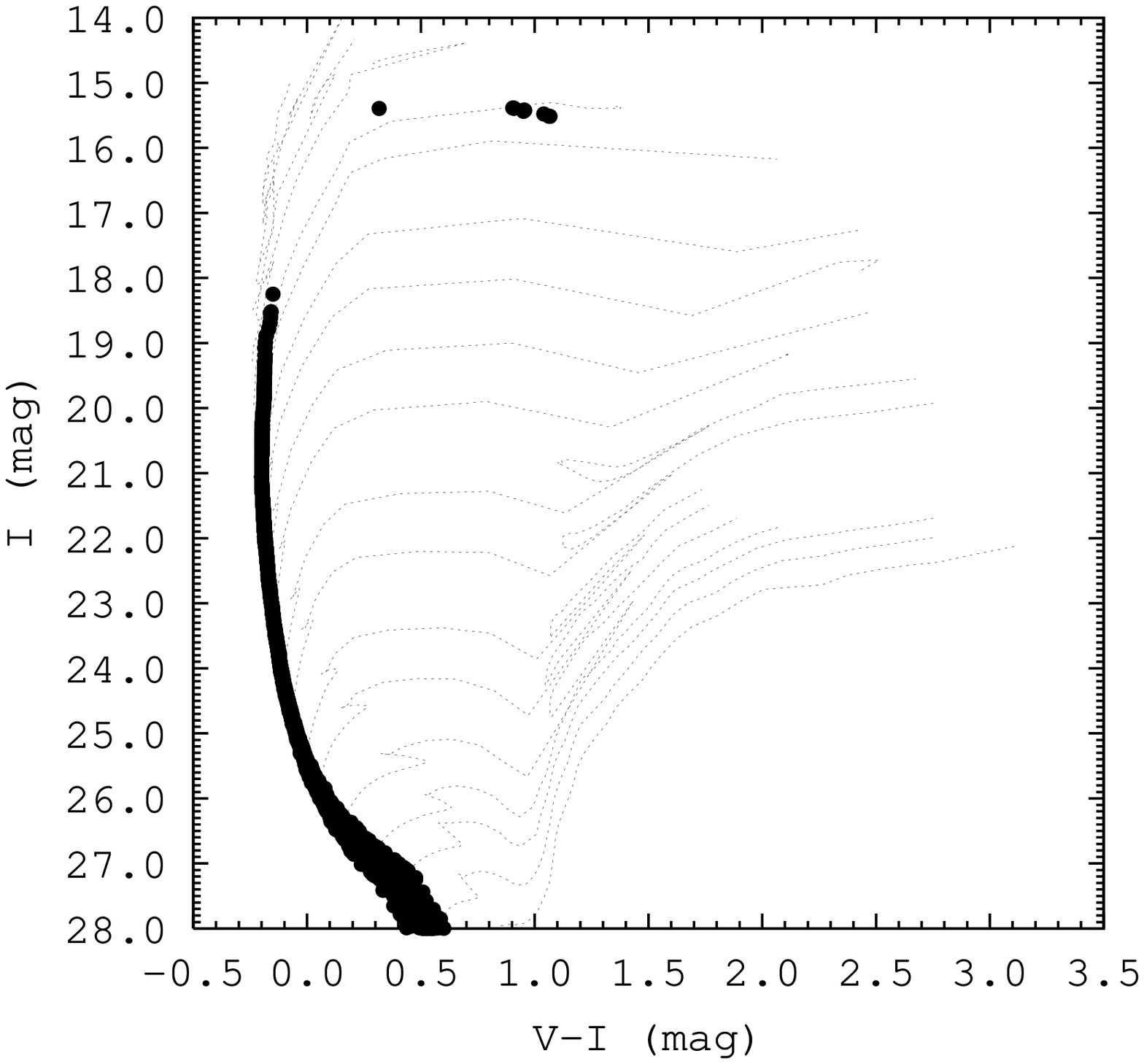}} 
\subfigure[log(age/yr)=7.00]{\includegraphics[angle=0,width=0.24\textwidth]{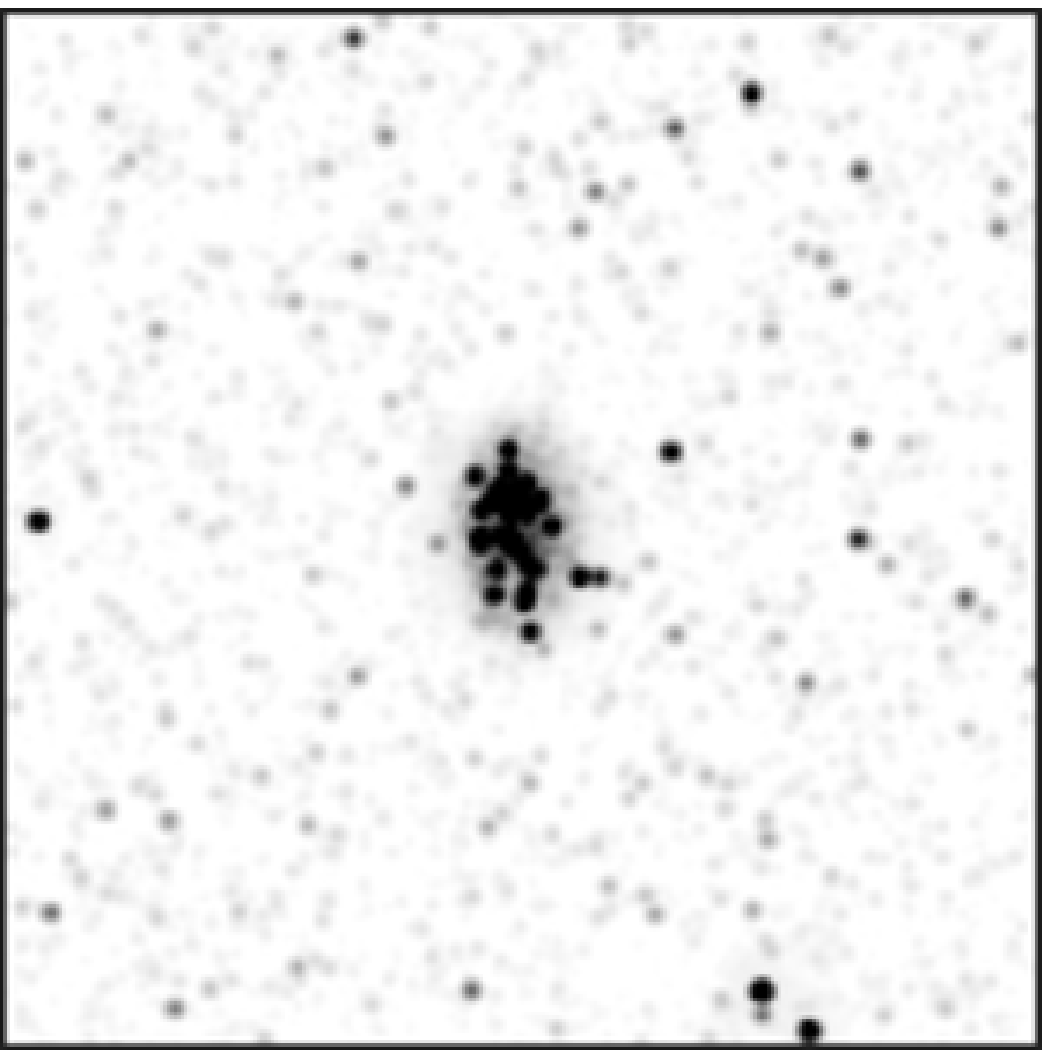}}
\subfigure[log(age/yr)=7.00]{\includegraphics[angle=0,width=0.24\textwidth]{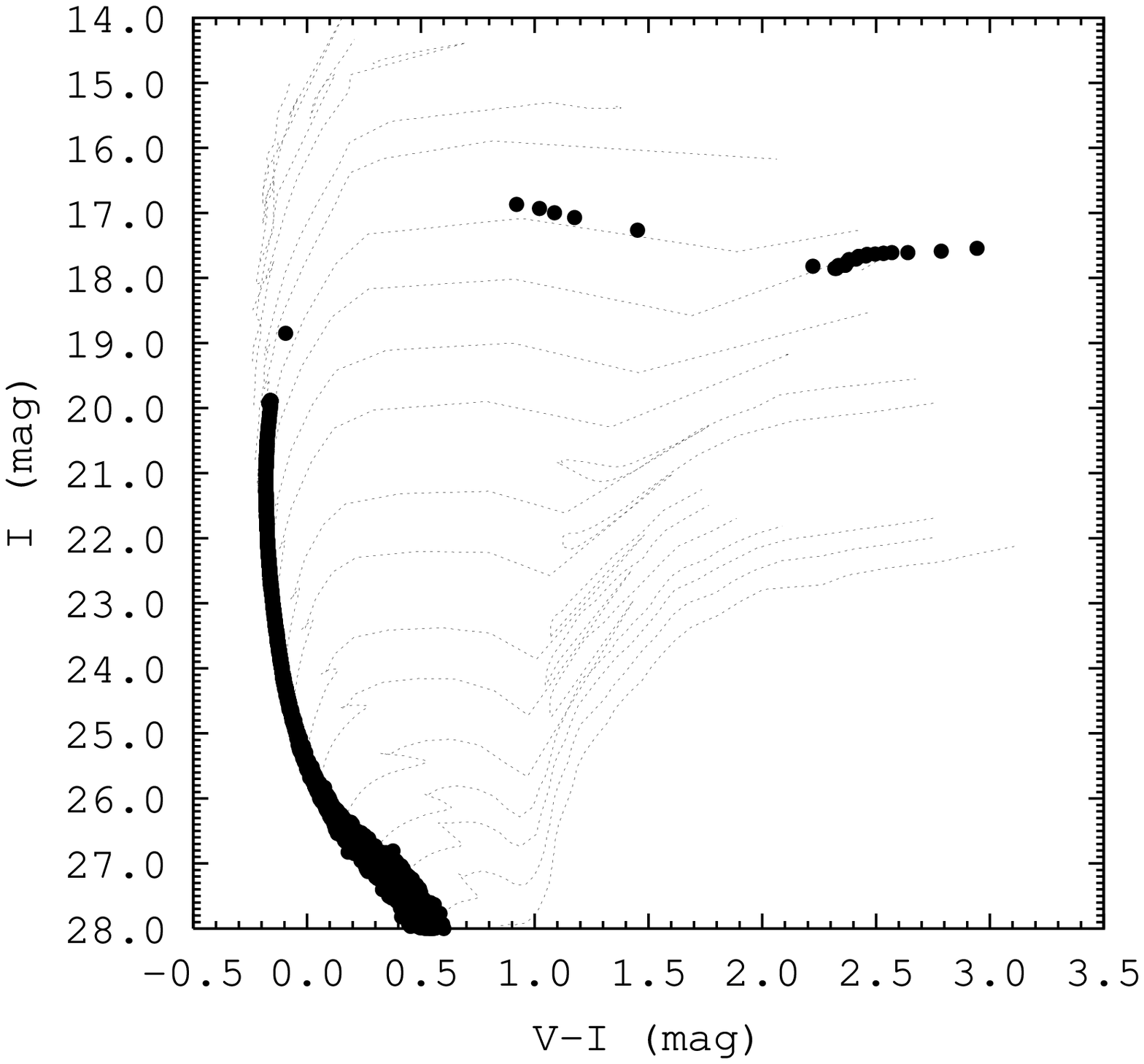}}
\subfigure[log(age/yr)=7.20]{\includegraphics[angle=0,width=0.24\textwidth]{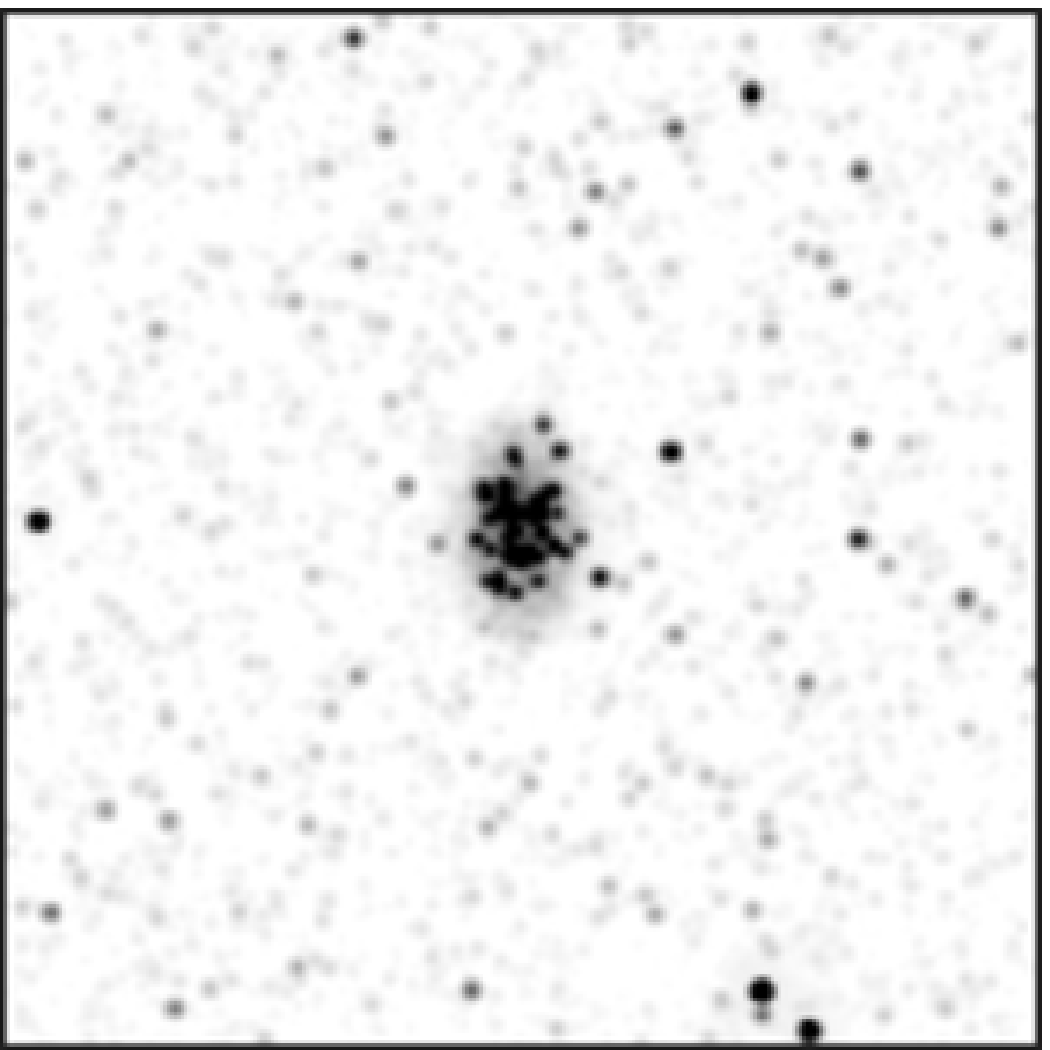}}
\subfigure[log(age/yr)=7.20]{\includegraphics[angle=0,width=0.24\textwidth]{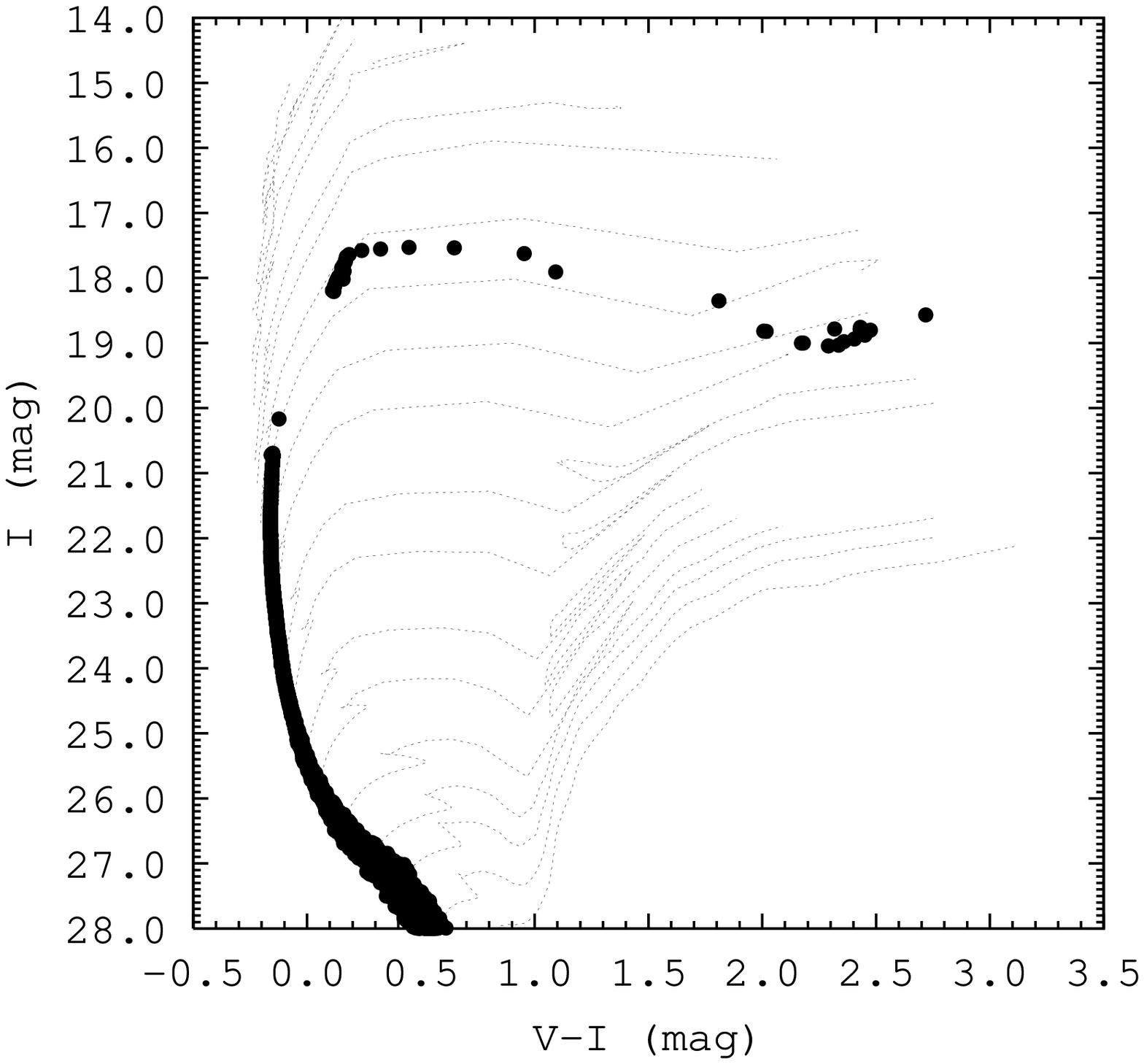}}
\subfigure[log(age/yr)=7.50]{\includegraphics[angle=0,width=0.24\textwidth]{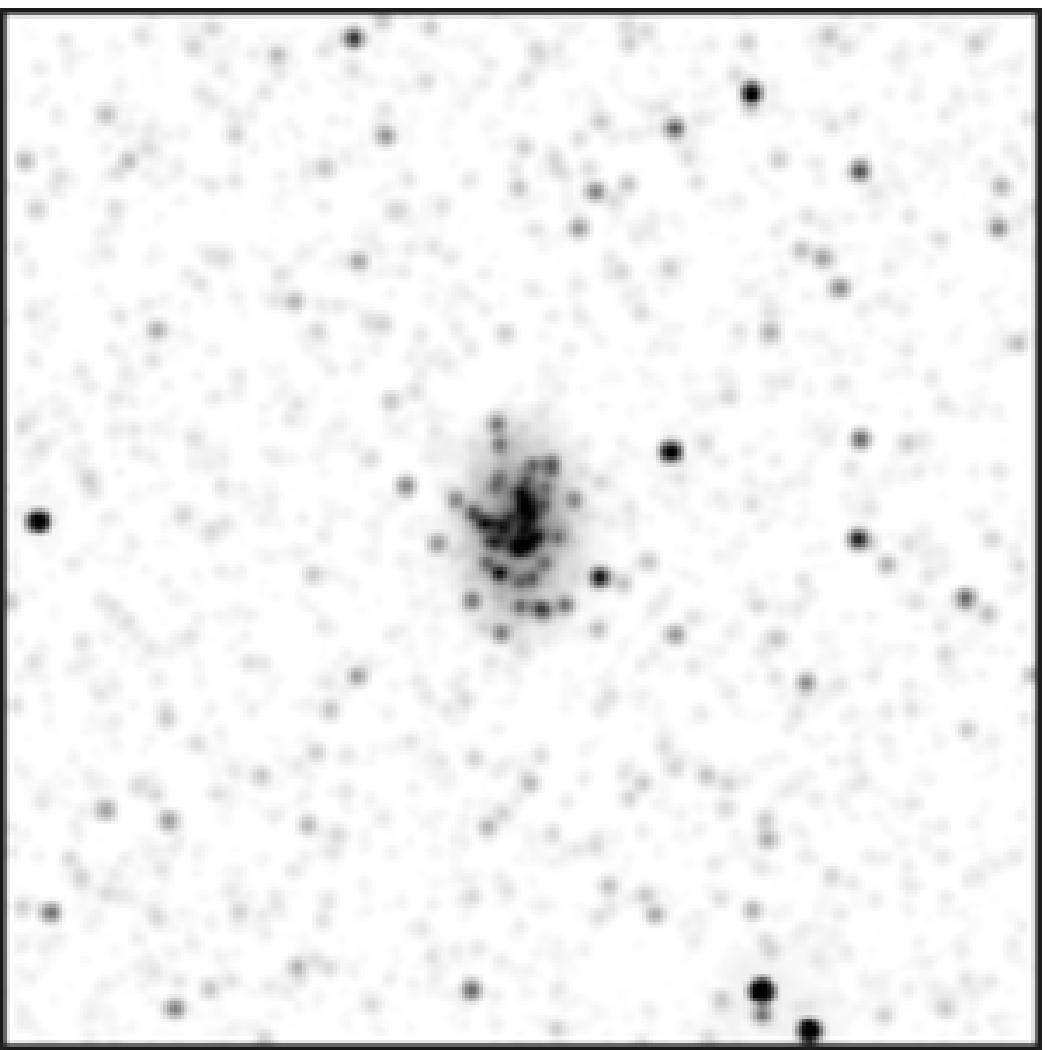}}
\subfigure[log(age/yr)=7.50]{\includegraphics[angle=0,width=0.24\textwidth]{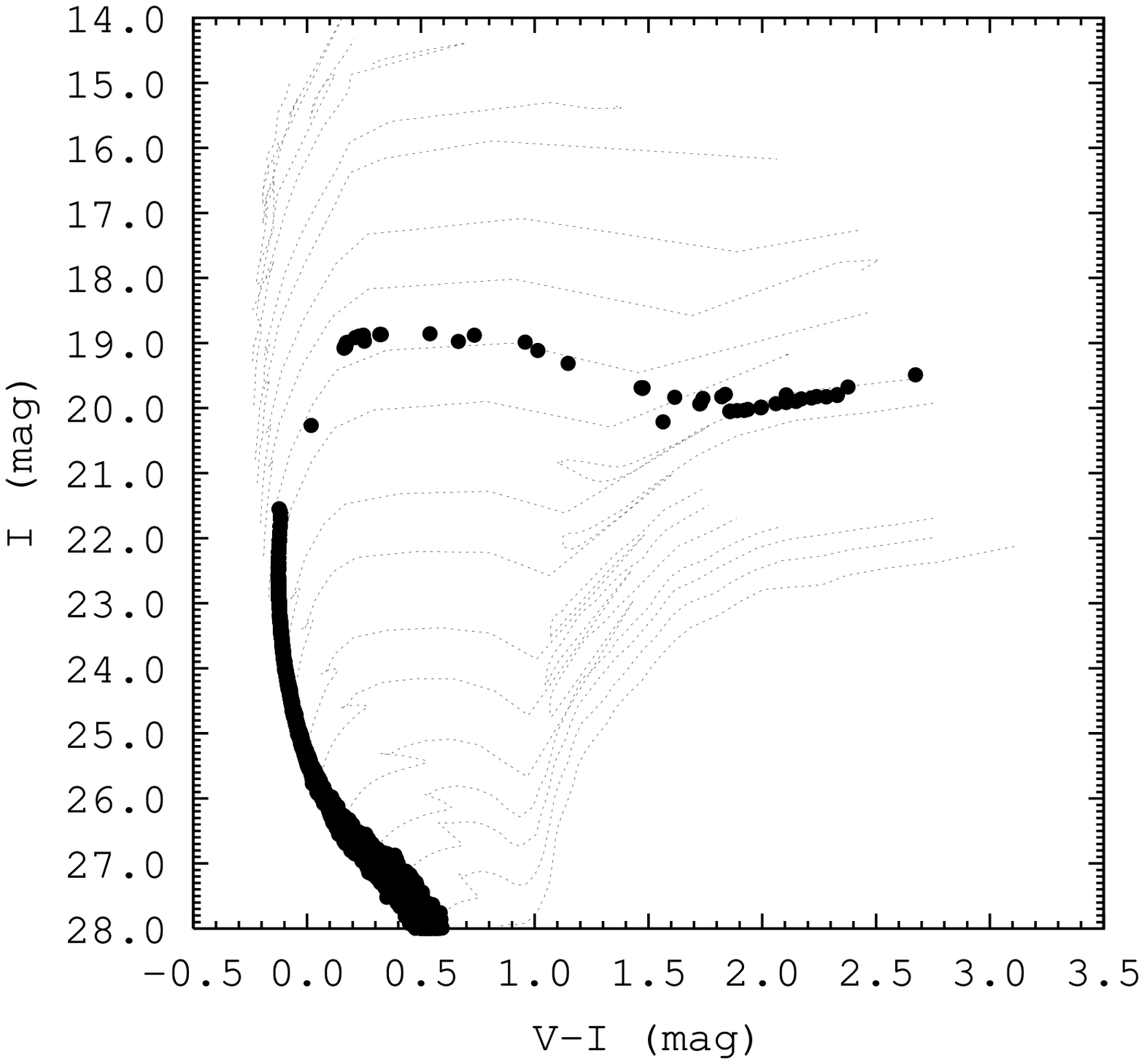}}
\subfigure[log(age/yr)=8.00]{\includegraphics[angle=0,width=0.24\textwidth]{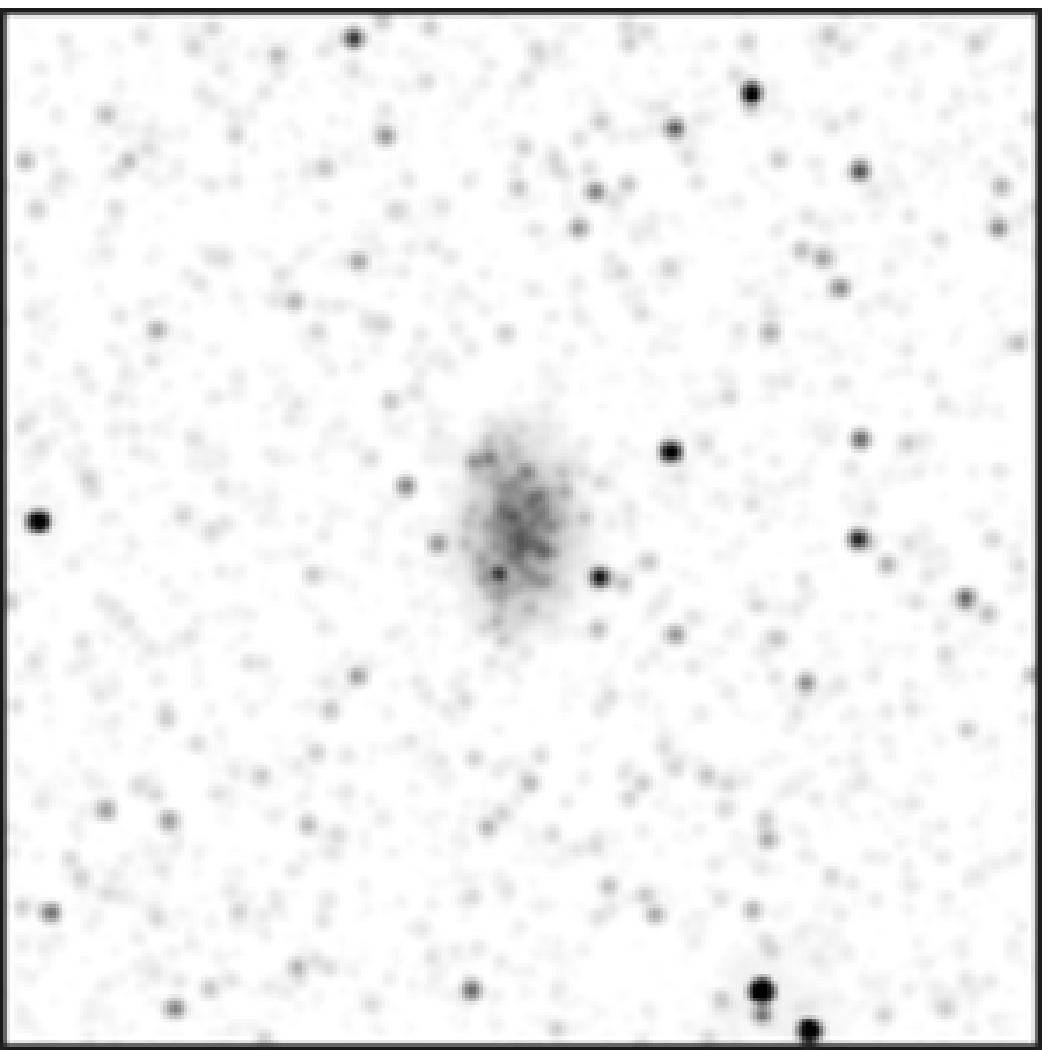}}
\subfigure[log(age/yr)=8.00]{\includegraphics[angle=0,width=0.24\textwidth]{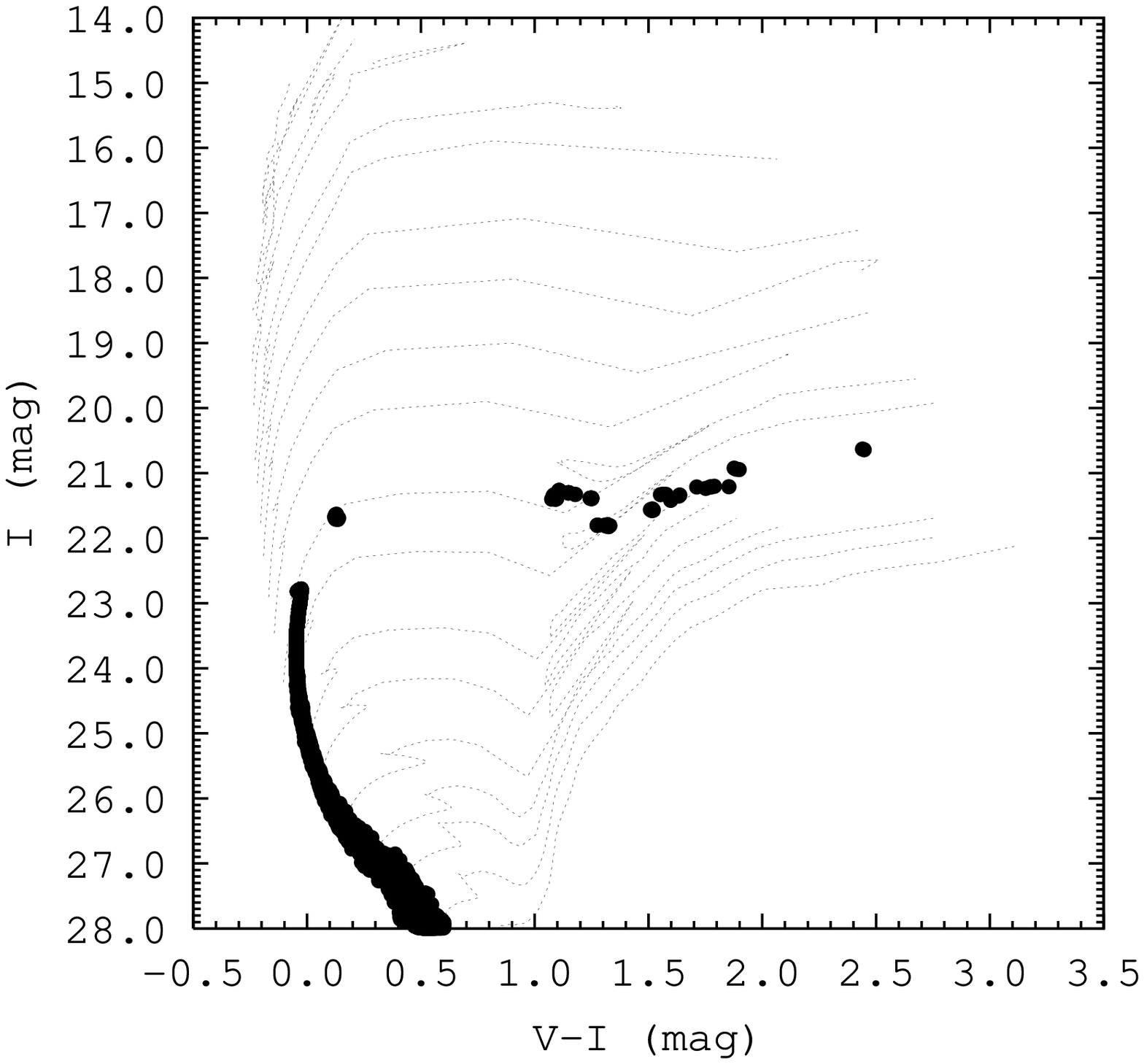}}
\caption{\small \textit{Westerlund 1} in M31 (I Band), simulated images and color-magnitude diagrams.\normalsize}\label{fig:tbd02}
\end{center}
\end{figure*}

\begin{figure*}[htbp] 
\begin{center}
\subfigure[log(age/yr)=6.00]{\includegraphics[angle=0,width=0.24\textwidth]{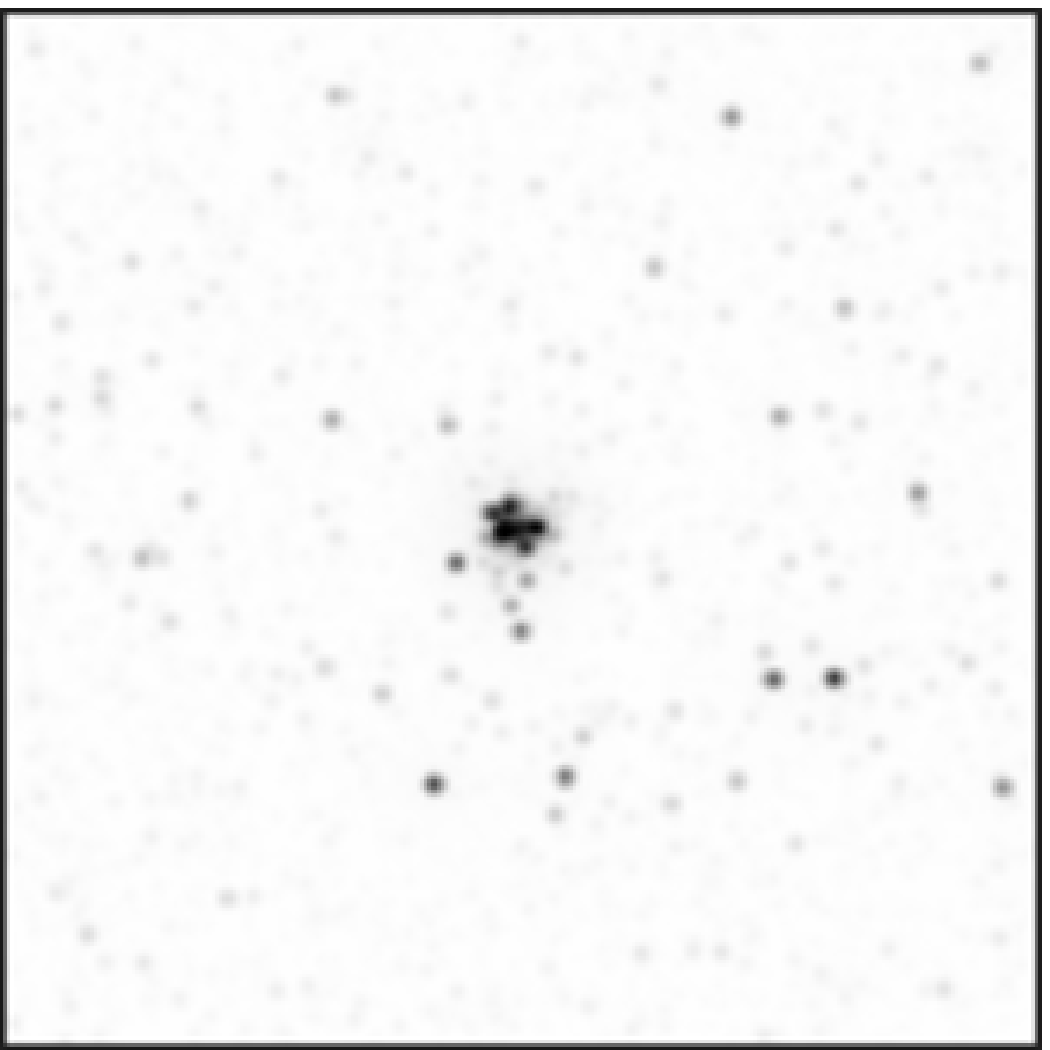}}
\subfigure[log(age/yr)=6.00]{\includegraphics[angle=0,width=0.24\textwidth]{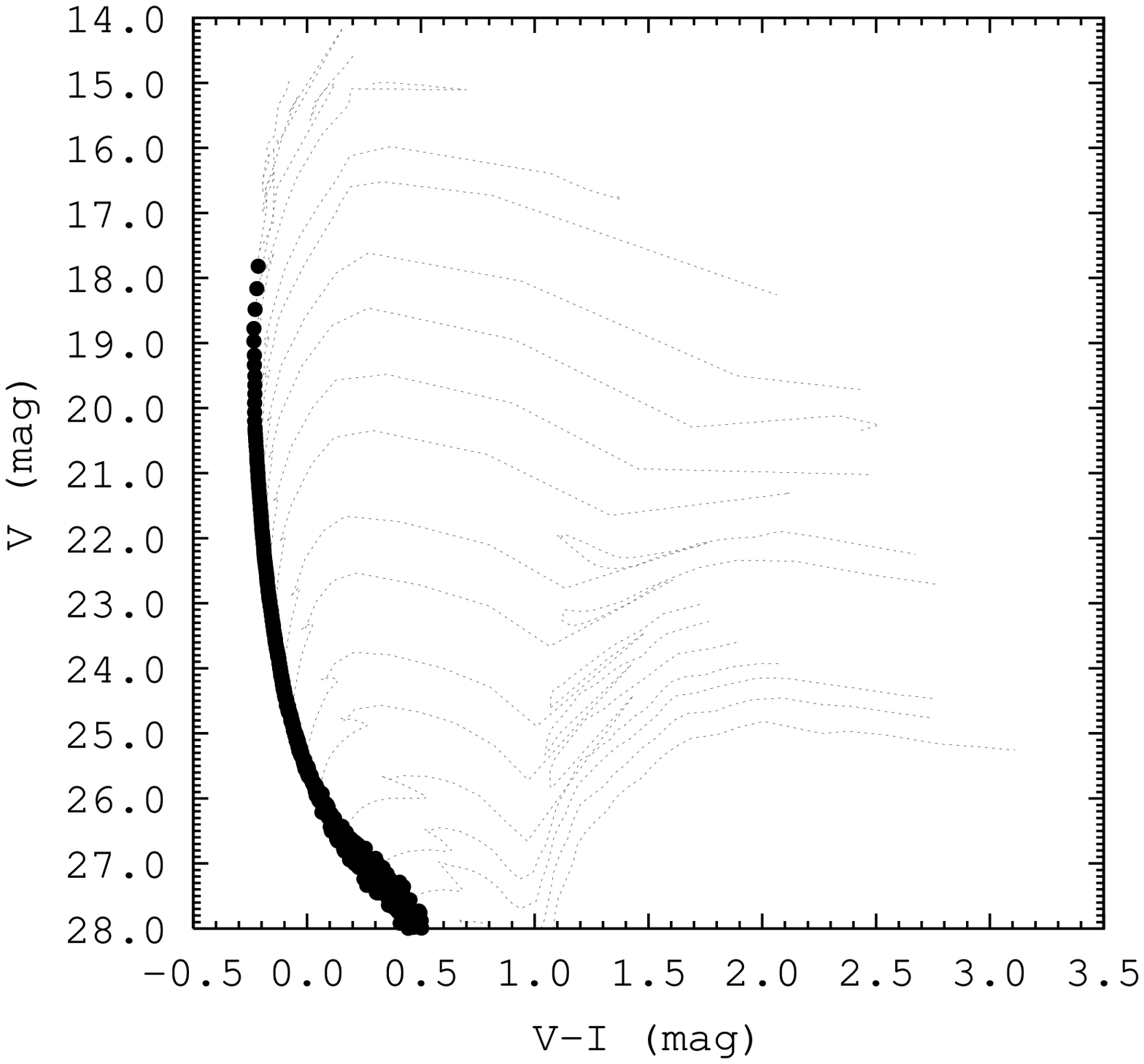}}    
\subfigure[log(age/yr)=6.50]{\includegraphics[angle=0,width=0.24\textwidth]{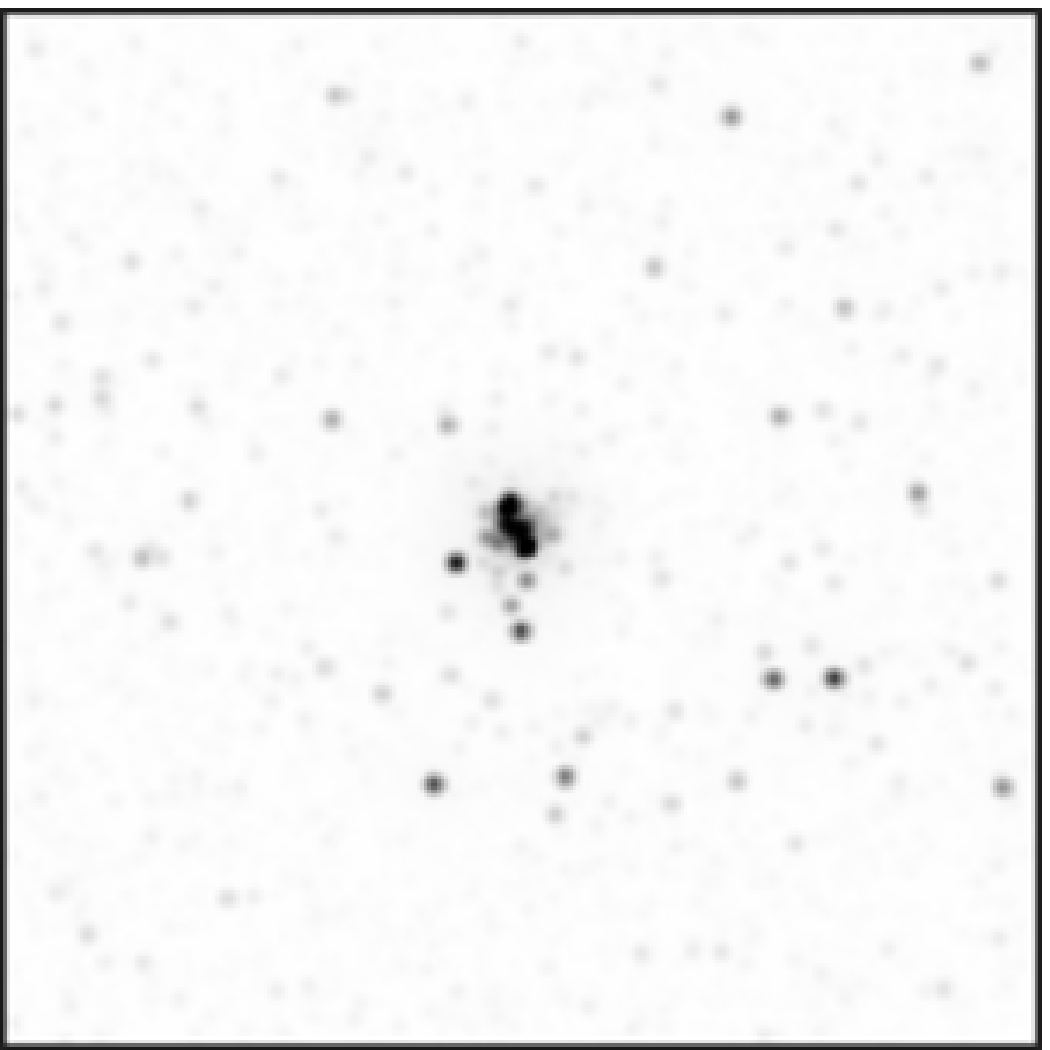}}
\subfigure[log(age/yr)=6.50]{\includegraphics[angle=0,width=0.24\textwidth]{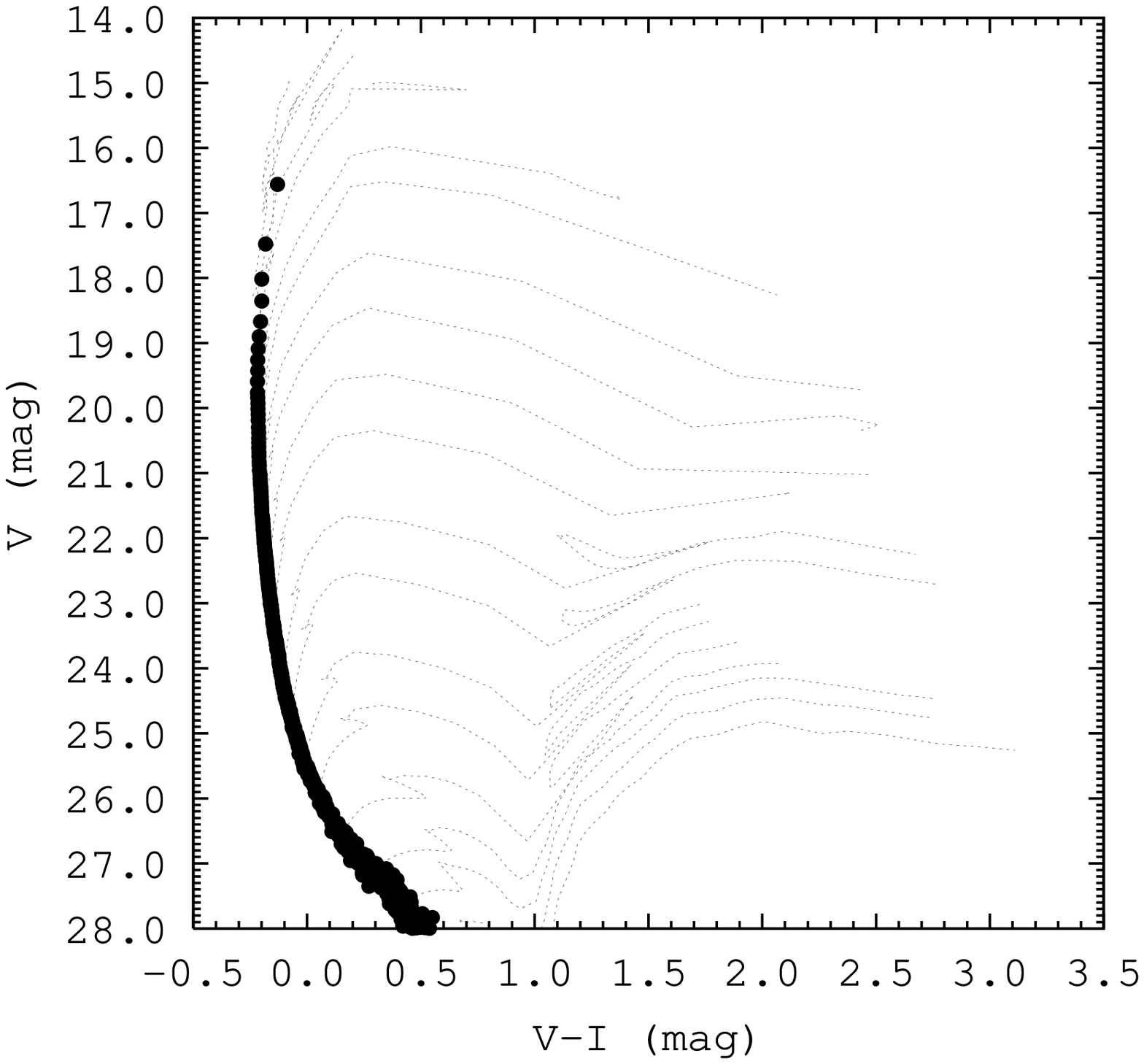}}
\subfigure[log(age/yr)=6.65]{\includegraphics[angle=0,width=0.24\textwidth]{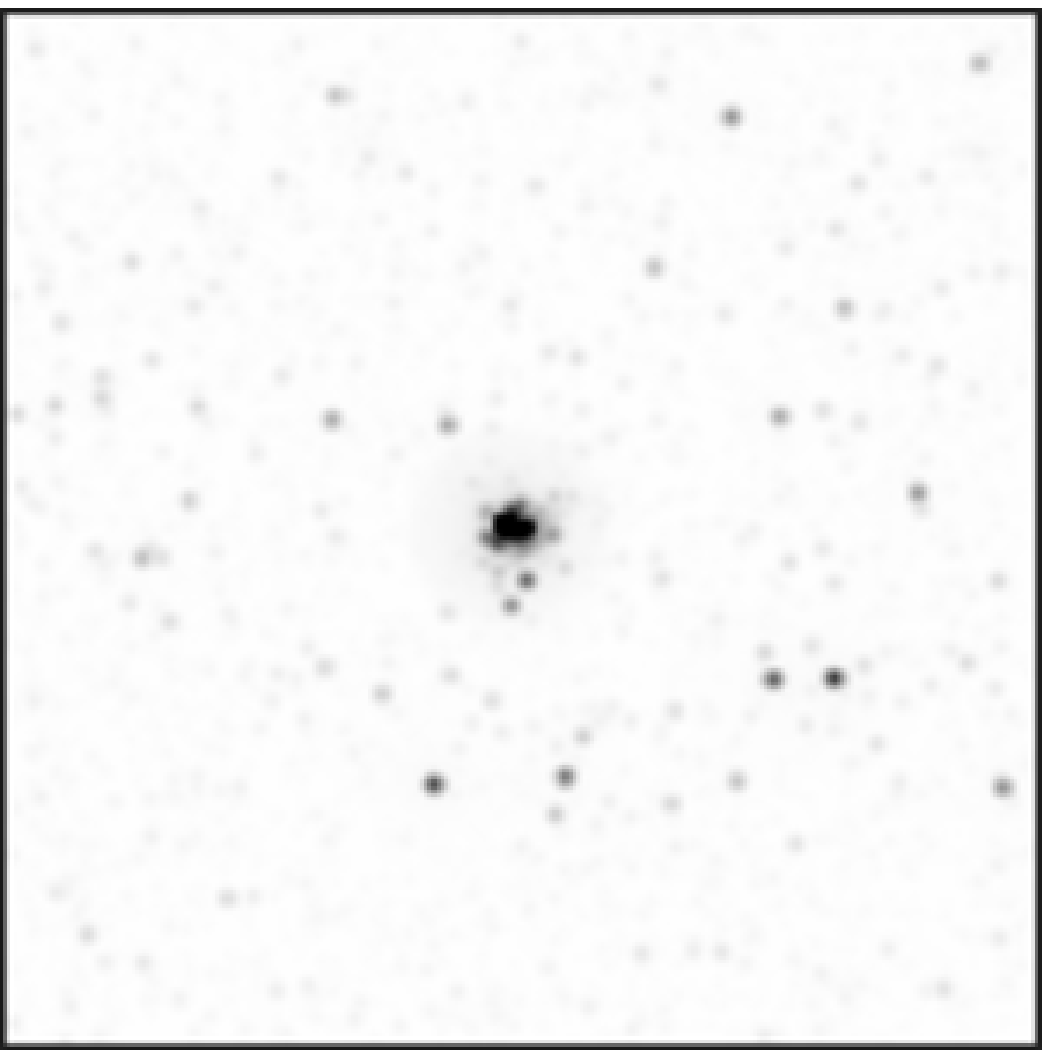}}
\subfigure[log(age/yr)=6.65]{\includegraphics[angle=0,width=0.24\textwidth]{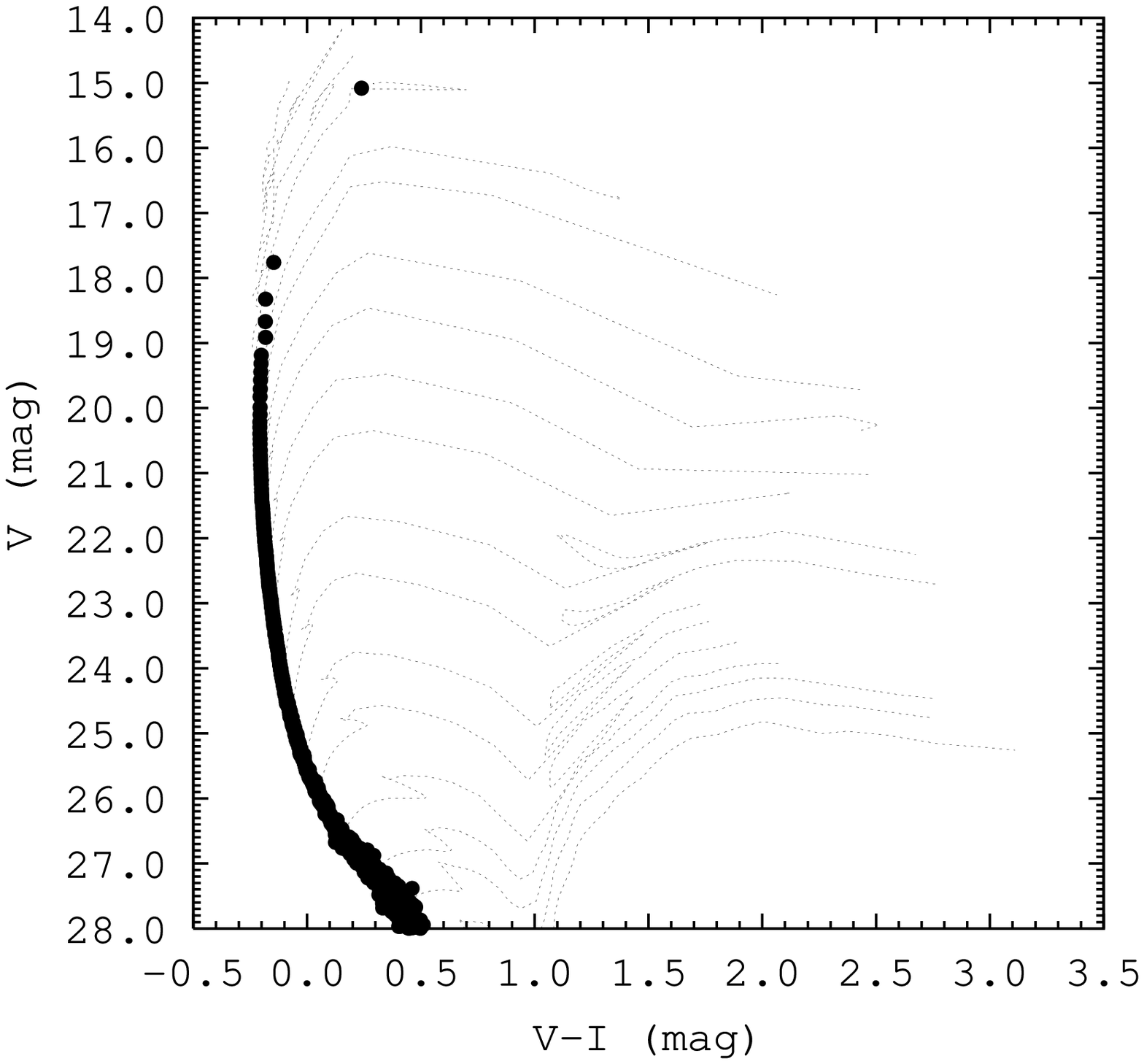}}
\subfigure[log(age/yr)=6.75]{\includegraphics[angle=0,width=0.24\textwidth]{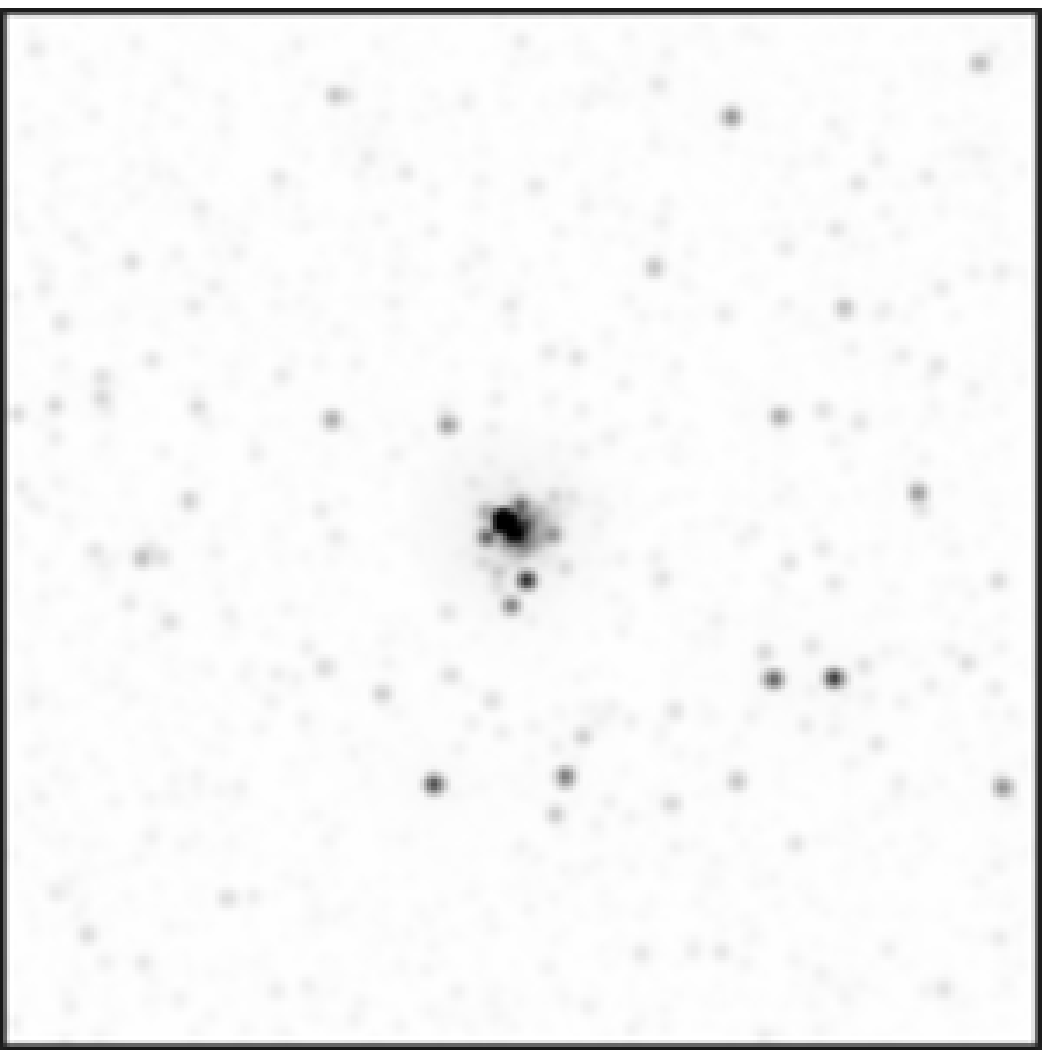}} 
\subfigure[log(age/yr)=6.75]{\includegraphics[angle=0,width=0.24\textwidth]{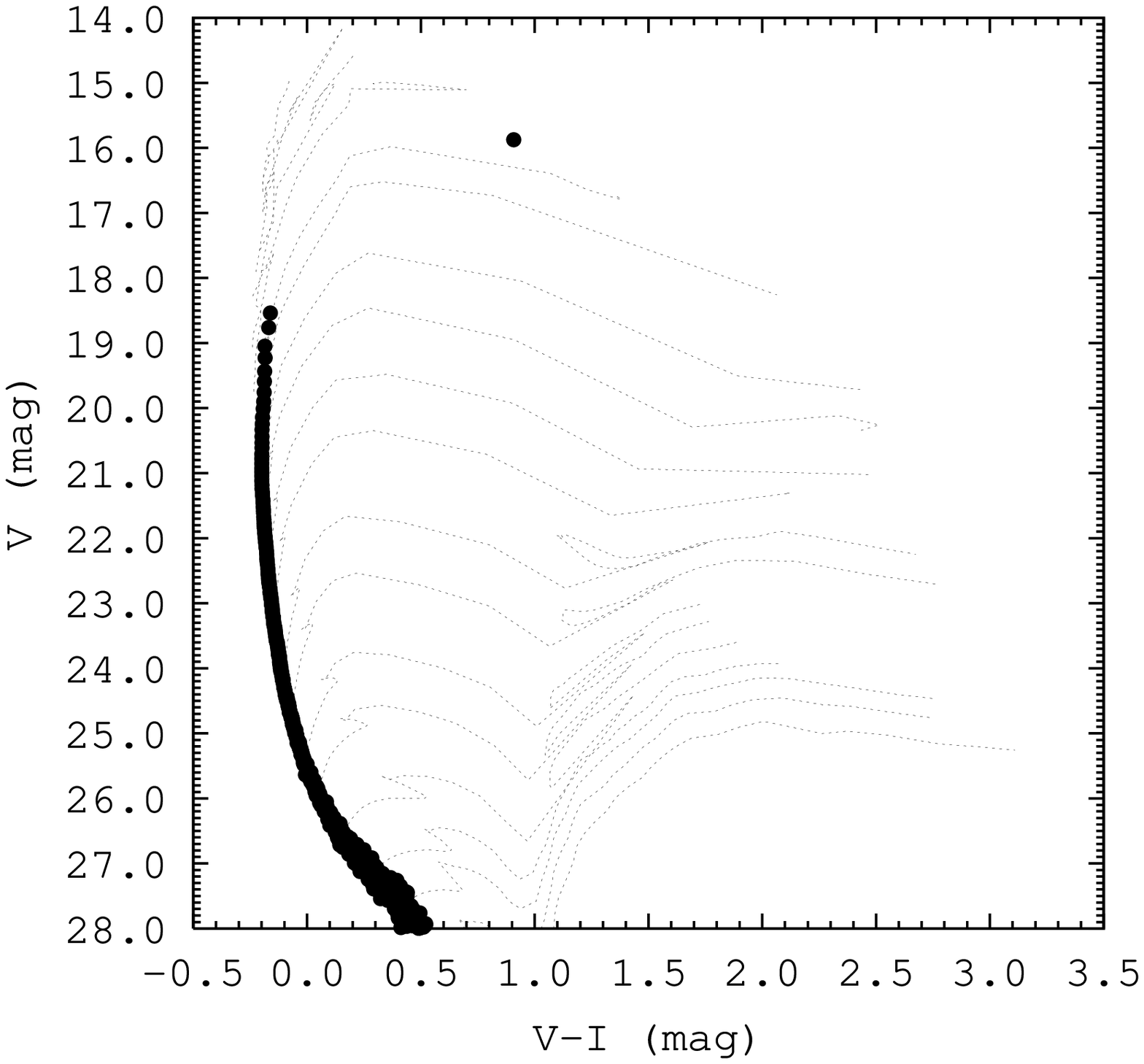}} 
\subfigure[log(age/yr)=7.00]{\includegraphics[angle=0,width=0.24\textwidth]{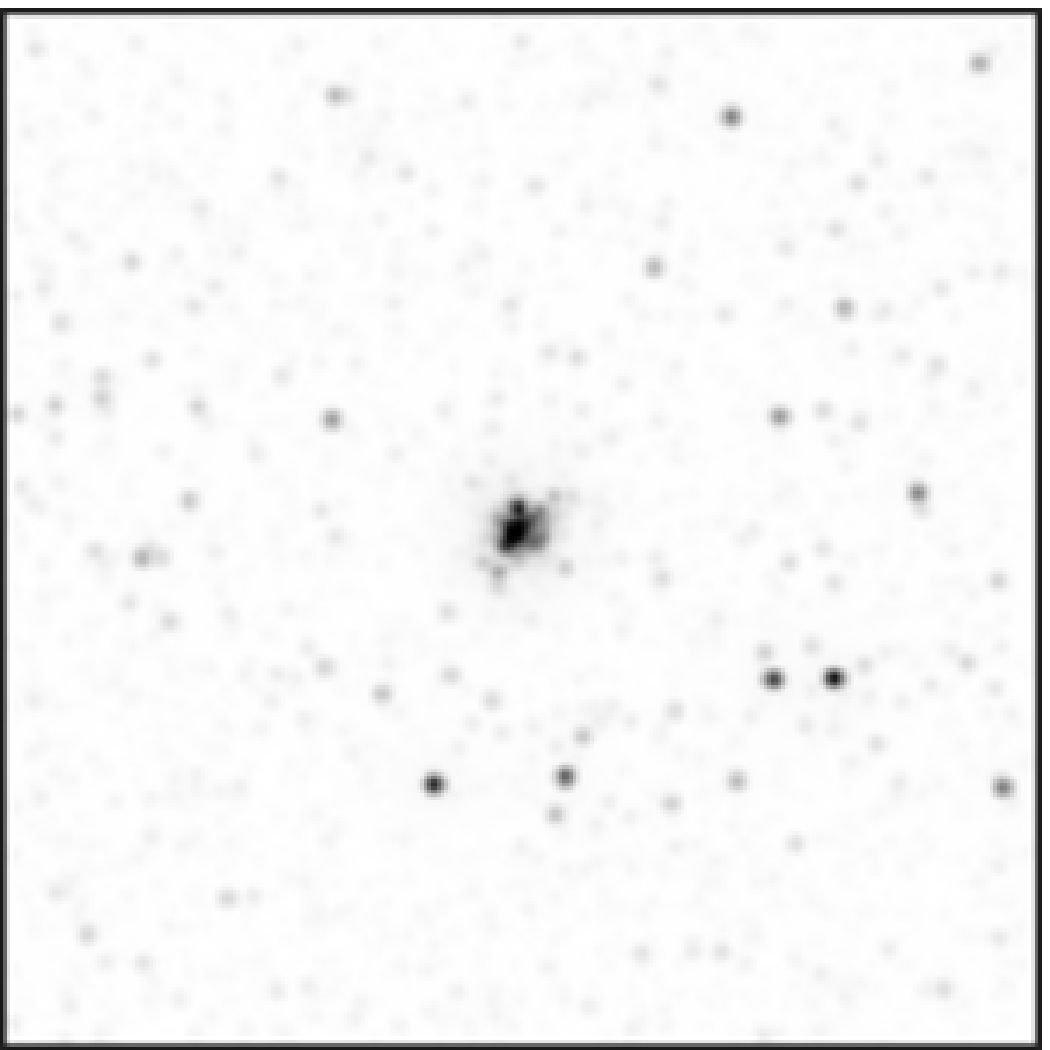}}
\subfigure[log(age/yr)=7.00]{\includegraphics[angle=0,width=0.24\textwidth]{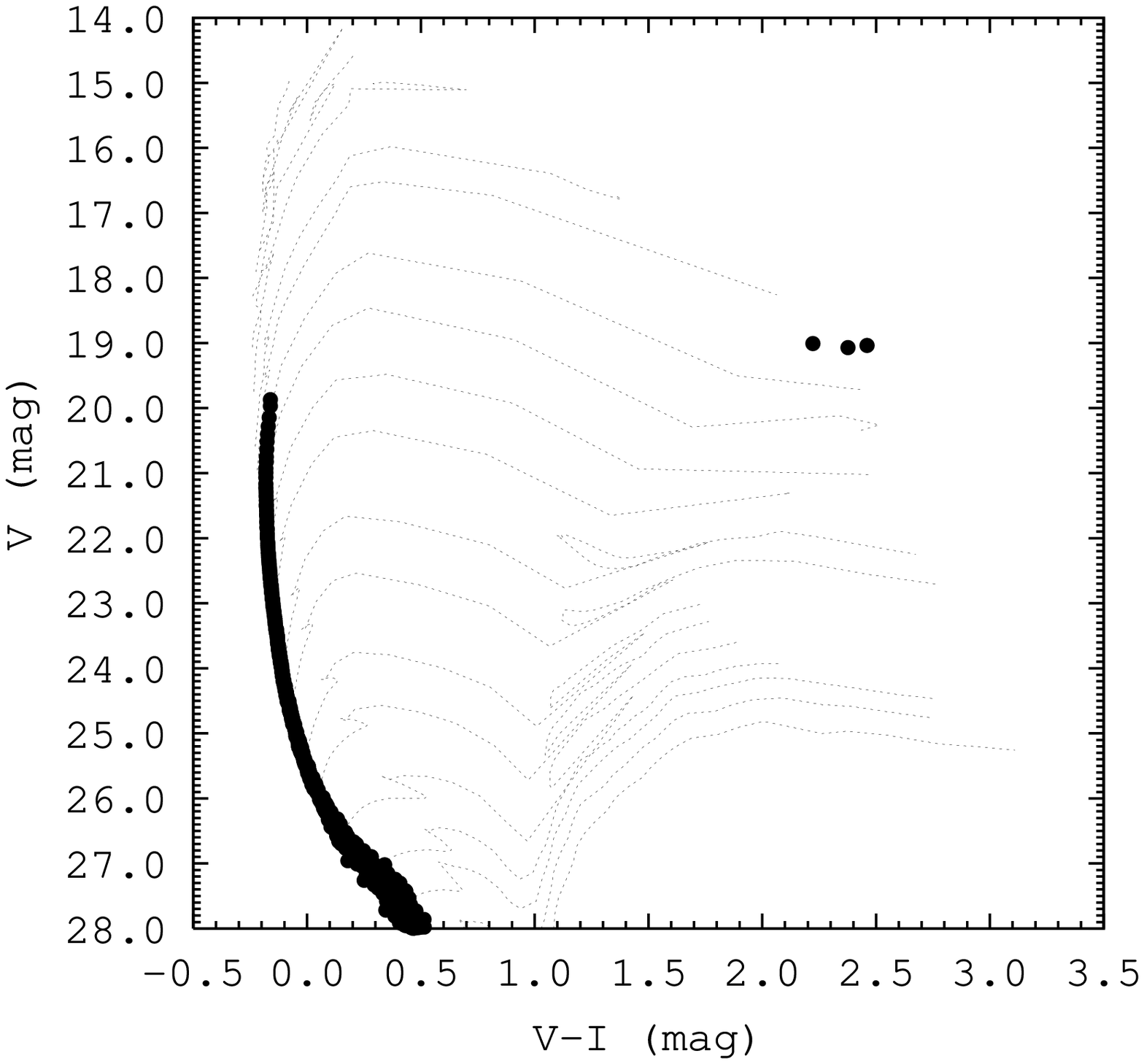}}
\subfigure[log(age/yr)=7.20]{\includegraphics[angle=0,width=0.24\textwidth]{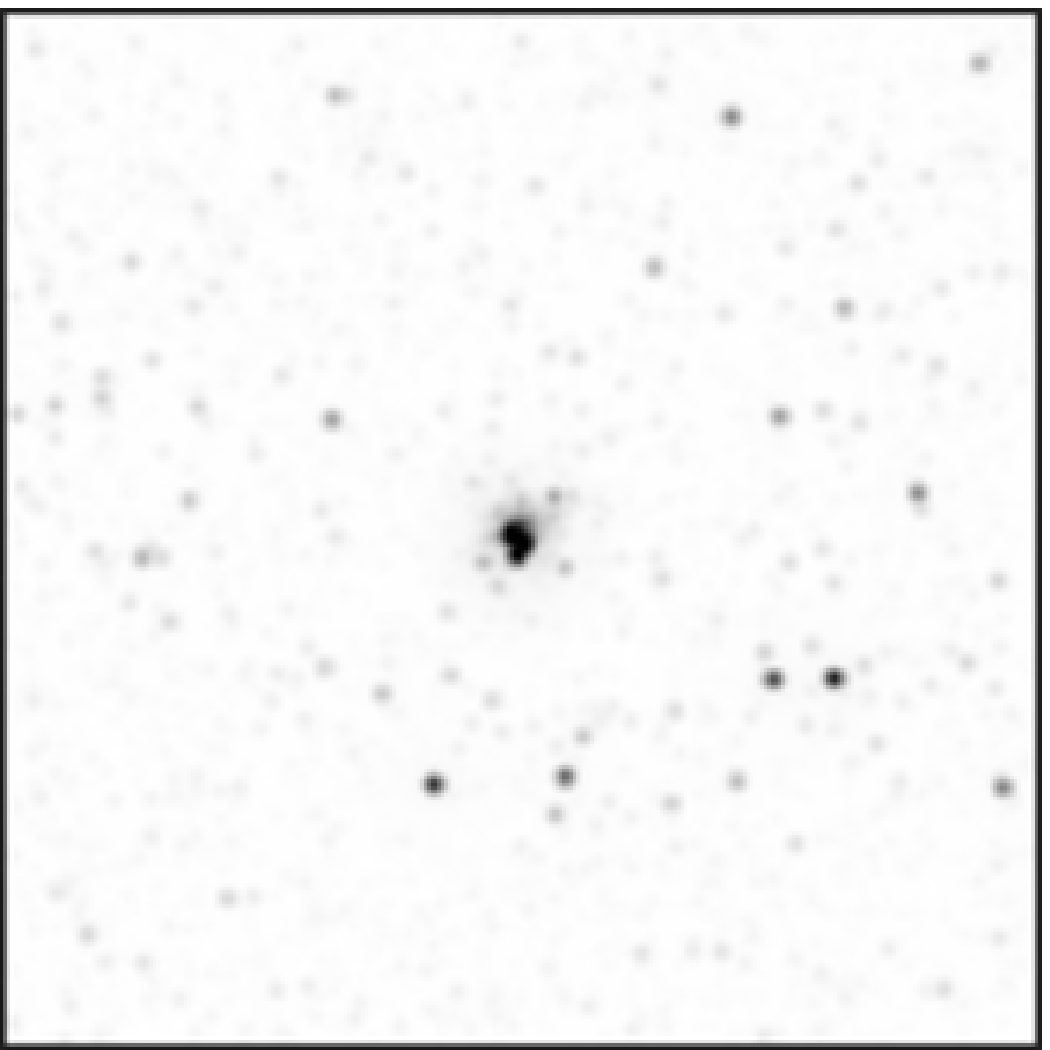}}
\subfigure[log(age/yr)=7.20]{\includegraphics[angle=0,width=0.24\textwidth]{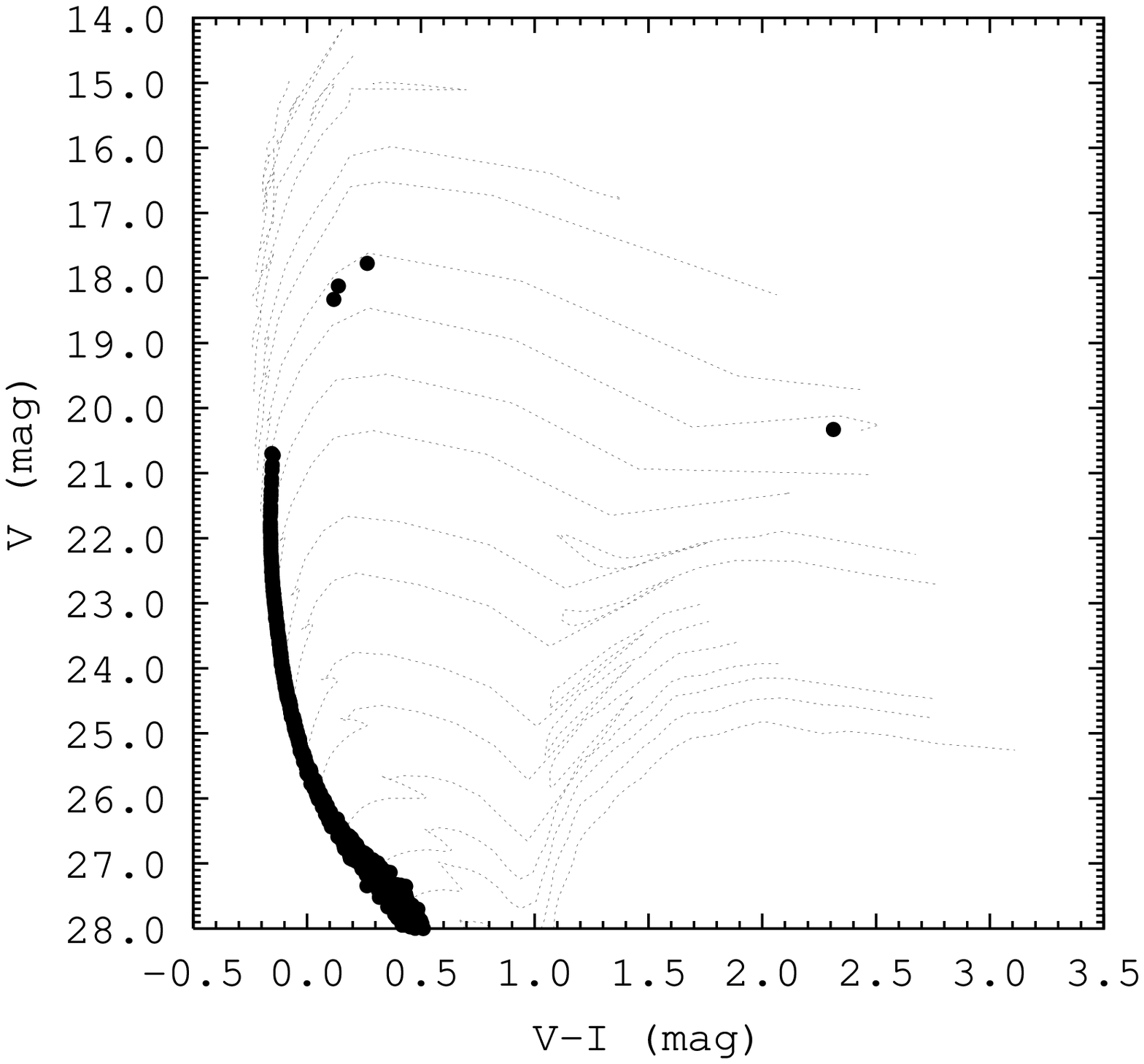}}
\subfigure[log(age/yr)=7.50]{\includegraphics[angle=0,width=0.24\textwidth]{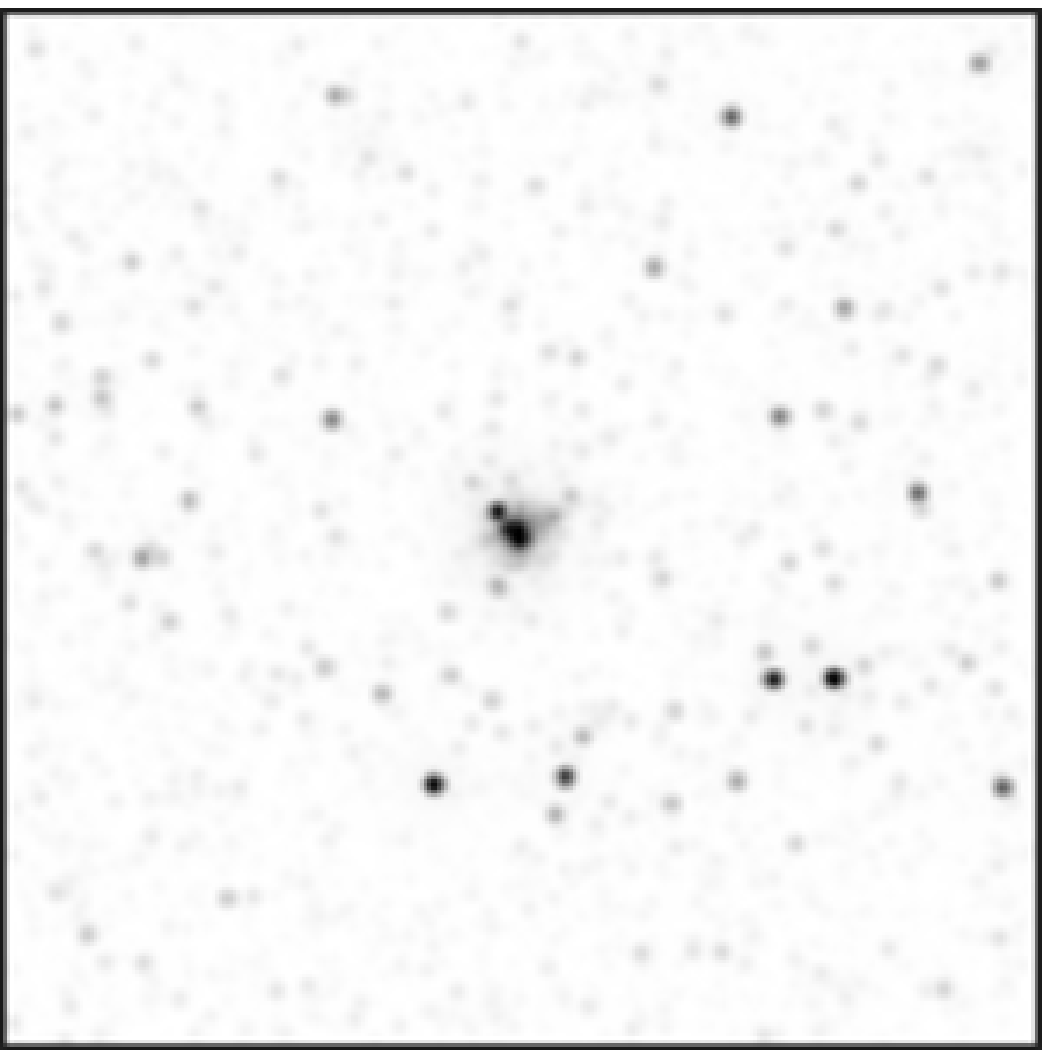}}
\subfigure[log(age/yr)=7.50]{\includegraphics[angle=0,width=0.24\textwidth]{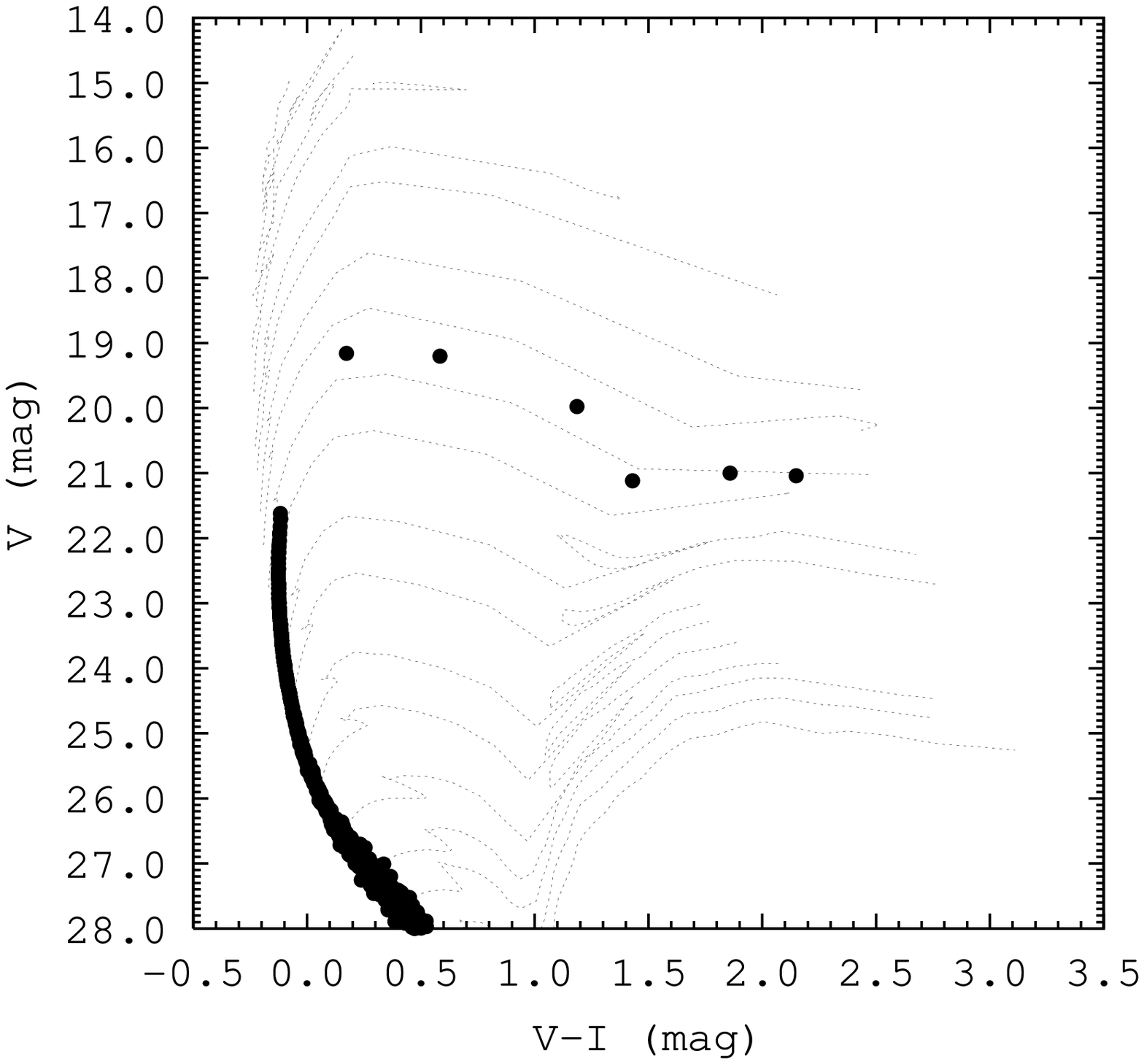}}
\subfigure[log(age/yr)=8.00]{\includegraphics[angle=0,width=0.24\textwidth]{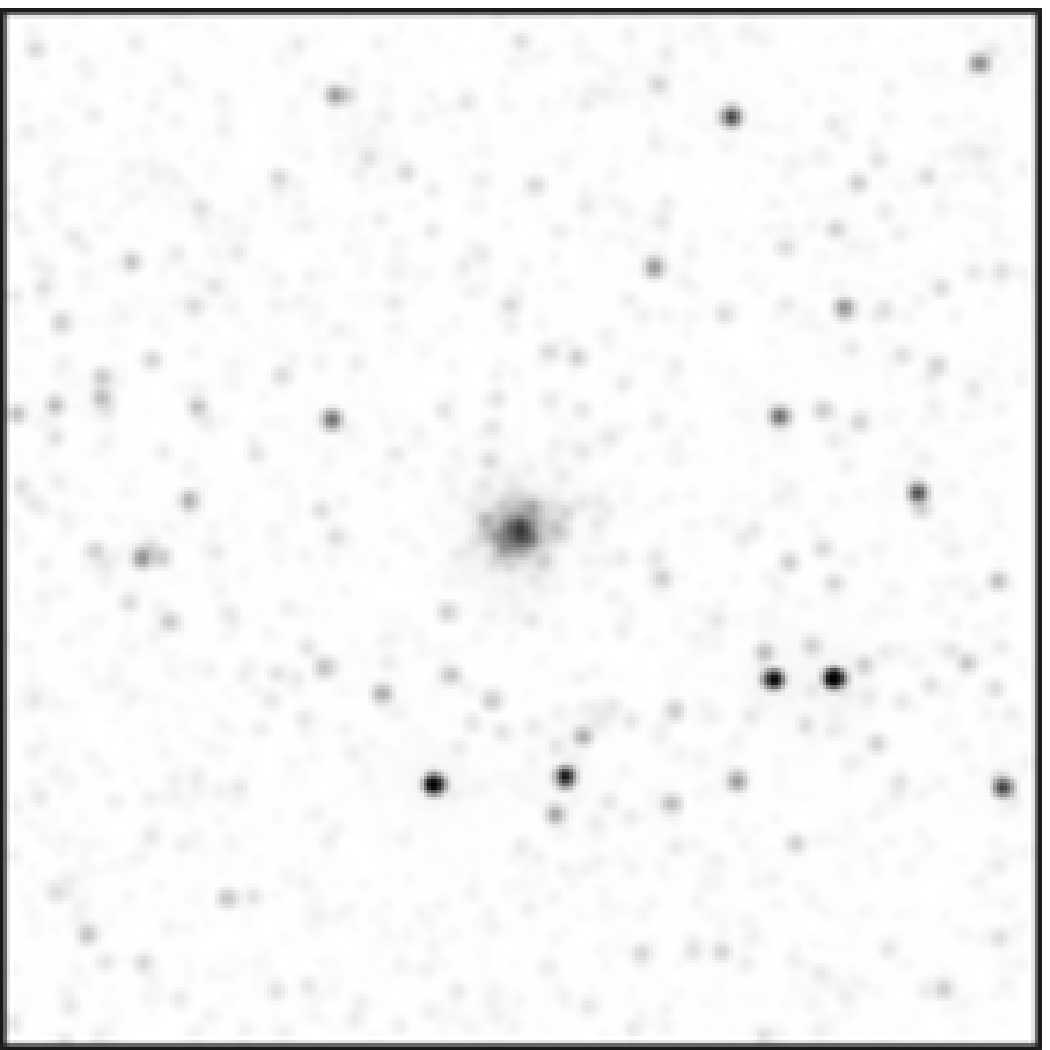}}
\subfigure[log(age/yr)=8.00]{\includegraphics[angle=0,width=0.24\textwidth]{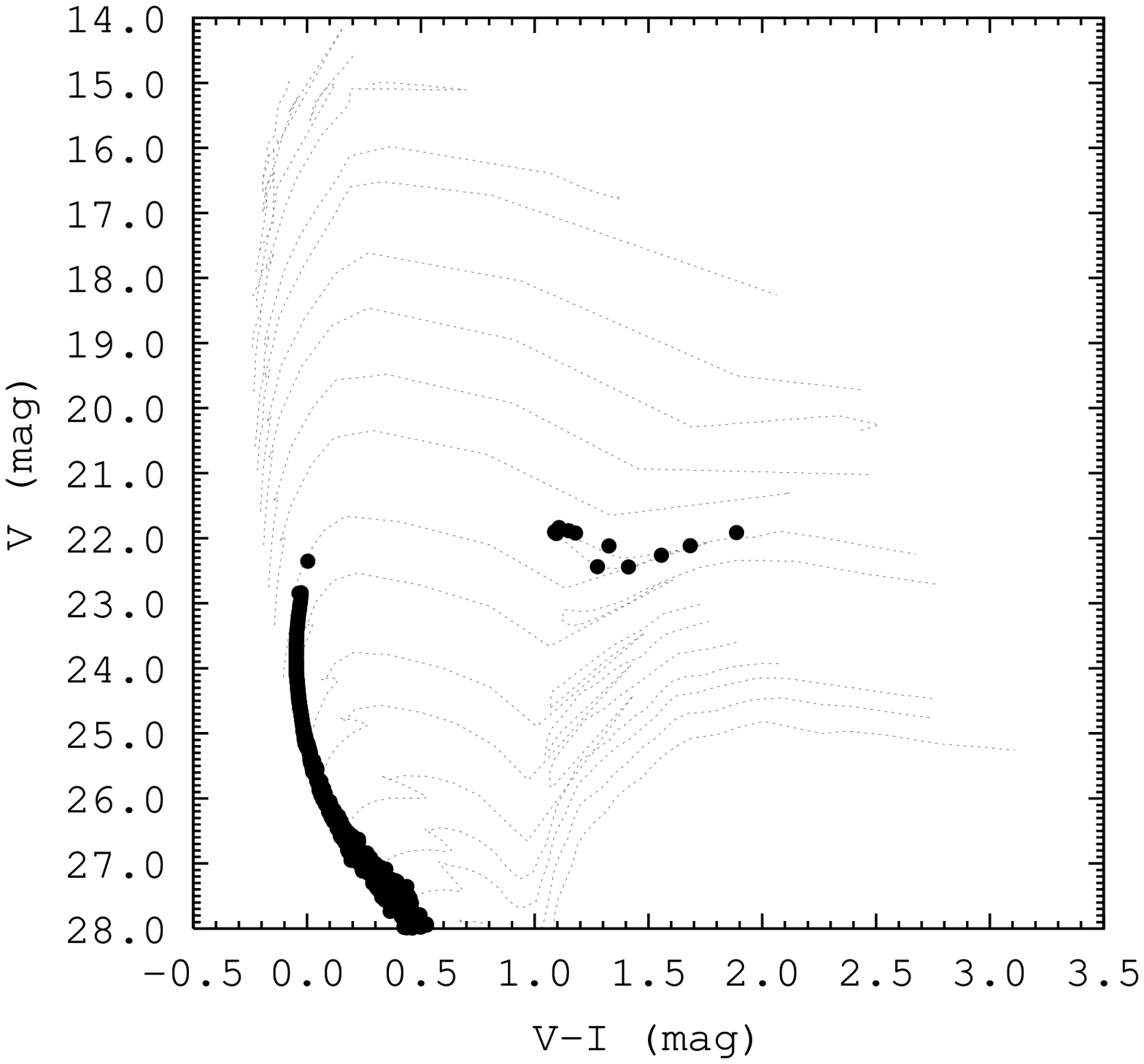}}
\caption{\small \textit{NGC 3603} in M31 (V Band), simulated images and color-magnitude diagrams.\normalsize}\label{fig:tbd03}
\end{center}
\end{figure*}

\begin{figure*}[htbp] 
\begin{center}
\subfigure[log(age/yr)=6.00]{\includegraphics[angle=0,width=0.24\textwidth]{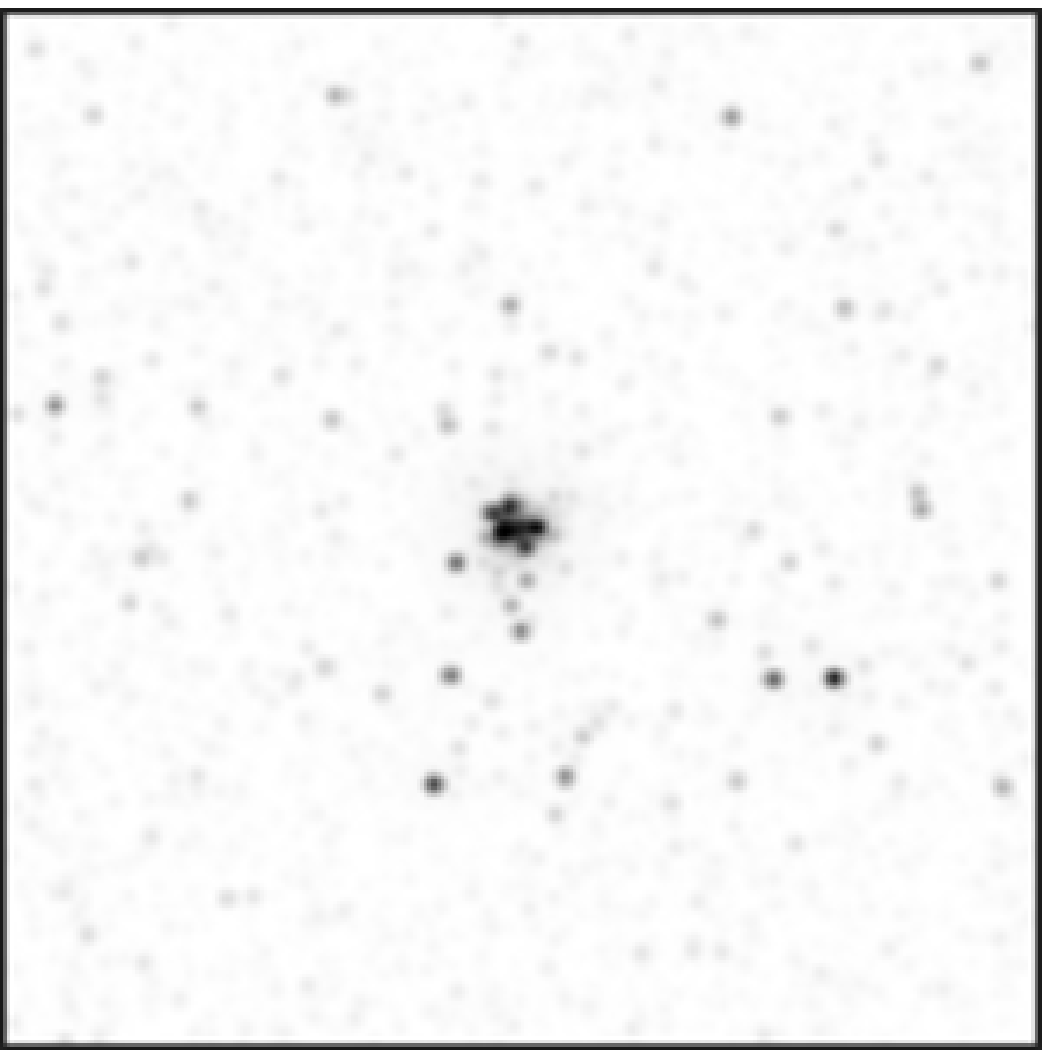}} 
\subfigure[log(age/yr)=6.00]{\includegraphics[angle=0,width=0.24\textwidth]{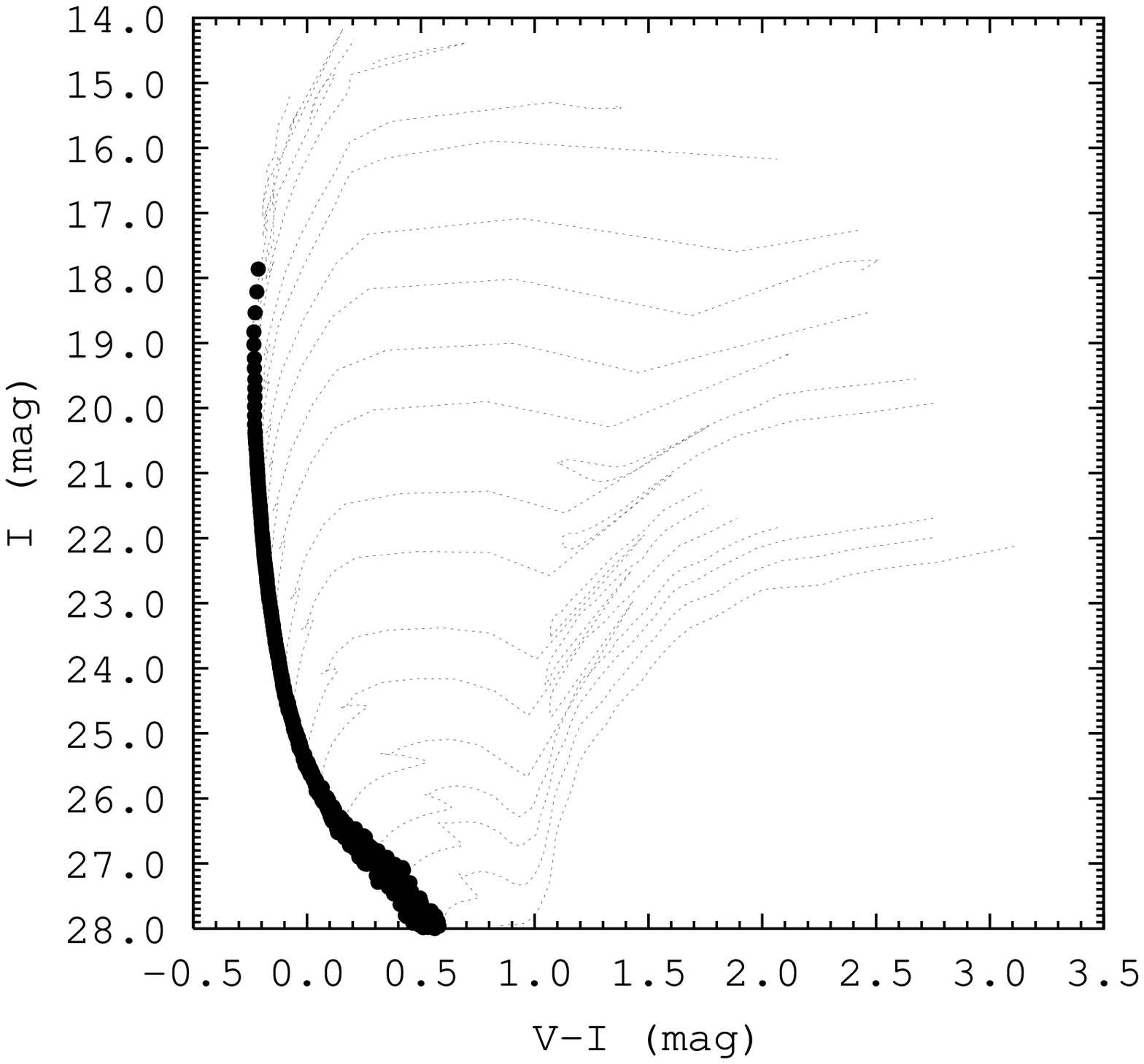}} 
\subfigure[log(age/yr)=6.50]{\includegraphics[angle=0,width=0.24\textwidth]{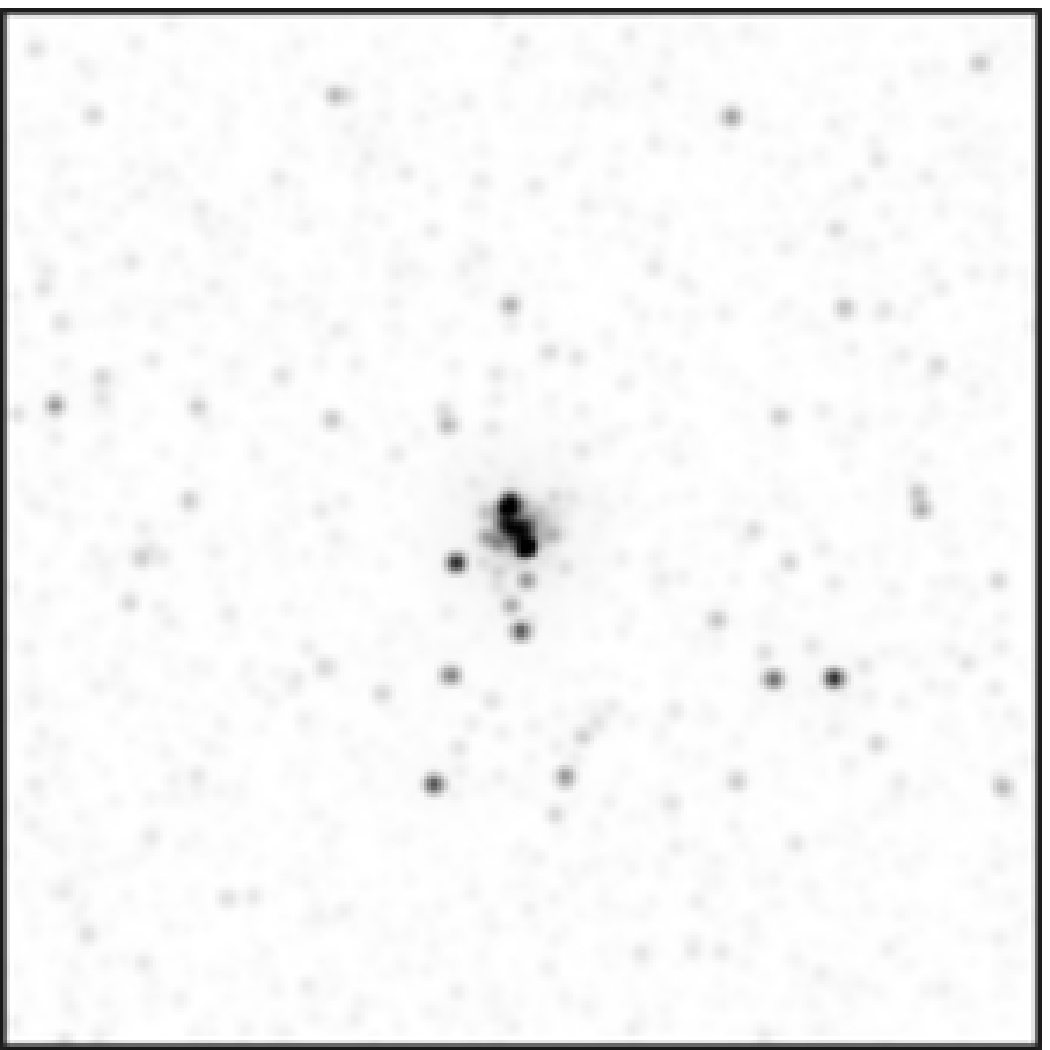}}
\subfigure[log(age/yr)=6.50]{\includegraphics[angle=0,width=0.24\textwidth]{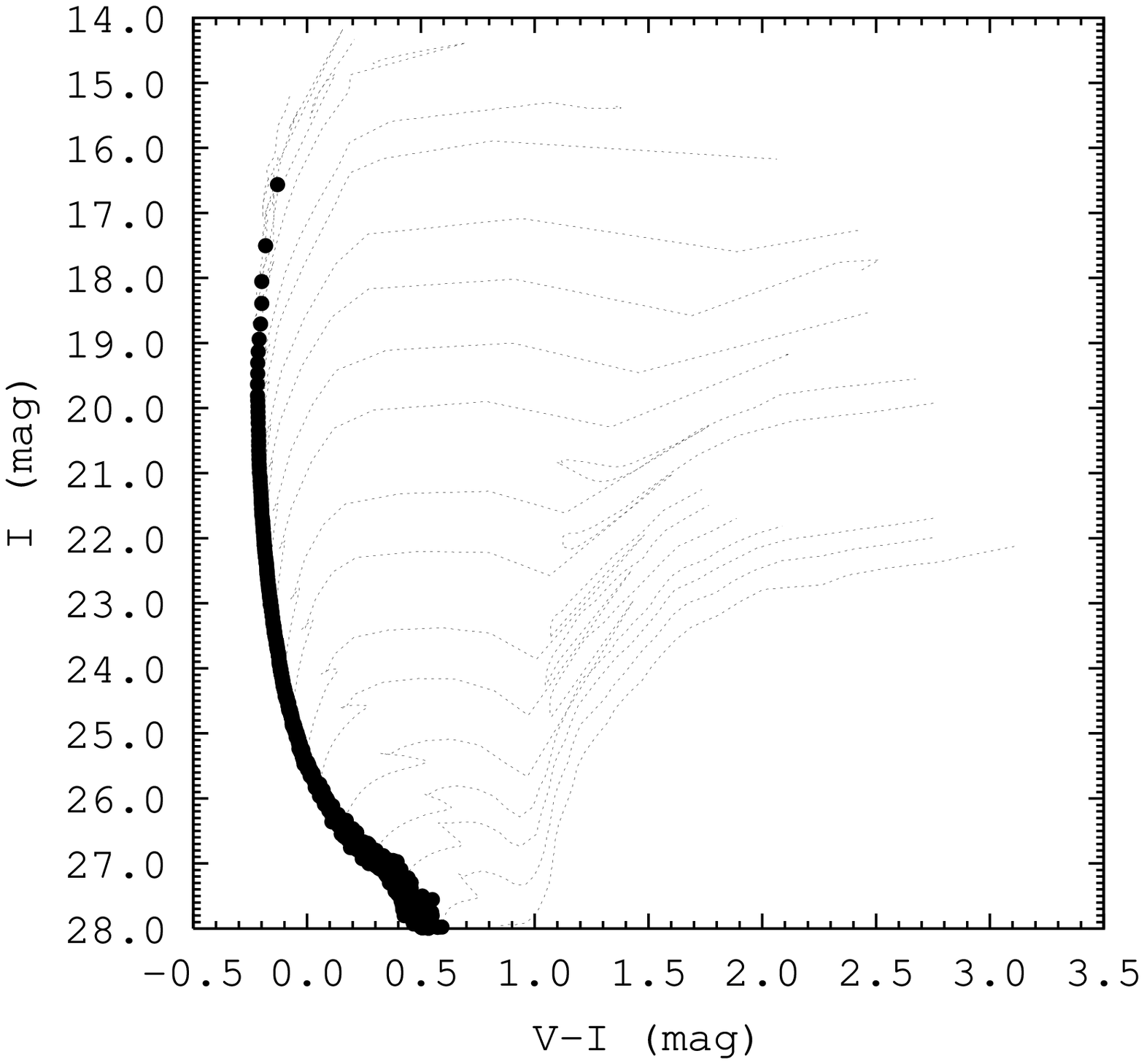}}
\subfigure[log(age/yr)=6.65]{\includegraphics[angle=0,width=0.24\textwidth]{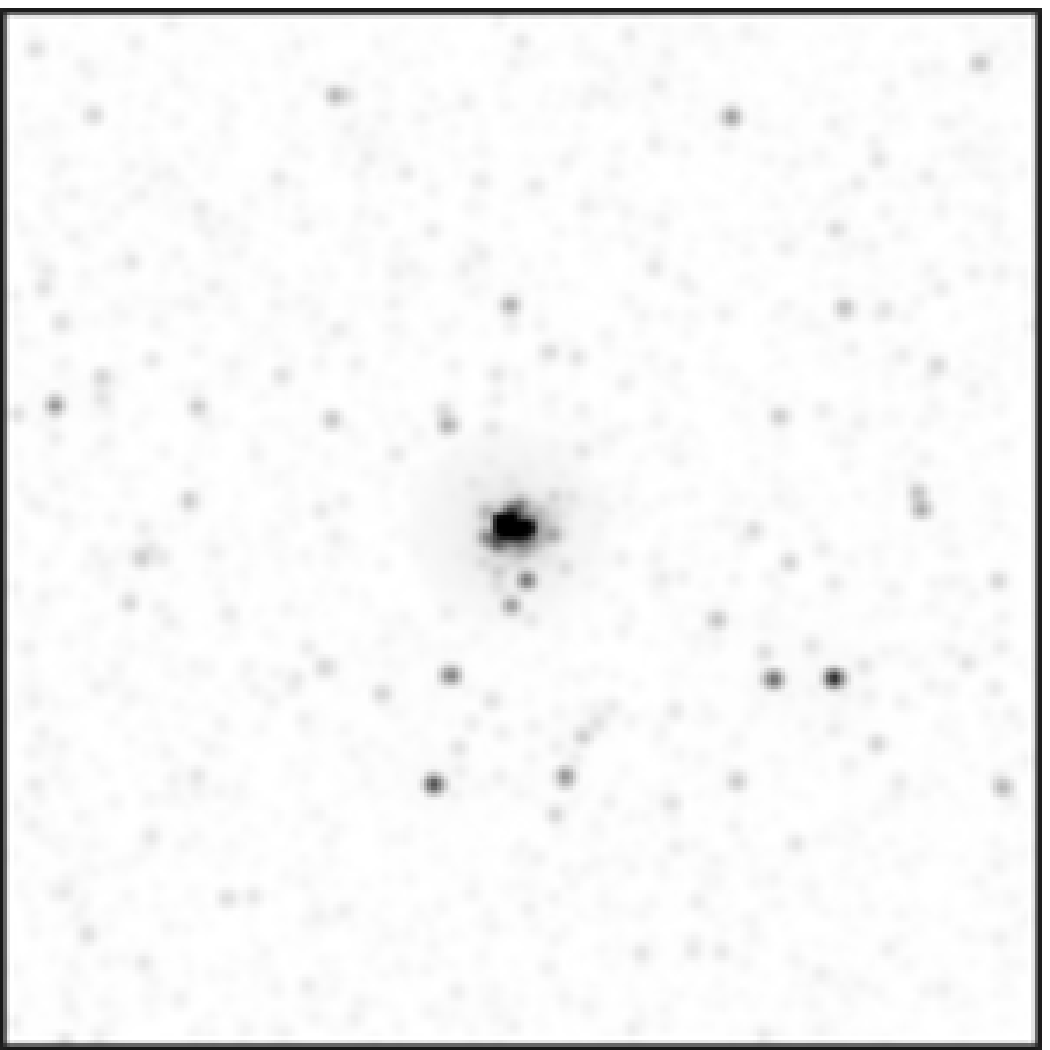}}
\subfigure[log(age/yr)=6.65]{\includegraphics[angle=0,width=0.24\textwidth]{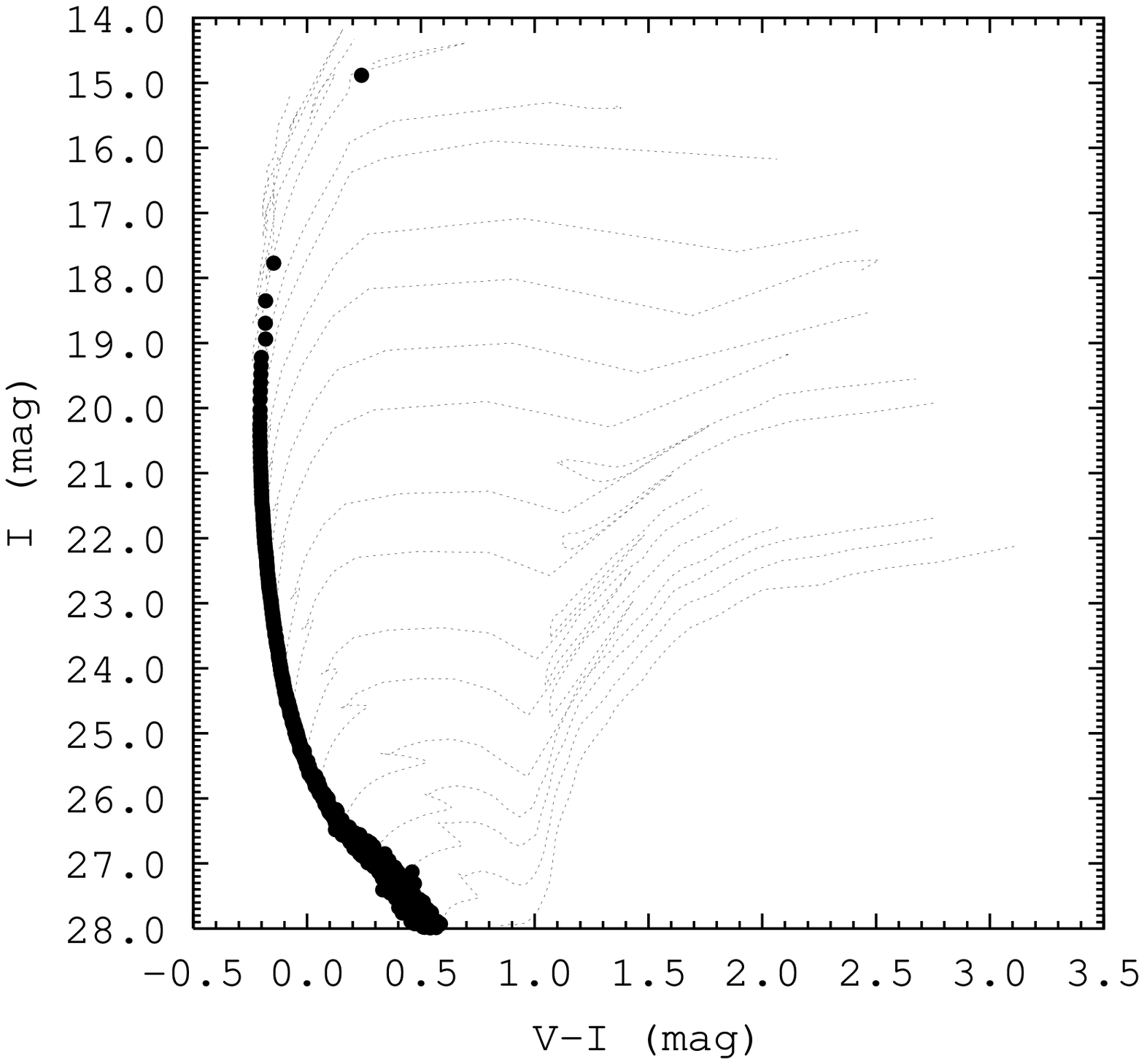}}
\subfigure[log(age/yr)=6.75]{\includegraphics[angle=0,width=0.24\textwidth]{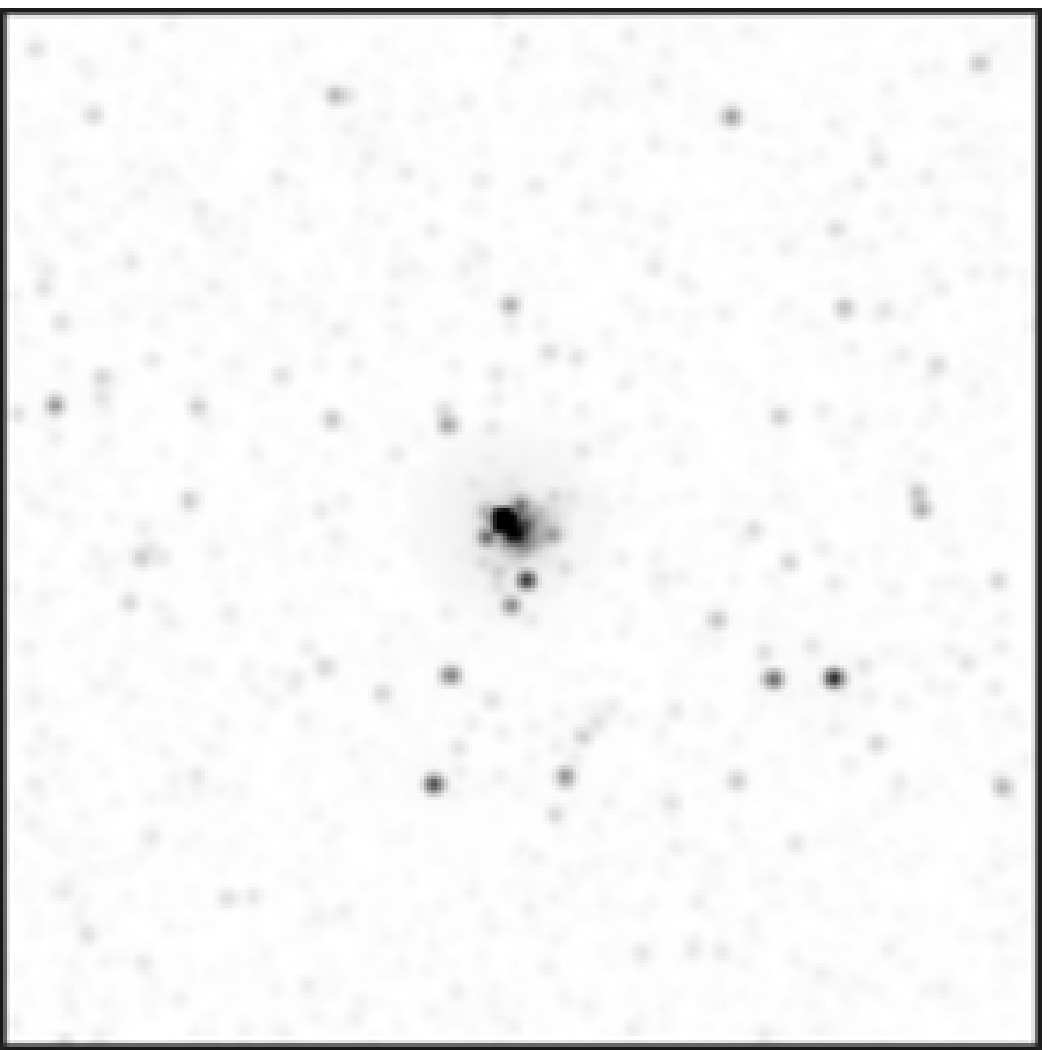}}
\subfigure[log(age/yr)=6.75]{\includegraphics[angle=0,width=0.24\textwidth]{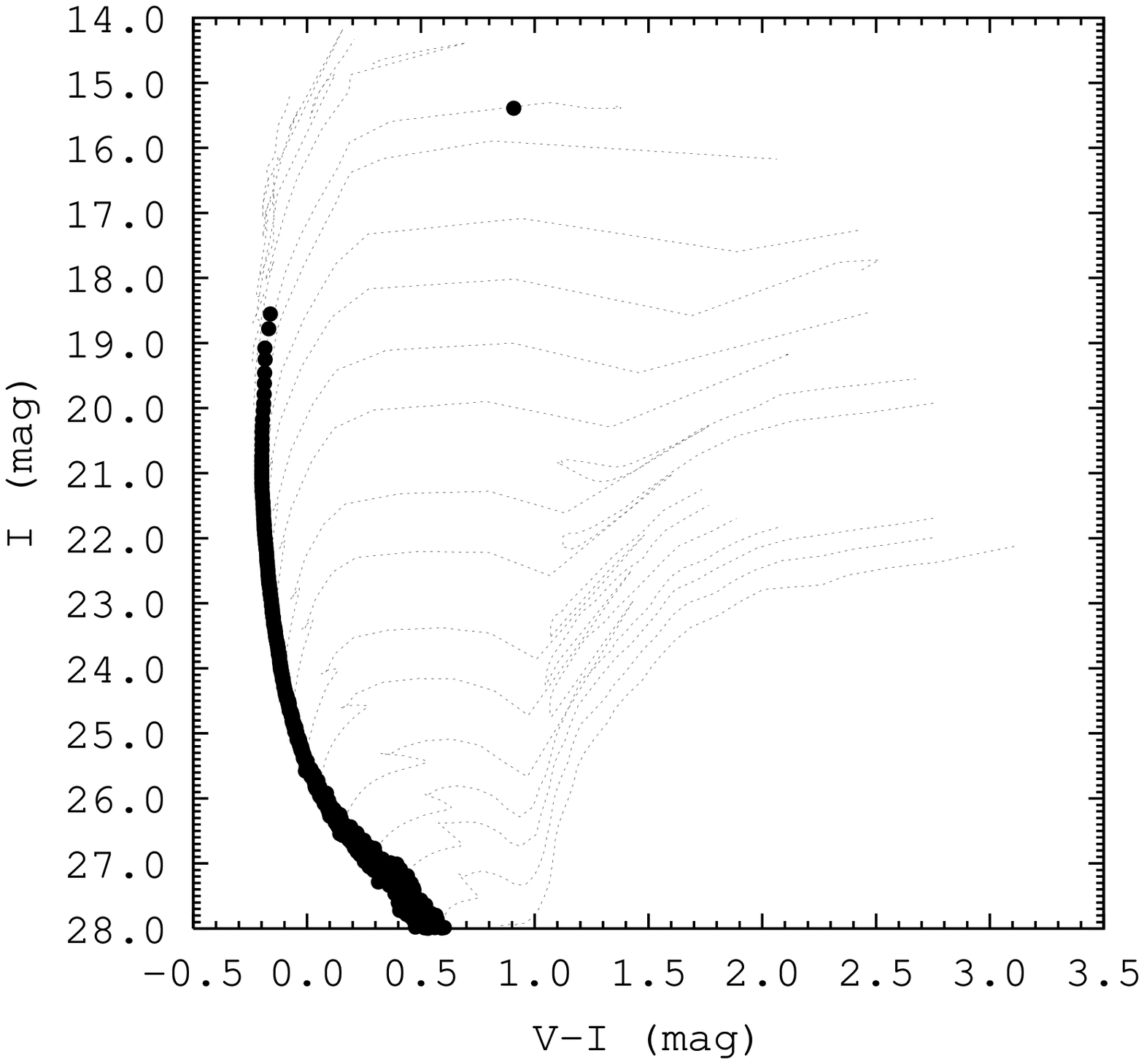}} 
\subfigure[log(age/yr)=7.00]{\includegraphics[angle=0,width=0.24\textwidth]{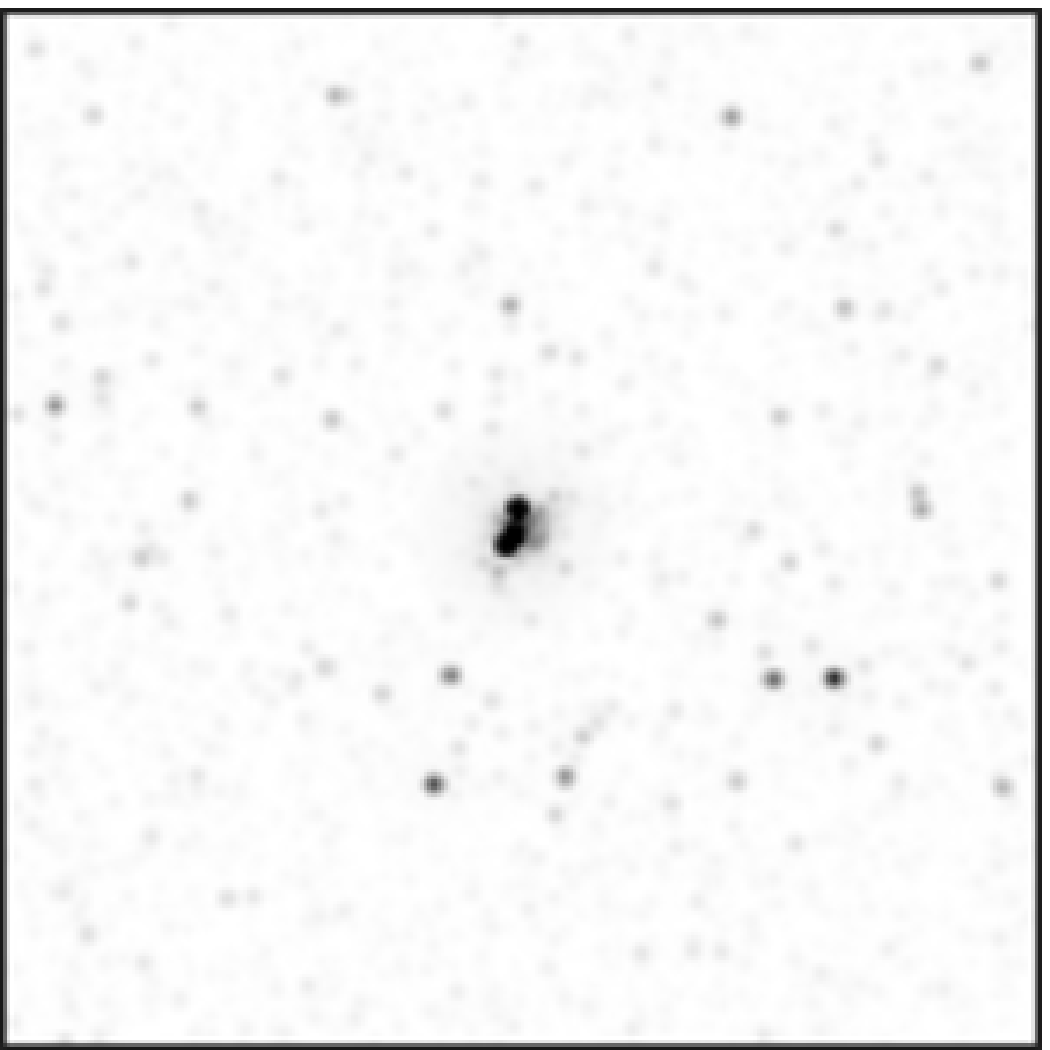}}
\subfigure[log(age/yr)=7.00]{\includegraphics[angle=0,width=0.24\textwidth]{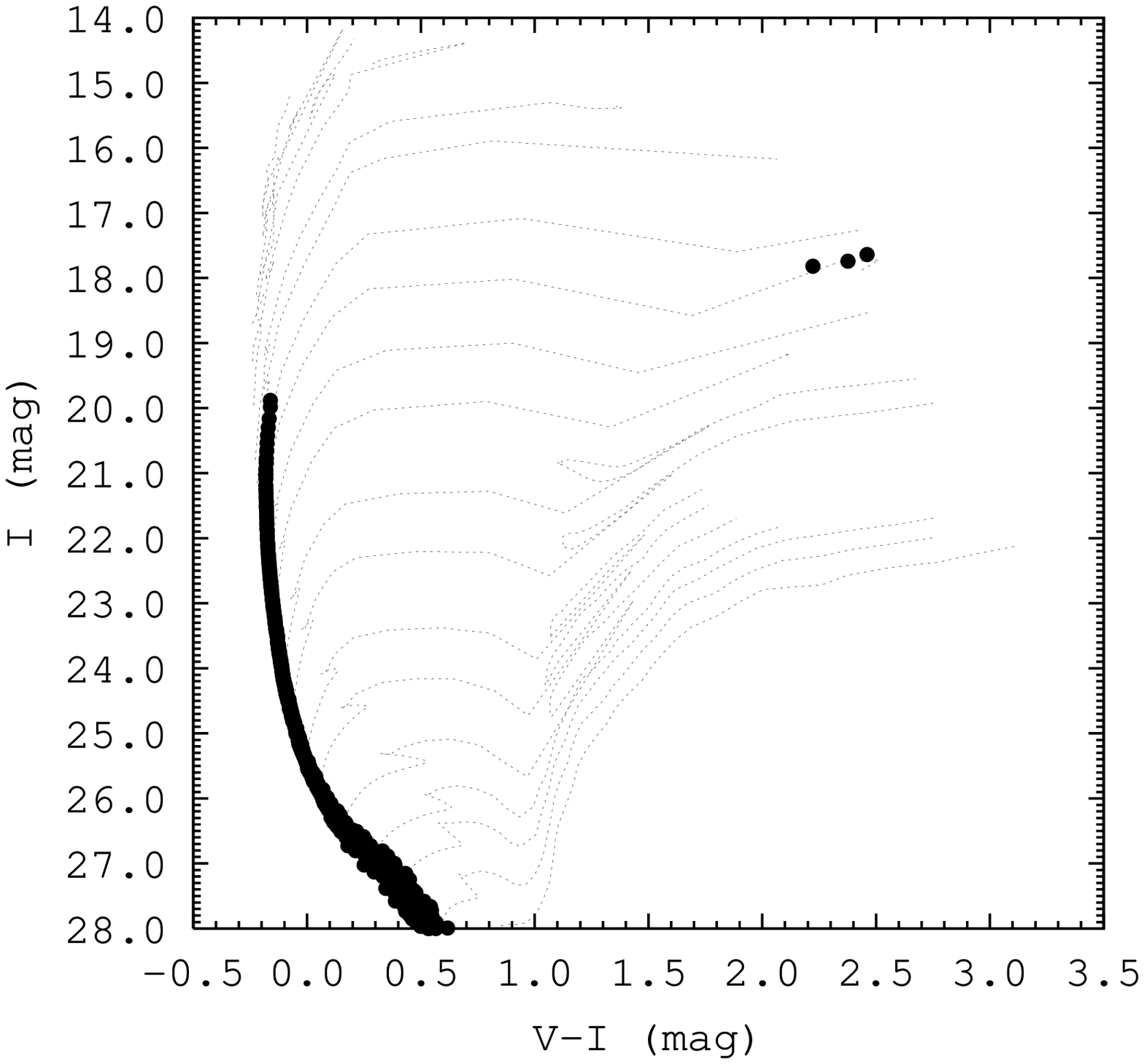}}
\subfigure[log(age/yr)=7.20]{\includegraphics[angle=0,width=0.24\textwidth]{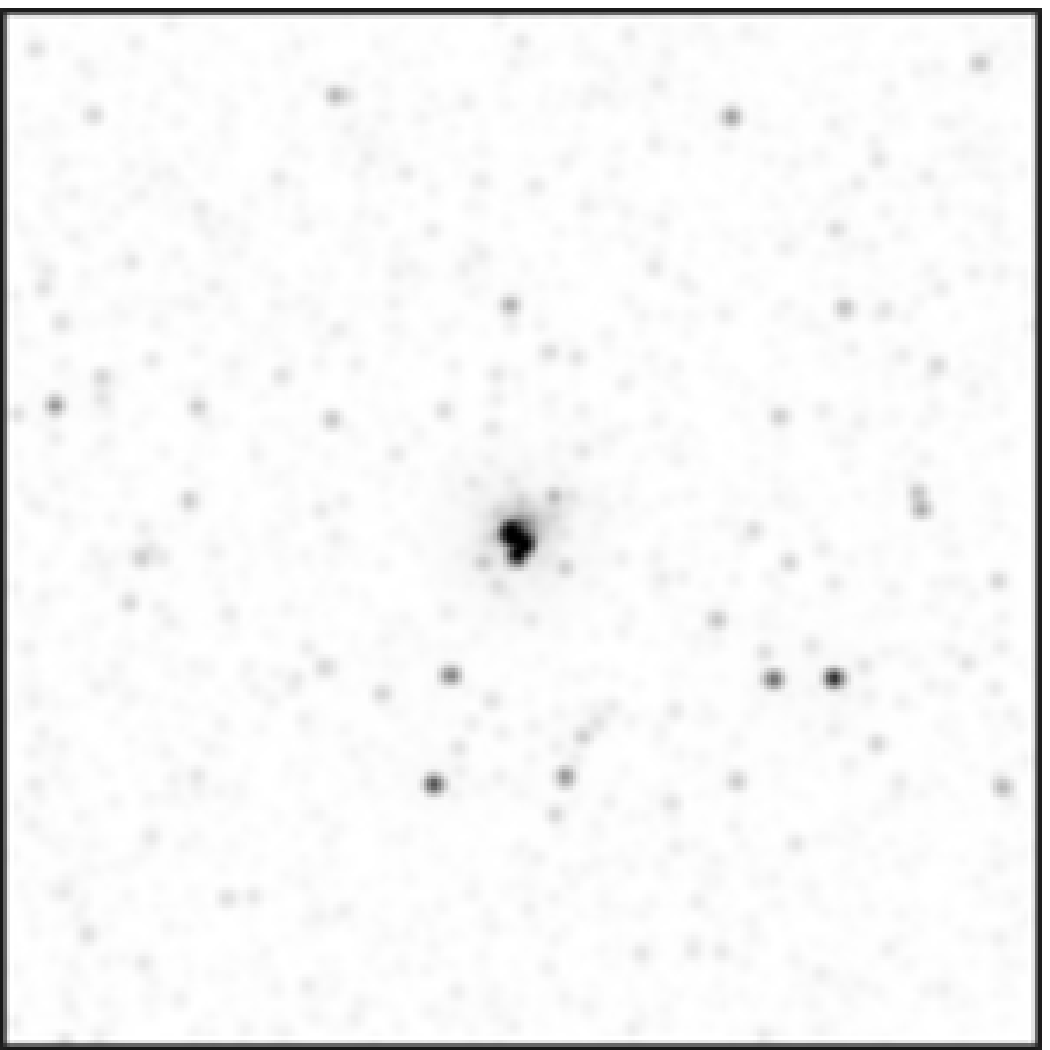}}
\subfigure[log(age/yr)=7.20]{\includegraphics[angle=0,width=0.24\textwidth]{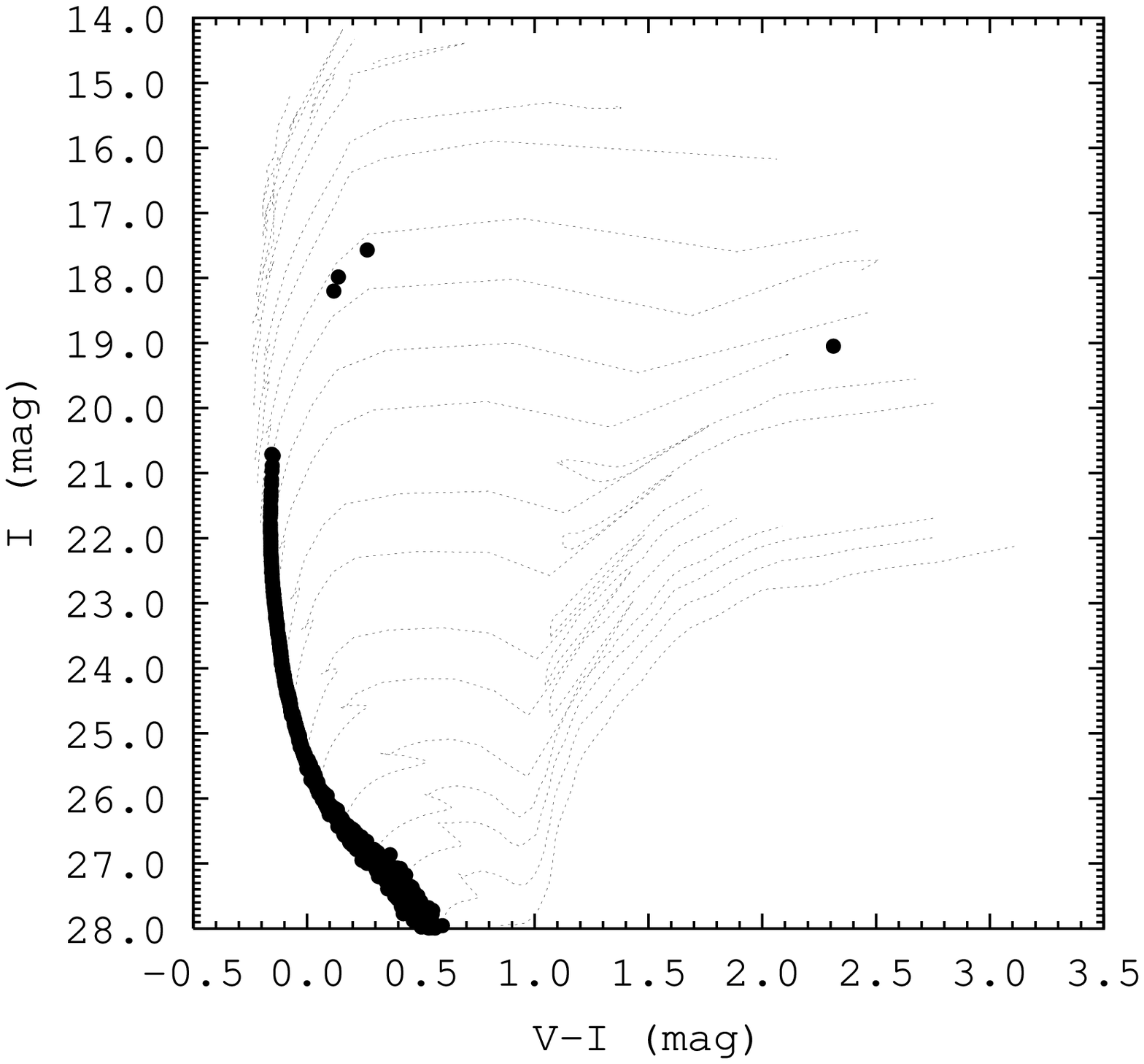}}
\subfigure[log(age/yr)=7.50]{\includegraphics[angle=0,width=0.24\textwidth]{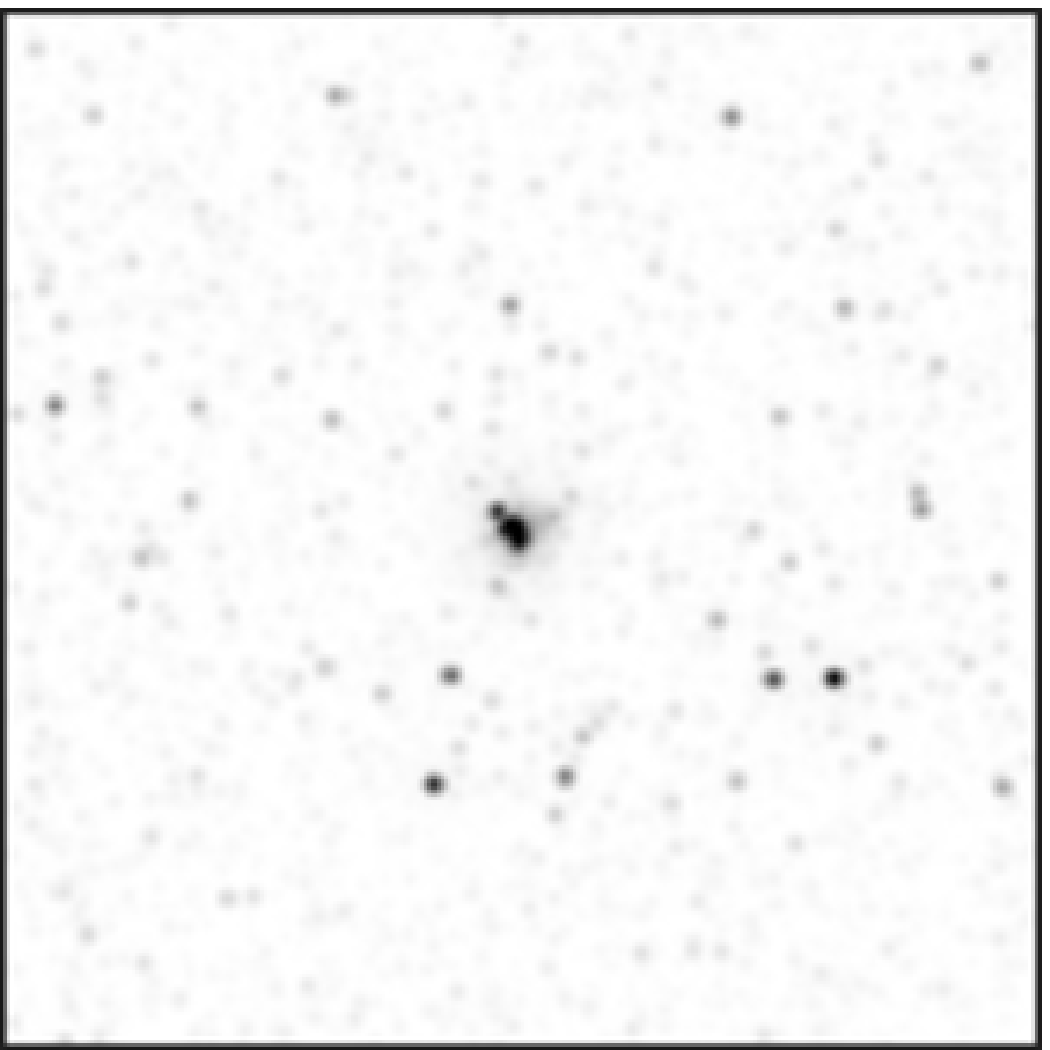}}
\subfigure[log(age/yr)=7.50]{\includegraphics[angle=0,width=0.24\textwidth]{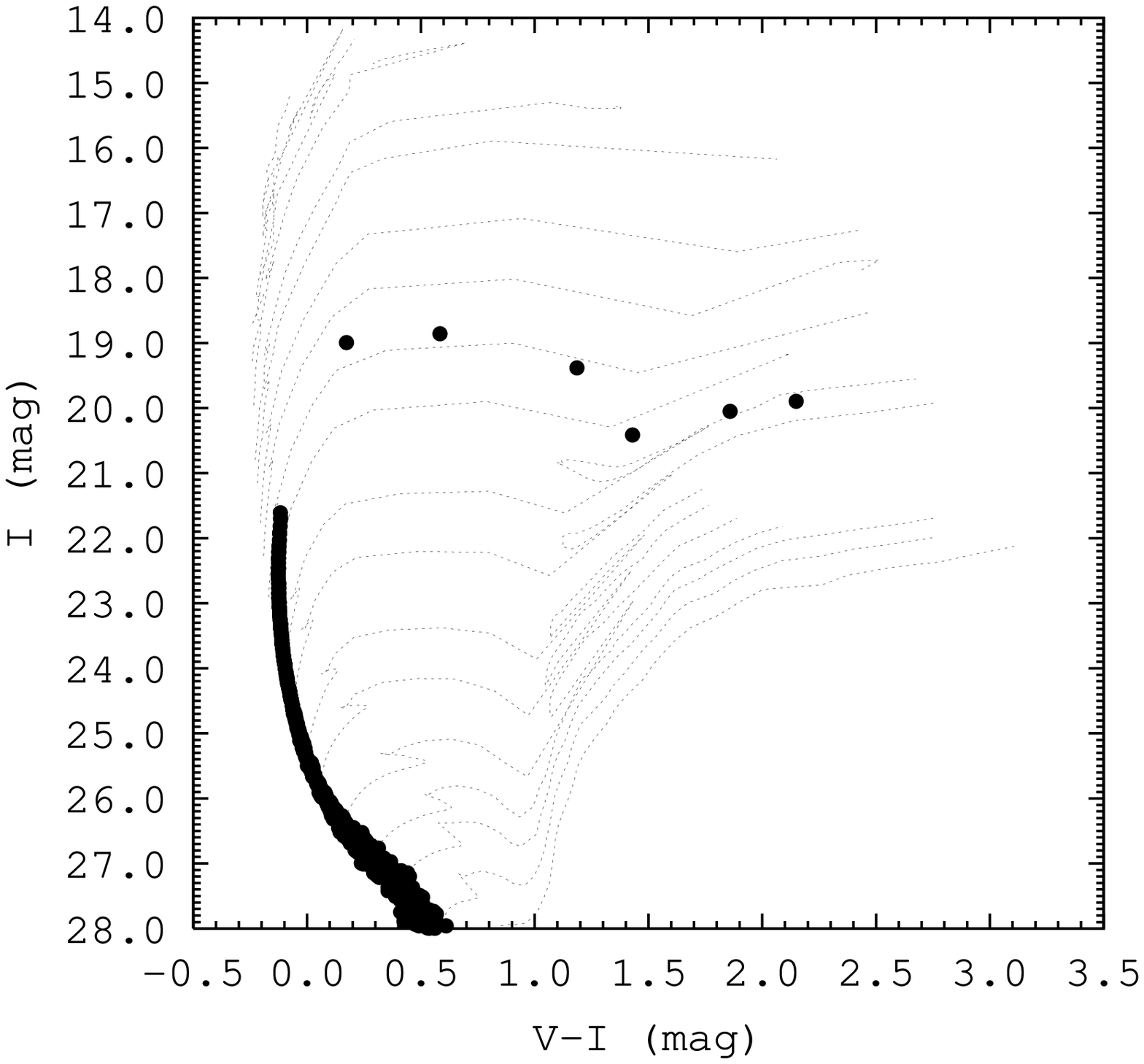}}
\subfigure[log(age/yr)=8.00]{\includegraphics[angle=0,width=0.24\textwidth]{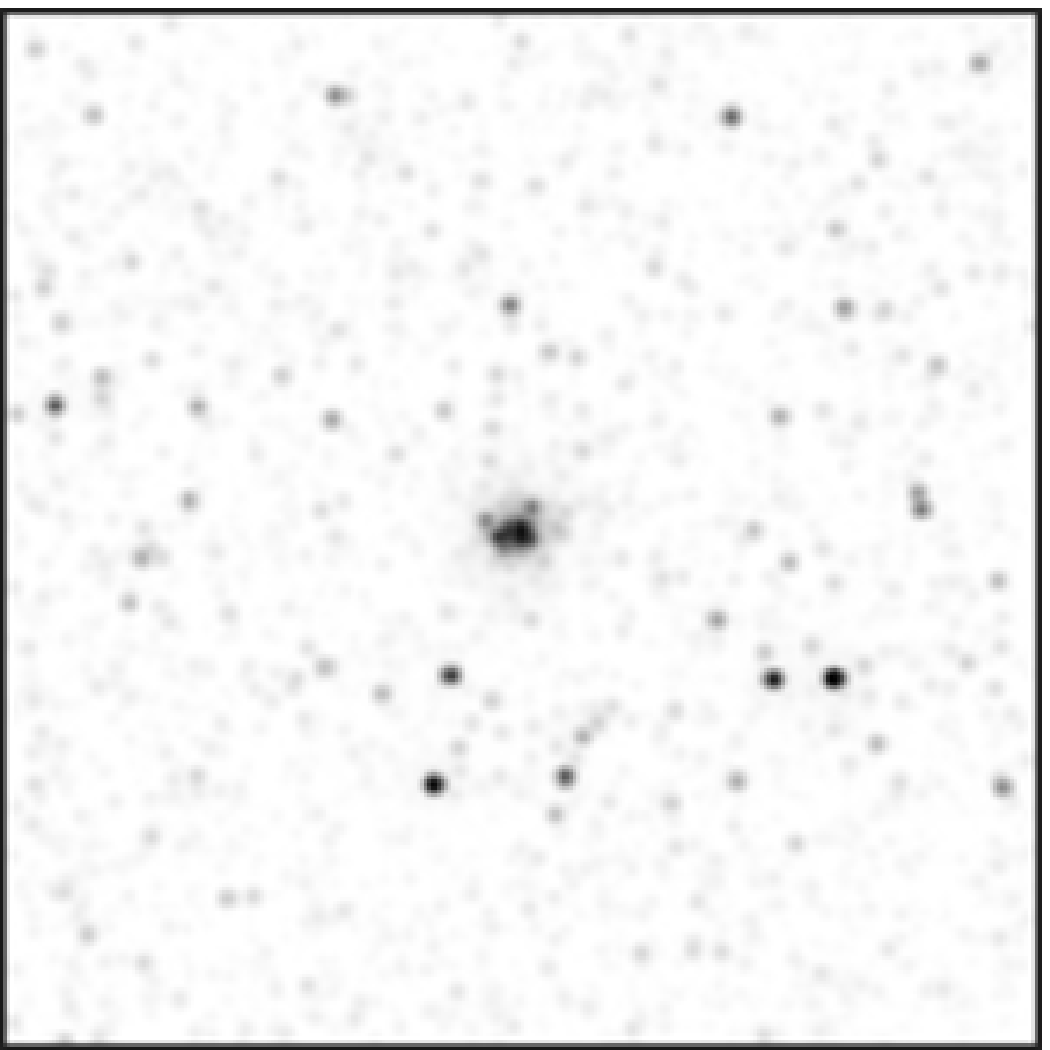}}
\subfigure[log(age/yr)=8.00]{\includegraphics[angle=0,width=0.24\textwidth]{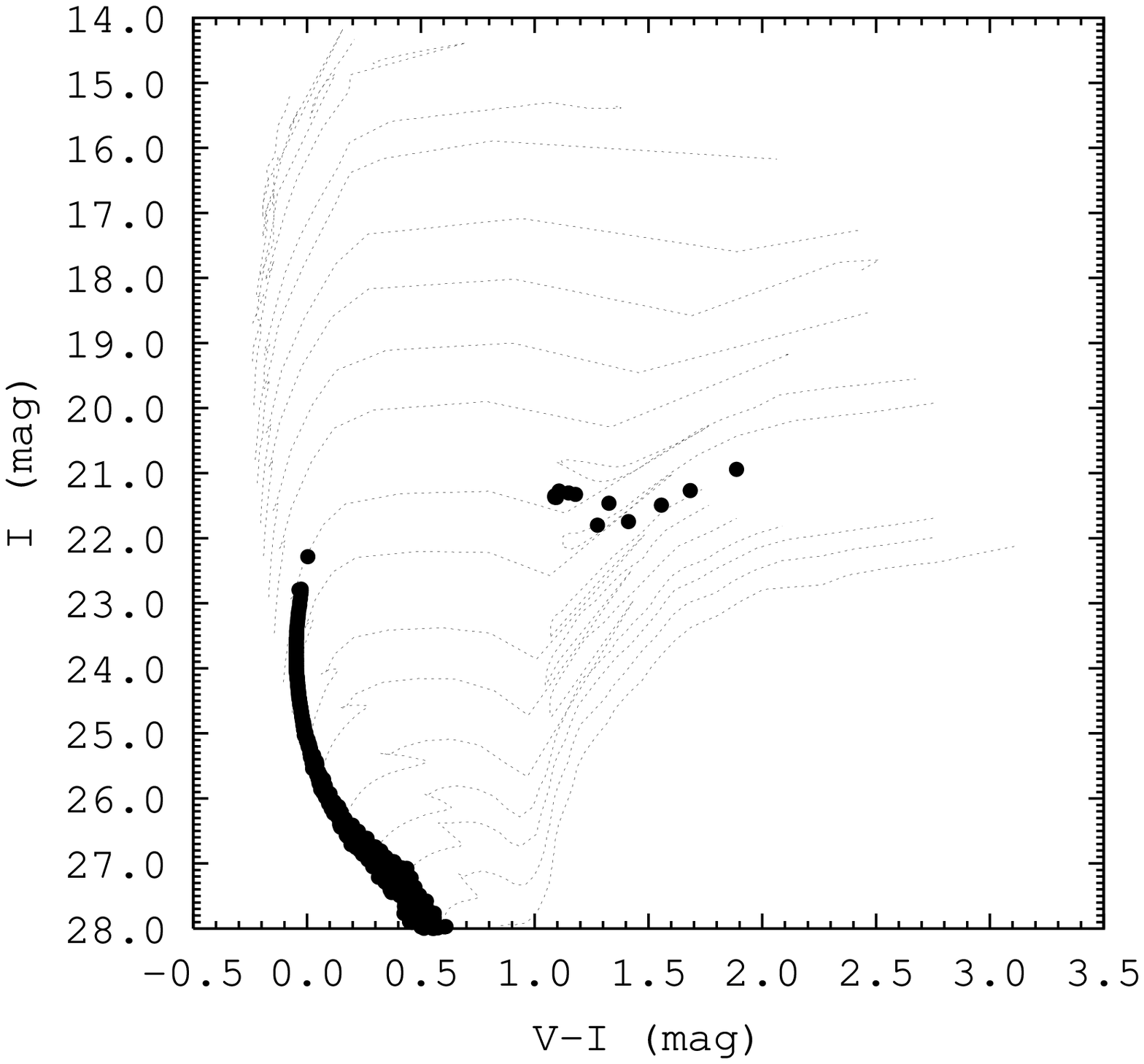}}
\caption{\small \textit{NGC 3603} in M31 (I Band), simulated images and color-magnitude diagrams.\normalsize}\label{fig:tbd04}
\end{center}
\end{figure*}

The images given in Figures \ref{fig:tbd01} through \ref{fig:tbd04} have been rendered from the original \texttt{MASSCLEAN} generated FITS files to pdf.  Thus considerable dynamic range of the magnitudes originally contained in those FITS images have been lost.  We have also been forced to select a single set of minimum and maximum brightness levels and a single slope greyscale solution to view the clusters.  This allowed us to show the relative change in the cluster's appearance with age, though the levels may not be optimal for all age periods.  This has all lead to highly degraded images as given in this paper.  However, one can recover the full dynamic range of the magnitudes given by the package by viewing the FITS files included in the electronic, on-line version of this paper.

In Figures \ref{fig:tbd01} through \ref{fig:tbd04}, we chose a static model for the spatial distribution of stars ((keeping the stars in the same location) in order to allow the viewer to easily compare the brightness of stars at different ages.  Naturally, clusters are prone to many kinematic disruptions, internal and external. Kinematics and disruption mechanisms, such as the N-body code STARLAB (\citeauthor*{starlab} \citeyear{starlab}), could be included in a future version of \texttt{MASSCLEAN} since the individual stellar masses are tracked with time.

\section{Applications of \texttt{MASSCLEAN} } \label{discussion}
\texttt{MASSCLEAN} is not meant to replace previous, widely used SSP models. In fact, in applications where spatial resolution is not available and spectral information is, \texttt{MASSCLEAN} will be an inferior choice.  However, we have identified a few astronomical questions of interest to us that drove us to create an image simulation routine, based on the tenants of the modern SSP models.  We will discuss just a few of these below.

\subsection{Selection effects in stellar cluster searches}

Researchers have been developing systematic surveys of large scale Galactic databases in the infrared to identify new, previously unknown stellar clusters.  Some of the earliest and most successful of those investigations were completed using the 2MASS (\citeauthor{2MASS} \citeyear{2MASS}) survey, first by \citeauthor*{dutra2003} \citeyear{dutra2003} and \citeauthor*{bica2003} \citeyear{bica2003}.  More recently an extensive study of the 2MASS database by \citeauthor*{Froebrich2007} \citeyear{Froebrich2007} has yielded more than 1000 new stellar cluster candidates.  The Spitzer Space Telescope GLIMPSE galactic plane survey (\citeauthor*{benjamin2003} \citeyear{benjamin2003}) and UKIDSS Galactic plane survey (\citeauthor*{lucas} \citeyear{lucas}) have recently gone public leading to additional candidate stellar cluster detections (\citeauthor*{mercer2005} \citeyear{mercer2005}).  

In uncovering the galactic open and globular cluster population, researchers have become well aware of the problem of false positives.  A sight-line with slightly reduced average extinction will peer deeper into the inner galaxy causing an observed stellar density enhancement, mimicking a cluster (\citeauthor*{cotera2003} \citeyear{cotera2003}, \citeauthor*{Froebrich2007} \citeyear{Froebrich2007}). But false positives are not the only errors important in obtaining a complete picture of the star clusters of our galaxy. There also lies the error of the second kind: the false negative.

As an example, it has been proposed that very massive clusters (mass $> 10^5 M_{\Sun}$) do not exist in the Milky Way Galaxy, either because the Milky Way is not capable of creating very massive clusters or the power law of the cluster mass function turns over (\citeauthor*{gieles2006a} \citeyear{gieles2006a},\citeyear{gieles2006b}).  Since we do not see clusters of this high a mass within our galaxy, do we accept that they do not exist?  And what if they do exist but are not detectable by present surveys?  Then we have made the error of failing to observe an object when in fact there is one. Such an error is called a false negative.  The extent of false negatives is impossible to derive directly from the clusters which are detected nor can it be estimated analytically.  An image simulation package such as \texttt{MASSCLEAN}, when used with a realistic simulation for the resolution and sensitivity of current Milky Way surveys, can be used to derive selection effects and expected yields from present search methods. 

\subsection{Constraining field-star contamination}

Beyond false negatives, efforts to identify and characterize even optically observed galactic clusters can suffer greatly from field star contamination (\citeauthor*{bonatto2006} \citeyear{bonatto2006}), particularly for low-contrast clusters. \citeauthor*{bonatto2007} \citeyear{bonatto2007} estimate that 10-20 percent of radii of open clusters may be underestimated due to the confusion brought on by high field star densities. Moreover, with time the dynamical evolution of these clusters, leading to increased mass segregation, further reduces the surface brightness at large radii.  Imaging simulations can begin to address the extent of these biases in various fields, in different bands and as a function of galactic latitude, stellar field properties and cluster mass segregation.

\subsection{\texttt{MASSCLEAN} as part of a Monte Carlo analysis}

\texttt{MASSCLEAN} determines integrated luminosities and colors, such as those shown in Figure \ref{fig:multiplot1}, based on the pure summing up of the luminosity of the individual objects in the simulation as individual objects are being tracked in the simulation.  This is quite different from the standard SSP codes which rely on a probabilistic mass function of stars to describe luminosities.  In the later method, there is no way to estimate the 'stochastic fluctuations' expected with a finite number of stars in a real cluster and the range of observed integrated colors which will naturally result (\citeauthor*{cervino2006} \citeyear{cervino2006}). The mean values derived from probabilistic models are correct on average, but they may not represent the actual values observed, particularly in clusters which lie below the Lowest Luminosity Limit as described by \citeauthor*{cervino2004} \citeyear{cervino2004} . \texttt{MASSCLEAN}, when used as part of a Monte Carlo analysis, can be used to characterize the expected distribution of integrated broad-band colors as a function of cluster mass.   We demonstrate this already in \S 3.1.

\section{Conclusion}

Our \texttt{MASSCLEAN} package has been introduced.  We provide a few first order checks that our simulations are consistent with other, well accepted, modern SSP codes.  We have demonstrated its features simulating various clusters over a mass range of a few $\times 10^{3}$ $-$ $10^{5} M_{\sun}$ and showing simulated CMDs and single-band images.  Although the package can simulate low mass clusters, we emphasize its use for massive clusters. In order to determine the mass, age and distance of a cluster, observational astronomers typically rely on the brightest stars. Perhaps no more than 10\% of the cluster is visible in typical images, particularly when extinction is high. Our approach complements the efforts of observers to take full advantage of all information obtained in their images and by simulating all of the stars (not only the visible ones). In a forthcoming paper, we will develop and present methods to derive the goodness of fit for a simulated versus a real cluster using cumulative distribution functions.


\acknowledgements
We are grateful to suggestions made to an early draft of this work by Rupali Chandar, Bruce Elmegreen, Deidre Hunter and Brad Whitmore.  Their ideas lead to significant improvements in the presentation. 
This material is based upon work supported by the National Science Foundation under Grant No. 0607497 to the University of Cincinnati. BP was supported by a graduate summer fellowship from the University of Cincinnati's Research Council.


\end{document}